\documentclass[11pt,a4paper,oneside]{article}

\usepackage{booktabs,subcaption,amsfonts,dcolumn}
\usepackage{geometry}
\usepackage[utf8]{inputenc}
\usepackage{setspace}
\usepackage{threeparttable}
\usepackage{booktabs}
\usepackage{tabularx}
\usepackage{datetime}
\newdate{date}{July}{2024}
\usepackage{caption}
\usepackage{graphicx}
\usepackage{amsmath}
\usepackage{amsthm}
\usepackage{amssymb}
\usepackage{tikz}
\usetikzlibrary{automata,arrows,positioning,calc}
\usepackage{amsthm}
\usepackage{array}
\usepackage{multirow}
\usepackage{bbm}
\usepackage{rotating}
\usepackage[round]{natbib}   
\theoremstyle{plain}

\theoremstyle{definition}

\usepackage{verbatim}
\usepackage[verbose]{placeins}
\usepackage{tikz}
\usepackage{amsmath}
\usepackage{istgame}

\theoremstyle{plain}
\usepackage{xcolor}
\usepackage[
bookmarksopen,
bookmarksdepth=2,
colorlinks=false,
citecolor    = blue,
urlcolor=blue]{hyperref}

\usepackage{subcaption}
\usepackage{cleveref}
\providecommand{\keywords}[1]{Keywords: #1}
\providecommand{\JEL }[2]{JEL Codes: #1}

\title{The Pass-through of Retail Crime\thanks{We thank Thorsten Schank, Thomas Otter, Guido Friebel, Markus Reisinger, Michael Pollmann, Frank Verboven, Sylvia Hristakeva, Aljoscha Janssen, Bart Bronnenberg and seminar participants at Tilburg University, UW Foster School of Business, NASMES, CAED, ESEM, EARIE, EMAC, and Marketing Science for helpful comments. We are grateful to Tim Haggerty, Chuck Groom, Erik Skaar, and Ian Eisenberg for invaluable insights on the cannabis industry and the traceability data.}
}
\author{{Carl Hase\thanks{Goethe University Frankfurt and JGU Mainz, carlhase@stud.uni-frankfurt.de} \quad\quad Johannes Kasinger\thanks{Tilburg School of Economics and Management - Tilburg University, j.kasinger@tilburguniversity.edu} }}

\date{October 2024}

\begin{document}

\maketitle
\onehalfspacing
\begin{abstract}

This paper shows that retailers increase prices in response to organized retail crime. We match store-level crime data to scanner data from the universe of transactions for cannabis retailers in Washington state. Using quasi-experimental variation from robberies and burglaries, we find a 1.5-1.8\% price increase at victimized stores and nearby competitors. This rise is not driven by short-to-medium-term demand changes but is consistent with an own-cost shock. Effects are larger for independent stores and less concentrated markets. We estimate that crime imposes a 1\% ‘hidden’ unit tax on affected stores, implying \$33.9 million additional social costs, primarily borne by consumers.

\end{abstract}

\noindent \keywords{Organized retail crime, public crime prevention, social costs of crime, pricing, market power, tax incidence} \\
\JEL{D22, H22, H32, H76, L11, L81 K42 }

\newpage
\section{Introduction}

Retail crime has surged to the forefront of public discourse in the United States, fueled by an apparent rise in organized retail crime.\footnote{In contrast to petty theft or shoplifting, organized retail crime involves large-scale theft of merchandise or cash with the intent to resell, often as part of a criminal enterprise. In line with this definition, we limit our analysis to robberies and burglaries. Throughout the paper, we use the terms ``organized retail crime" and ``retail crime" interchangeably. More details on organized retail crime can be found in Section \ref{sec:retail}.} 
Retail crime imposes costs on businesses, individuals, and society. 
A comprehensive assessment of these costs is central for determining the optimal level of public crime prevention \citep{stigler1970optimum, owens2021economics}. 
One factor often overlooked when considering the social 
costs of retail crime is its impact on market outcomes. In particular, retail crime-induced price changes, i.e., the pass-through of retail crime, have critical distributional implications and can introduce an excess burden by distorting firms' and consumers' decisions. Yet, evidence of a causal link between retail crime and market prices is nonexistent. 

This paper investigates the impact of organized retail crime on prices and the resulting welfare implications using the Washington state retail cannabis industry as a natural laboratory.\footnote{The US cannabis industry generates over \$25 billion in annual revenue. More than half of Americans live in states with legal retail cannabis markets, and about one-third of adults regularly consume cannabis in those states \citep{pew2024,statista2024}.} 
We choose this context for three main reasons. 
First, similar to other retail sectors, retail crime is a common occurrence at cannabis stores and has attracted the attention of policymakers \citep{ferguson2022}. Second, we use rich scanner data that captures the universe of upstream and downstream transactions for every cannabis store, matched with store-level retail crime data. This combination of industry-wide scanner and crime data, which is unique to our setting, enables a clean identification of the effects of retail crime on store-level outcomes and their underlying mechanisms. The detailed data further allows for a comprehensive welfare analysis using a sufficient statistics approach based on insights from the tax incidence literature \citep{weyl2013pass}. Third, online sales and interstate transactions are not permitted in the Washington state cannabis industry, which implies that cannabis retailers compete locally. This segregation enables us to investigate the spillover effects of retail crime from victimized stores to competitors and examine the impact on aggregate market outcomes.

Exploiting quasi-experimental variation from the timing of store-level crime incidents, we use a difference-in-differences framework and compare store-level outcomes at victimized stores to outcomes in unaffected local markets.\footnote{
In contrast to other studies exploiting spatial treatment variation, a major advantage in our setting is that we observe the universe of vertical transactions between producers and retailers. This allows us to estimate retailers' spatial sensitivity to competitors' unit cost shocks and identify a valid control group that is not contaminated by strategic price competition \citep{hollenbeck2021, muehlegger2022pass}.} We estimate that victimized stores increase prices by 1.8\% four months after a crime incident. The price increases persist and remain stable twelve months after an incident. 
We find no immediate effect on quantity sold in the months following a crime, which speaks against fear-induced demand substitution on the part of the consumers. However, quantities sold fall over a longer time horizon, consistent with consumers responding to higher prices. Additionally, we only find a negligible effect of crime incidents on wholesale costs at victimized stores.

We also investigate effects at nearby rival stores, e.g. due to strategic price responses, or an own-cost shock. We find that rivals of victimized stores---stores within a 5-mile radius of a victimized store---increase prices by a similar amount (1.5\% after four months). However, in contrast to victimized stores, this price increase materializes with a lag of around two months. Again, we find no medium-term effects on wholesale costs or quantities sold. This finding makes demand substitution an unlikely mechanism for driving the price response at rival stores. To assess whether rivals' price response is strategic, we leverage the fact that---in addition to retail transactions---we observe the universe of vertical transactions between producers and retailers. This allows us to estimate the extent to which changes in stores' own wholesale costs and the costs of their competitors are passed through to retail prices. 
Our results indicate that stores primarily adjust prices in response to their own costs rather than rivals' costs. We find a marginal cost pass-through rate of 1.67, i.e. a \$1 increase in a product's wholesale price corresponds to a \$1.67 increase in its retail price. The influence of rivals' cost changes is statistically significant but economically negligible at \$0.02 (from a \$1 increase in wholesale price). 
Thus, rivals' substantial price hikes due to nearby retail crimes cannot be attributed to a strategic price response. 


We further identify heterogeneity in price effects conditional on store and local market characteristics. 
First, we find that crime pass-through is higher at independent stores as compared to chains. 
Our findings suggest that retail crime pass-through is particularly relevant for mom-and-pop shops, while owners of multiple stores can offset costs across locations, mitigating retail crime's impact on prices. Second, we find the largest price increases for stores operating in markets with comparably low market concentrations. This finding aligns with the idea that pass-through rates and market power are inversely related, as increased competition makes prices more sensitive to marginal costs---a theoretical prediction that relies on the curvature of firms' cost functions \citep{ritz2024does} and on whether pass-through rates exceed unity \citep{miller2017}.  


Our main specification builds on the stacked difference-in-differences (DiD) estimator introduced by \cite{cengiz2019} with treatment timing defined as the store-specific month of a crime incident. Like other DiD frameworks, stacked DiD estimates the causal treatment effect under the assumption of parallel trends and no anticipation. 
To account for biases from heterogeneous treatment timing \citep{de2020two, gb2021}, stacked DiD creates crime incident-specific sub-experiments comparing treated stores to ``clean" control stores, i.e., never-treated or not-yet-treated stores for a particular crime incident. 
We choose stacked DiD over related estimators \citep[e.g.,][]{callaway2021a, Borusyak2021} because the rules for defining clean controls can be readily extended to geographic criteria. In addition to the timing-based criteria, we require incident-specific clean control stores to be located between 30 to 60 miles from the respective victimized store (which we refer to as unaffected local markets). These conditions account for confounding effects from treatment spillovers to nearby stores \citep{muehlegger2022pass} while ensuring a sizeable control group that is comparable to our treatment group.\footnote{We conduct several robustness checks that show that our main findings are not sensitive to our choice of estimator or the definition of unaffected local markets.} 
Our paper is one of the first to extend the stacked DiD framework to spatial criteria, contributing to the advancing literature on this methodology \citep{cengiz2019, deshpanade2019, butters2022, wing2024}.\footnote{Using a stacked DiD estimator, \cite{deshpanade2019} investigate spillover effects of social security field office closings to nearby offices. However, they use timing-based clean control inclusion criteria that do not incorporate geospatial distance. Our estimator combines both timing and geospatial distance in defining clean controls. \cite{butters2022} investigate spillover effects of state excise taxes on chain stores in nearby states. Their approach is similar to ours in that they exclude chain stores whose parent company is affected by a tax hike in another state within the event window.}

In our welfare analysis, we postulate that retail crime incidents constitute a positive marginal cost shock to affected stores that can be understood as a hidden crime tax. While we abstract from fixed cost shocks in our analysis, as they do not affect pricing decisions according to standard theory, we remain agnostic about the source of the marginal cost shock. This shock could arise from direct marginal costs, proportional to quantity sold, such as increased spending on security and insurance, as suggested by various industry surveys, reports, and news articles documenting retailers' efforts to combat crime \citep{uscc2023,rila2021,nrf_retail_security_2022,hs2022}. However, it may also reflect retailers' subjective perceptions of their marginal costs that could be influenced by total costs or other factors.

To derive the welfare implications of retail crime pass-through, we draw from the imperfect competition model by \citet{weyl2013pass}, which nests standard imperfect competition models, as well as the monopoly and perfect competition cases. The implications of standard tax theory carry over to our case: i) the burden of the hidden crime tax falls disproportionately on the more inelastic side of the market; ii) under imperfect competition, there is an excess burden that increases with firms' market power; iii) the pass-through rate serves as a sufficient statistic for tax incidence and the excess burden.  

Next, we employ the model---together with our marginal cost and retail crime pass-through estimates---to quantify the welfare effects of the crime-related price increases in the Washington state cannabis industry. At the average unit price, our estimated retail crime pass-through rate implies a \$0.50 unit price increase in affected markets. This unit price increase corresponds to a hidden unit tax of approximately \$0.30, or about 1\% of the average unit price. We estimate an annual decrease in consumer surplus of about \$22.8 million and a negative effect on retailers of around \$11.1 million. These results indicate an incidence of the hidden crime tax on consumers of around 67\% and a total social cost of retail crime pass-through of \$33.9 million. The estimated annual excess burden is around \$20.2 million, assuming that the entire fictional tax revenue stays within the Washington state economy.

The implications of our findings extend well beyond the cannabis industry. Organized retail crime is common in many retail sectors (see Section \ref{sec:retail}). Numerous industry surveys suggest that retailers across different industries invest heavily in strategies to prevent retail crime and pass these additional costs through to consumers \cite{hs2022,rila2021}. This conjecture is supported by the findings of a study by \citet{jackson2020}, which shows that increased felony larceny thresholds are associated with higher prices for automobiles and computers. The similarities between the retail cannabis industry and other retail settings in terms of cost structures and demand elasticities, as well as the absence of a prevalent black market competing with legal cannabis sales, further support the generalizability of our findings \citep{hollenbeck2021}. A naive extrapolation of our annual excess burden estimate, using relative sales shares and assuming retail crime impacts other retail industries similarly, suggests that the additional U.S.-wide social costs of retail crime pass-through exceed \$80 billion. While effects are likely to be less pronounced in other sectors due to different market structures and retail crime incidence, the estimate highlights that the social costs of retail crime are significantly underappreciated when not accounting for pass-through effects. 

We contribute to two broad strands of literature. First, our study contributes to the extensive literature on the economic impact of crime and the related policy discussion on the optimal level of crime prevention.
Much of this literature focuses on estimating the social costs of crime and the trade-offs between the costs and benefits of crime prevention \citep{stigler1970optimum, owens2021economics}.
We are the first to document a causal relationship between retail crime and market prices. Additionally, we provide a thorough discussion of the associated welfare implications highlighting an often neglected aspect of retail crime’s social costs. 

Previous research has linked property and violent crimes to a range of economic outcomes: increased property prices \citep{lynch2001measuring, gibbons2004costs, linden2008estimates}, economic growth \citep{fenizia2024organized}, urban depopulation \citep{cullen1999crime}, elevated savings levels \citep{de2008does}, changes in working time \citep{hamermesh1999crime}, land use and crop yields \citep{dyer2023fruits}, and even increased physical activity \citep{janke2016assaults}.
More specific to our focus, several studies examine crimes' impact on consumer behavior and business dynamics. \citet{mejia2016crime} highlight how crime reduces the consumption of conspicuous goods. \citet{fe2022} observe a crime-related decline in local food and entertainment consumption, while others identify a connection between crime and overall business activity and entrepreneurial decisions \citep{greenbaum2004impact, hipp2019fight, rosenthal2010violent}. Using increased public security expenditures as an instrument for violent crime in Columbia, \citet{rozo2018} shows that increased violence leads to lower output and prices, along with a rise in firms exiting the market. \citet{stolkin2023paying} links organized drug trafficking to higher consumer prices in Mexico, and \citet{jackson2020} find larceny thefts are positively associated with prices for computers and cars. In contrast to existing studies, we focus on organized retail crime in the U.S. and identify underlying factors behind the rise in retail prices. 

Second, we contribute to the extensive literature studying the pass-through of cost shocks, the underlying drivers, and their implications \citep{weyl2013pass, miravete2018, miller2017}. In our study, we examine a distinctive type of cost shock---retail crime---where costs are endogenous to firms updating their beliefs about the probability of future crime risks and outcomes. Our analysis suggests that rivals' price adjustments are less a strategic response and more a reaction to their own increased marginal costs (e.g. due to precautionary security or insurance expenditures). This pattern could reflect a kind of spatial knowledge or learning spillover \citep{thornton2001learning, audretsch2004knowledge, bloom2013}. 
Such spillovers are not only relevant in the context of retail crime but also in other settings where firms compete in local markets and learning from competitors is central to firms' costs. In these cases, price changes driven by spillovers could be mistakenly interpreted as strategic pricing. 

Our heterogeneity analyses also contribute to the growing literature on asymmetric strategies and market outcomes between chain and independent stores \citep[e.g.][]{jia2008happens, hollenbeck2017economic, hollenbeck2022winning, janssen2023retail, klopack2024one}. We find that the pass-through of retail crime is considerably smaller for chain stores, aligning with studies that show a lack of within-chain price adjustments to local conditions \citep{hitsch2021prices, dellavigna2019uniform}. From a welfare perspective, this pattern suggests a potential benefit to increasing the presence of chain stores within an industry. Additionally, our finding of higher pass-through rates in less concentrated markets adds to the ongoing discussion about the relationship between market competition and pass-through rates \citep{miller2017, cabral2018larger, genakos2022competition, ritz2024does}.

This paper proceeds as follows. Section \ref{sec:instution} describes the institutional context of our study, including retail crime in the United States and the retail cannabis industry. Section \ref{sec:data} details our data and Section \ref{sec:strategy} describes our main empirical strategy. Section \ref{sec:results} presents our main findings. Section \ref{sec:robustness_checks} discusses potential endogeneity concerns and robustness checks. Section \ref{sec:policy} outlines our policy analysis and derives the welfare implications. Section \ref{sec:discussion} concludes.

\section{Institutional Background}\label{sec:instution}

\subsection{The cannabis industry in Washington state}

Approximately 50\% of U.S. states have legal recreational cannabis markets. 
Washington state's cannabis market opened in July 2014 for adults 21 years and older.
Cannabis has since become one of the largest agricultural industries in the state, contributing \$1.85 billion to gross state product \citep{wsu2020}. 
Cannabis consumption is widespread, with approximately 30\% of Washington adults consuming cannabis on a monthly basis \citep{wadoh}. 
Consumption is relatively equal across race/ethnicity, education, and gender, but decreases monotonically with income and age. 
We detail the demographic characteristics of cannabis consumers in Appendix \ref{sec:industry}. 

The industry is regulated by the Washington State Liquor and Cannabis Board (LCB) which offers separate business licenses for upstream and downstream establishments  \citetext{Washington State Legislature, Title 314, Chapter 55}. Producer-processors (i.e. upstream establishments) can cultivate, harvest, and process cannabis but cannot sell to end consumers. Retailers (i.e., downstream establishments) can purchase fully packaged and labeled products from producer-processors and sell them in retail stores. Producer-processors cannot own retail licenses and vice versa, creating complete vertical separation along the supply chain. 
Retailers can only buy from producer-processors located in Washington state and producer-processors can only sell to retailers in the state. This seals off the core of the supply chain from other U.S. states with legal recreational markets. Retailers cannot sell online, meaning retailers operate brick-and-mortar stores. Appendix \ref{sec:industry} contains additional information on the cannabis supply chain. 

Cannabis business licenses are capped by the LCB at 556 retailers and 1,426 producer-processors \citep{lcb2020}. Licenses are granted at the establishment level so that a single firm can own several licenses. However, a firm can only own licenses of the same type. Approximately 65\% of retail stores belong to one- or two-store firms; 25\% of stores belong to 3-5 store chains; less than 11\% of stores belong to chains with more than 5 stores.\footnote{When the market was created in 2014, the LCB allocated licenses according to a lottery. Since a single firm could apply for more than one license, the lottery created exogenous variation in firm size \citep[see e.g.][]{hollenbeck2022winning}.} Not all licenses are actively in business, meaning that some license holders have not opened an establishment and have no reported sales activity, especially at the producer-processor level. During our sample period, there were 511 active retailers and 692 active producer-processors. 

The LCB distributes retail licenses to counties according to population density but there are no restrictions on where producer-processors can be located. Retailers are located in 37 of the 39 counties in Washington state. The average cannabis consumer in Washington is located approximately five miles from the nearest retailer \citep{ambrose2021}. The geographic distribution of retail stores is illustrated in Figure \ref{fig:locations}, Panel A. 

Retail sales are subject to a 37\% sales tax but there is no tax on upstream sales. Per month, retailers sell approximately 14,500 units and earn about \$280,000 in (tax-inclusive) revenue (see Table \ref{tab:product_desc_stats}).\footnote{For context, the average cannabis retailer generates about twice the revenue of an average convenience store or one-fifth of an average supermarket in the United States \citep{statista2022,statista2024}.} 
For more information on the distribution of store characteristics, see Appendix \ref{sec:sales_across_stores}.

Table \ref{tab:product_desc_stats} provides an overview of the cannabis product market. Retail stores sell a variety of cannabis products---around 470 distinct products per month on average. The LCB classifies products according 12 categories \citep{lcb2015}. Usable marijuana (i.e. unprocessed dried flower) and concentrate for inhalation (e.g. for use in vape pens) account for more than 80\% of all retail sales. Another 14\% of retail sales comes from solid edibles (chocolate bars, gummies, etc), liquid edibles (soda and other infused drinks), and infused mix (e.g. pre-roll joints infused with concentrates). The remaining categories make up less than 2\% of total revenue; these are topical products (e.g. creams and ointments), packaged marijuana mix (e.g. pre-roll joints), capsules, tinctures, transdermal patches, sample jar, and suppository. 

\begin{table}[!htbp]
\centering
\caption{Product market descriptive statistics}
\begin{subtable}[h]{0.8\textwidth}
\centering
\caption{ Retail store characteristics}
\renewcommand{\tabcolsep}{1pt}{
\def\sym#1{\ifmmode^{#1}\else\(^{#1}\)\fi}
\begin{tabular*}{\hsize}{@{\hskip\tabcolsep\extracolsep\fill}l*{2}{c}}
\toprule
            
&\multicolumn{1}{c}{\parbox{4cm}{\centering Monthly average \\ per store}}&\multicolumn{1}{c}{\parbox{2cm}{\centering Sample \\ total}} \\
\midrule

Establishments      &     & 511  \\

\addlinespace
Units sold      &  15,157  & 263 million \\
    &  (5,829) &   \\
\addlinespace
Distinct products    &   470   &   210,842 \\
                & (305) & \\
\addlinespace

Sales      &   \$282,857  &  \$5.07 billion  \\
            & (\$273,082) & \\

\addlinespace

\bottomrule
\addlinespace
\end{tabular*}

\begin{minipage}[h]{\textwidth}
\footnotesize \emph{Notes:} Column 1 reports monthly averages at the store level during the sample period. Standard deviations are in parentheses. Column 2 reports totals across all stores and months in the sample period. Sales are tax-inclusive.
\end{minipage}

}
\end{subtable} \\
\begin{subtable}[h]{0.8\textwidth}
\caption{Retail product categories}
\renewcommand{\tabcolsep}{1pt}{
\def\sym#1{\ifmmode^{#1}\else\(^{#1}\)\fi}
\begin{tabular*}{\hsize}{@{\hskip\tabcolsep\extracolsep\fill}l*{2}{c}}
\toprule
            
\multicolumn{1}{l}{Product category} &\multicolumn{1}{c}{\parbox{4cm}{\centering Monthly sales \\ (in millions of \$)}} &\multicolumn{1}{c}{\parbox{2cm}{Market share}} \\
\midrule

Usable marijuana   & \$58.77   &     0.52   \\

\addlinespace

Concentrate for inhalation  & \$34.70    &   0.31   \\
\addlinespace
Solid edible    & \$8.45
 &   0.08    \\
\addlinespace
Infused mix   & \$5.40  & 0.05      \\
\addlinespace
Liquid edible  & \$2.96     &   0.03    \\
\addlinespace
Other      & \$2.16 &   0.02  \\

\bottomrule

\end{tabular*}

\begin{minipage}[h]{\textwidth}

\footnotesize \emph{Notes:} Column 1 reports the average monthly retail cannabis sales (in millions of dollars) across Washington state for the major product categories during the sample period; Column 2 shows the corresponding market shares. Sales are tax-inclusive.
\end{minipage}
}    
\end{subtable}
\label{tab:product_desc_stats}
\end{table}

\subsection{Organized retail crime}\label{sec:retail}

\subsubsection*{Organized retail crime in the United States}

Organized retail crime is defined as large-scale theft of retail merchandise or cash, typically by two or more people, often as part of a criminal enterprise. While the exact definition varies across law enforcement agencies and jurisdictions, a defining characteristic of organized retail crime is that merchandise is not stolen for personal use but is instead resold through third-party outlets. This contrasts with petty theft (e.g. stealing cosmetics because one cannot afford it), which is not the focus of our study. For ease of notation, we refer to \emph{organized retail crime} and \emph{retail crime} interchangeably throughout the paper. 

In 2022, the U.S. Chamber of Commerce declared organized retail crime a ``national crisis" \citep{uscc2022}. According to the National Retail Security Survey (NRSS), organized retail crime is responsible for nearly \$5 billion in annual losses \citep{nrf2020}. Organized retail crime is pervasive in scope, with criminals targeting a variety of stores and goods 
\citep{hs2022}. 
Incidents are often violent in nature and threaten employee and customer safety.\footnote{In 2023, 81\% of NRSS respondents reported an increase in retail crime-related violence against employees \citep{nrf2023}. In a separate survey, more than three-quarters of retailers stated that a criminal had threatened to use a weapon against an employee, while 40\% of Asset Protection Managers reported incidents where an organized retail criminal used a weapon to inflict harm on an employee \citep{rila2021}.} Many retailers invest heavily in strategies to prevent retail crime, including third-party guard services, enhanced surveillance technologies, locking cases, and employee safety and de-escalation training \citep[see e.g.][]{target2023,cnbc2019,cnbc2022}.\footnote{In 2023, for example, 34\% of National Retail Security Survey respondents increased payroll to support security and 45\% increased the use of third-party security personnel as a measure of crime prevention \citep{nrf2023}.} 

Many states have enacted legislation specifically targeting retail crime, including stiffer penalties for people caught stealing from stores with the intent to resell merchandise, and adding language targeting organized groups that rob multiple retail outlets \citep{mp2023}. 
In addition to legislation, numerous local, state, and federal law enforcement agencies have created retail crime task forces across the United States \citep{ferguson2022,hs2022}.

\subsubsection*{Organized retail crime in cannabis}

\begin{figure}[!htbp]
\centering
		\caption{Comparing overall robberies to cannabis retail crime in Washington state}

	    \includegraphics[width=.5\linewidth]{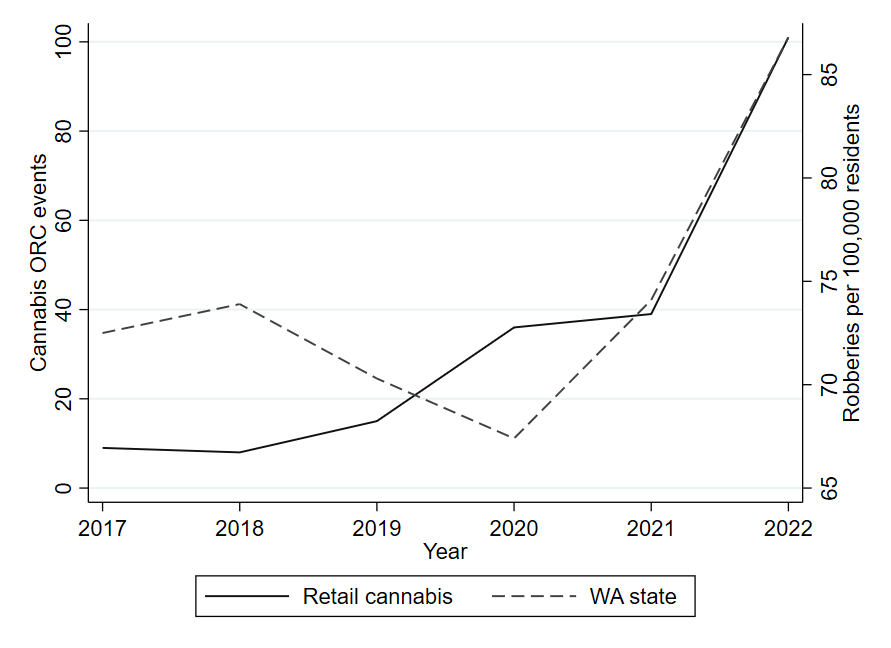}
\label{fig:wa_crime}
\par 
\rule{\textwidth}{0.4pt}
\begin{minipage}[h]{\textwidth}
\medskip
\footnotesize \emph{Notes:} The figure displays annually reported organized retail crime incidents in the Washington state cannabis industry (solid line) and robberies per 100,000 inhabitants in Washington state (dotted line) from 2017 through 2022. Data sources: Uncle Ike's Robbery Tracker for cannabis industry incidents and the FBI Crime Data Explorer for state-wide robbery statistics.
\end{minipage}
\end{figure}

Similar to other retail sectors, retail crime is pronounced at cannabis stores. Between 2017 and 2023 there were 210 reported robberies and burglaries at cannabis retailers in Washington state. Figure \ref{fig:wa_crime} illustrates that retail crime in cannabis has increased over time and that the trend largely tracks overall robberies in Washington state. This trend has prompted policymakers to take action. Retail cannabis was cited by the Washington state attorney general when creating the state's retail crime task force \citep{ferguson2022}, and in 2023 a senate bill was introduced in the state legislature aimed at providing tax relief for security improvements at cannabis retail stores \citep{wa2023}.

Organized retail crimes in cannabis are predominantly armed robberies that occur during store hours. Perpetrators typically extract cash from the register, demand that employees give them access to the safe, or clear shelves of merchandise. Since cannabis remains illegal at the federal level, cannabis stores do not have access to electronic payment processing services like Visa and Mastercard and nearly all retail transactions are cash-based.\footnote{Cannabis stores have access to other financial services such as commercial bank accounts \citep{dfi2022}.} Stores thus handle a large amount of cash, making them prime targets for retail crime. 

Robberies are often violent: in several instances, consumers or employees have been temporarily taken hostage, physically assaulted, shot at, or even killed. Victimized stores often respond to retail crime incidents by increasing security expenditures. According to the state's Organized Retail Crime Task Force, interviews with store owners, and numerous news reports, the most common expenditure involves hiring additional security guards \citep{lcb2022,st2022a,komo2022}. Since contracted security guards often cost between \$75-100 per hour in Washington state, this amounts to a sizeable labor cost shock for cannabis retailers \citep{st2022a,komo2022}. Further preventative measures include installing panic buttons, two-way door systems, and investing in employee de-escalation training.\footnote{Stores are required by law to have comprehensive round-the-clock video surveillance.}

\section{Data}\label{sec:data}

\subsection{Price data}

To monitor developments in the cannabis market, legalization came with stringent data reporting and sharing requirements for cannabis businesses. Retailers must record all sales and regularly upload data feeds to the LCB. Compliance with data reporting is strictly enforced by the LCB. When a business is issued a violation, it can receive a fine, a temporary license suspension, or both. In cases of repeated violations, a license can be revoked by the LCB board. Given such strict enforcement, violations are uncommon. In 2022 for example, the LCB issued 63 violations among over 1,100 active licensees \citep{lcb_violations}. 

The data, which is reported weekly, contains detailed information on the price and quantity of each product sold by a producer-processor to a retailer, and the subsequent price and quantity of that very same product sold at the retail level. We refer to the price charged by a producer-processor as the \emph{wholesale} price, and the price charged by retailers as the \emph{retail} price. This reflects that producer-processors resemble wholesalers when viewed from the perspective of retailers. The data captures granular product differentiation. For example, in our data, a 1.0-gram package and a 2.0-gram package of the same usable marijuana strain produced by the same producer-processor are treated as different products in the data.

The LCB switched providers for its traceability system in October 2017 and again in December 2021, creating two structural breaks in the price data. Our sample period lies between these breaks and spans March 2018 through December 2021. We obtained the data from Top Shelf Data, a data analytic firm that ingests the raw tracking data from the LCB and matches it with additional product information. The data covers the universe of sales from all 511 active retail establishments during the sample period. Per month, retailers sell approximately 14,500 units and earn about \$280,000 in tax-inclusive sales revenue (see Table \ref{tab:product_desc_stats}). Over the entire sample period, the data contain \$5.07 billion in tax-inclusive retail sales. All retail prices and revenues reported in this paper are tax-inclusive. 

To estimate pass-through semi-elasticities, we follow previous studies \citep[e.g.,][]{renkin2020,leung2021,lindner2019} and use as our dependent variable the natural logarithm of the monthly store-level price index. The log price index measures the price inflation for store $j$ in month $t$, and is denoted as $\pi_{j,t}$:
\begin{equation}
    \pi_{j,t} = \ln I_{j,t}, \, \text{with} \, I_{j,t} = \prod_c I_{c,j,t}^{\omega_{c,j,y(t)}}
\end{equation}
 $I_{j,t}$ is an establishment-level Young price index that aggregates price changes across product subcategories $c$, where each subcategory is a unique category-unit weight combination. The index weight $\omega_{c,j,y(t)}$ is the revenue share of subcategory $c$ in establishment $j$ during the calendar year of month $t$.\footnote{Price indexes are often constructed using lagged quantity weights. Since product turnover is high at retail stores, lagged weights would limit the number of products used in constructing the price indexes \citep{renkin2020}. Thus, contemporaneous weights are used.} The dependent variable is equivalent to the first difference of the log of the weighted store price level between month $t$ and $t-1$. 
 
 Store-level indices are common in the literature on retail price movements and carry several advantages over more disaggregated product-level price data. First, retail crime occurs at the store level, making the store a natural unit of analysis. Second, a store-level index allows the researcher to weigh products by their importance for each store. Finally, entry and exit occur at a much higher frequency for products compared to stores, and a product-level time series would contain frequent gaps. Since the vast majority of cannabis businesses have succeeded at staying in business, the store-level panel is much more balanced. We describe the store-level price index in more detail in Appendix \ref{sec:indexes}.

Besides prices, we are also interested in the effect of retail crime on other store-level outcomes. 
First, we consider the possibility that consumers substitute out of victimized and into nearby rival stores following crime incidents (e.g. due to increased prices or fear of physical harm). To investigate this, we construct a store-level quantity index measuring the monthly percent change in quantity sold. The quantity index is constructed similarly as the price index with index weights based on annual revenue shares (see Appendix \ref{sec:indexes}).

It is also possible that stores adjust their wholesale expenditures to offset the costs of crime. Wholesale costs are particularly important since the Cost of Goods Sold (COGS) accounts for 80\% of cannabis retailers' variable costs (see Appendix \ref{sec:industry}), similar to other retail sectors \citep{renkin2020}. An advantage in our setting is that we directly observe wholesale prices and quantities at the store-product-month level. We construct a wholesale cost index that captures the month over month percent change in wholesale prices paid by a store. The index weights are based on retailers' annual wholesale expenditures (rather than annual revenue as with the price index).\footnote{In a robustness check, we construct monthly (rather than annual) expenditure weights which capture potential wholesale substitution patterns on the part of retailers. Results are similar to our main specification that uses annual weights. See Section \ref{sec:robustness_checks} for details.} 

Table \ref{tab:dep_vars} provides descriptive statistics for our indices. The price and wholesale cost indices are centered around zero and have means close to zero. Standard deviations for these indices range from 0.002 to 0.029. The quantity index has a larger standard deviation than the other indices. This is a common characteristic of quantity indices constructed from store-level scanner data, and is similar to what is found elsewhere in the literature \citep[see e.g.][]{renkin2020}. The distributions of the indices can be found in Appendix Figure \ref{fig:index_distributions}.

\begin{table}[!htbp]
\centering
\caption{Dependent variable descriptive statistics}
\centering
\renewcommand{\tabcolsep}{1pt}{
\def\sym#1{\ifmmode^{#1}\else\(^{#1}\)\fi}
\begin{tabular*}{\hsize}{@{\hskip\tabcolsep\extracolsep\fill}l*{3}{c}}

\midrule
              & {\parbox{2cm}{\centering Mean}} & {\parbox{2cm}{\centering SD}} & {\parbox{2cm}{\centering Median}} \\ 
\midrule
\addlinespace
Log price index & -0.001 & 0.029 & 0 \\ 
Log quantity index & -0.008 & 0.21 & -0.012 \\ 
Log wholesale cost index & 1.18e-05 & 0.002 & 0 \\ 

\addlinespace

\bottomrule
\addlinespace
\end{tabular*}
\begin{minipage}[h]{\textwidth}
\footnotesize \emph{Notes:} This table presents the mean, standard deviation, and median of our main dependent variables: Log price index, Log quantity index, and Log wholesale cost index.
\end{minipage}

}
\label{tab:dep_vars}
\end{table}

\subsection{Retail crime data}

Similar to other industries, crime at retail cannabis stores is not formally tracked or aggregated by law enforcement agencies. However, on behalf of the market at large, Uncle Ike’s, a retailer and member of Washington’s Craft Cannabis Coalition of over 70 small businesses, maintains a public database of retail crime incidents reported by businesses, law enforcement, and the news media. While the database offers a reasonably accurate list of crimes reported to local police departments, it is not an official census of retail crime at cannabis stores. We view this as unproblematic as each incident in our sample period is cross-referenced with either a police report, police case number, or a news article that references a police investigation.\footnote{ As an example, the database's tally aligns perfectly with the Bellevue Police Department's data on armed robberies at cannabis retailers in the city of Bellevue \citep{st2022a}.} This makes it highly unlikely that treatment is falsely assigned to a store that is, in fact, untreated. Conversely, it is possible that not all crimes are reported to the police or news media and, hence, do not appear in the database. In that case, treated stores would potentially be misclassified as controls, leading to attenuation bias and conservative treatment effect estimates. 
However, since the control group is much larger than the treatment group in our setting, misclassified controls will be dominated by correctly classified controls. As a result, we expect attenuation bias from misclassified controls to be minimal.

\begin{figure}[!htbp]
\caption{Retail cannabis stores in Washington state}
    \centering
	\begin{subfigure}{\textwidth}
	\centering
	    \includegraphics[width=.75\linewidth]{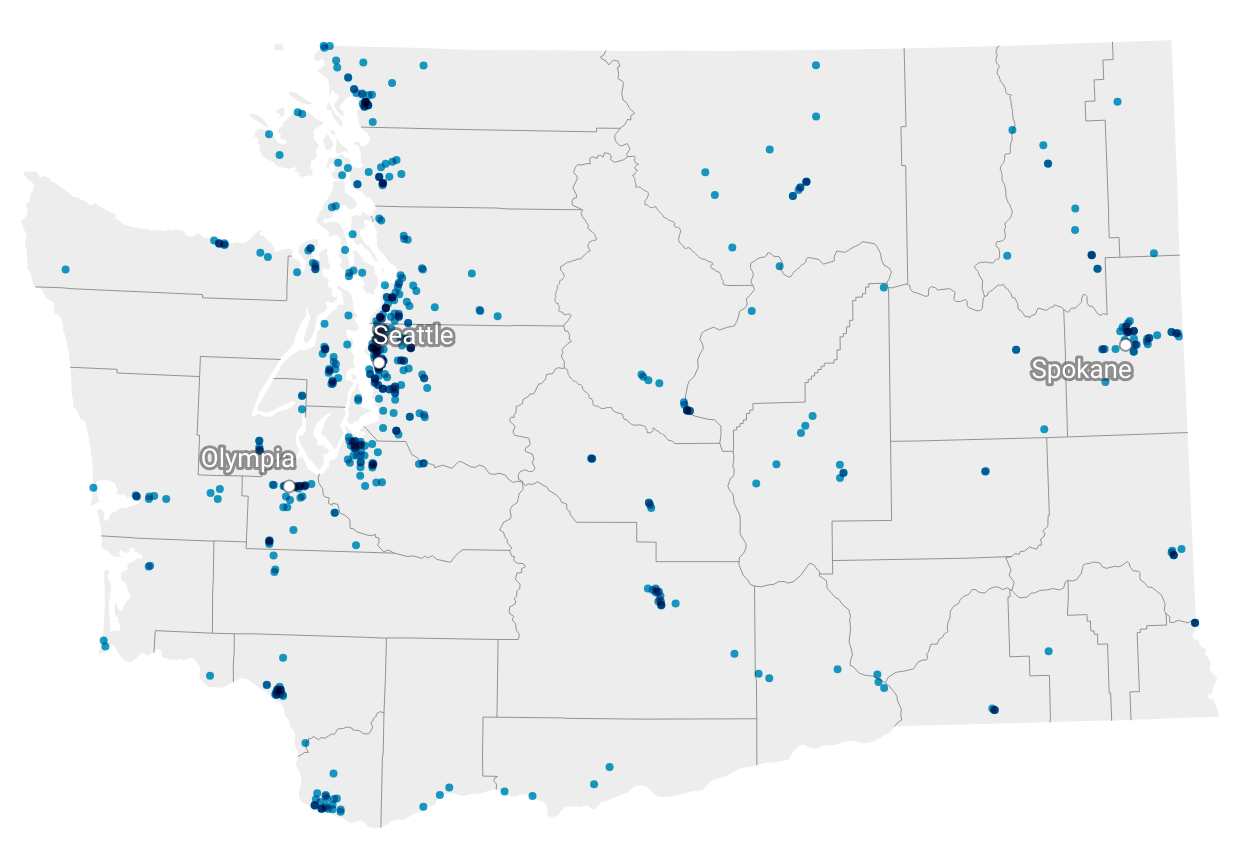}
	 \caption{Retail stores}
    \end{subfigure} \\
    
    \centering

	\begin{subfigure}{\textwidth}
	\centering
	    \includegraphics[width=.75\linewidth]{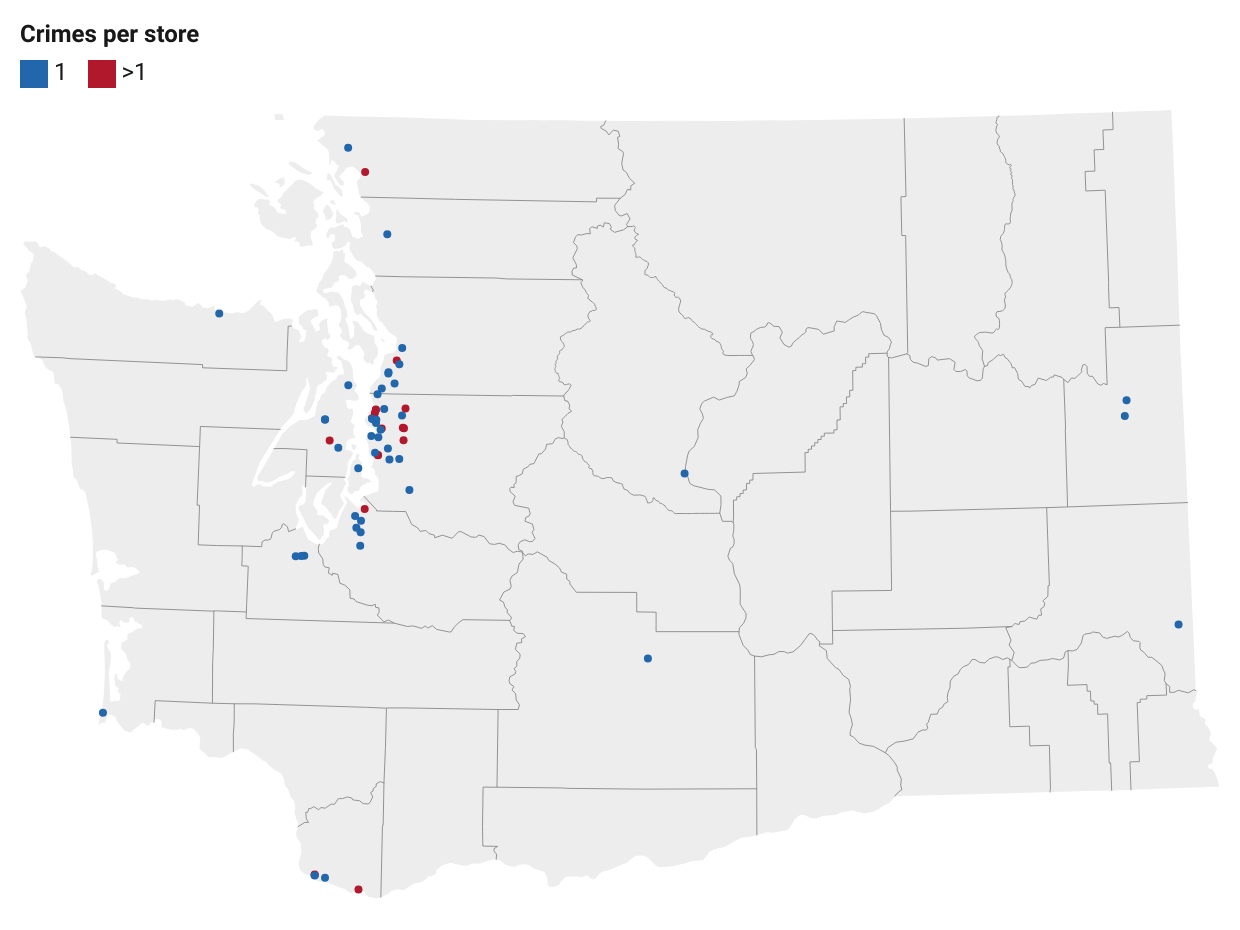}
		 \caption{Stores with reported crimes}
	\end{subfigure}
\label{fig:locations}
\par 
\rule{\textwidth}{0.4pt}
\begin{minipage}[h]{\textwidth}
\medskip
\footnotesize \emph{Notes:} Panel A illustrates the geographical distribution of cannabis retail stores in our sample. Panel B shows the geographical distribution of victimized stores, with stores reporting one crime incident marked by blue dots and those with at least two reported crime incidents marked by red dots.
\end{minipage}
\end{figure}

Figure \ref{fig:locations} shows the locations of stores with a reported crime during our sample period. There were 74 reported crimes at 57 different stores (46 stores were victimized only once and 11 stores were victimized more than once).\footnote{One additional store was victimized during our sample period, but it has no clean control stores so we omit the store from our estimation sample. We describe our criteria for defining clean control stores in Section \ref{sec:strategy}.} Of these crimes, 62 were armed robberies and 12 were burglaries. Approximately two-thirds of these crimes occurred at stores in the Seattle area and the neighboring cities of Tacoma and Bellevue, which roughly corresponds to the share of stores in those cities compared to the rest of the state. In general, crime incidence at cannabis stores reflects the geographic distribution of stores. 

There is some evidence of seasonality in crime incidents at cannabis retailers, with more crimes occurring in the fourth quarter compared to other quarters (see Appendix Figure \ref{fig:crime_seasonality}). Over the entire sample period, crime incidents at cannabis retailers increased, which roughly tracks the increase in reported crime in Washington state more generally (see Figure \ref{fig:wa_crime}). 


\section{Empirical strategy}\label{sec:strategy}

We are interested in the effect of retail crime on store-level prices. To identify a causal effect, we exploit the quasi-experimental variation in the time and location of retail crimes using a (stacked) difference-in differences (DiD) estimator. The following subsections describe our empirical strategy in more detail.

\subsection{Treatment and control groups}\label{sec:groups}
\subsubsection*{Two treatment groups: victimized and rival stores}

We define treatment timing as the store-specific month of a robbery or burglary. We distinguish between two treatment groups based on their exposure to retail crime: i) victimized stores, i.e., all stores that directly experience a robbery or burglary, and ii) rival stores, i.e., non-victimized stores located within a 5-mile radius of a victimized store, excluding any victimized store. The first treatment group allows us to analyze the direct impact of crime on victimized stores' pricing strategies, while the second measures the treatment effect at nearby stores. Such effects at nearby stores could result from demand substitution, where customers shift their demand from victimized stores, e.g. out of fear for their own safety or temporary crime-induced store closures, or from strategic pricing responses. Alternatively, they may stem from cost shocks at rival stores. For instance, upon learning about a crime at a victimized store, nearby retailers might adjust their own expectations of being targeted, leading them to invest in precautionary security measures or additional insurance. In assigning treatment to a larger group of nearby units, our strategy resembles those from the spatial treatment literature \citep[see e.g.][]{shapiro2019,mobarak2017}.

We choose the 5-mile radius for two reasons. First, the 5-mile radius is sufficiently small so as to reasonably capture local market characteristics that may influence how rival stores react to a nearby store being victimized. Second, victimized stores have an average of twenty competing stores within this radius, which ensures that a consumer at the average victimized store can choose between a sufficient number of alternative stores. To the extent that crime induces changes in consumer behavior, we view it as plausible that these nearby stores may be affected. 
Results are robust to adjusting the boundary or using a rank-based rival definition (see Appendix \ref{sec:het_spatial_robust}).  

\subsubsection*{Control group: unaffected local markets}

We compare both treatment groups to a common control group: stores in unaffected local markets. In defining unaffected local markets, we face a tradeoff. On the one hand, we want to ensure that stores in the control group are in markets that are comparable to (and hence near) the treated group. At the same time, if there are spillovers from the treated group to nearby stores, then including nearby stores in the control group leads to biased estimates.

These spillovers could arise from various factors. One concern is strategic complementarity in pricing. \citet{muehlegger2022pass} show that in the presence of imperfect competition, the strategic response of (untreated) competitors may disqualify them as a valid control group. In practice, this implies that the price a firm sets is a function of not just its own costs, but also those of its rivals. 

A major advantage in our setting is that---in addition to retail prices and quantities---we also observe prices and quantities for the universe of vertical transactions between producer-processors and retailers. This allows us to measure retailers’ spatial sensitivity to competitors’ costs and identify the distance at which retailers no longer compete in prices. We estimate the pass-through of the average wholesale costs of competing stores within a certain distance from store $j$ on store $j$'s retail prices. We find that sensitivity to competitors’ unit costs dissipates by the 30-mile mark under a variety of specifications (see Appendix \ref{sec:pt_figure} for more details). This suggests that stores located more than 30 miles from a victimized store will not have a strategic price response to the victimized store’s crime pass-through, and hence serve as valid controls from a strategic pricing standpoint.

Spillovers can also arise if consumers substitute demand out of victimized stores and into nearby competitors. 
While demand substitution is distinct from strategic complementarity in pricing, at a mechanical level, both entail consumers choosing between stores from the same choice set. Therefore, we consider the 30-mile boundary sufficiently distant to prevent demand substitution from biasing our estimates.

Finally, spillovers can occur if crime incidents induce a cost shock at rival stores. These spillovers could emerge through increased precautionary security measures or higher insurance expenditures, similar to the responses at treated stores. While the geographic scope of such spillovers is difficult to quantify, it is plausible that stores located close to a victimized store are more likely to face increased costs, e.g. due to local market characteristics. Therefore, we expect these spillover effects to operate within a similar geographic radius as strategic complementarity in pricing and demand substitution. Additionally, if control stores experience rising marginal costs, we likely underestimate the treatment effects.


While potential spillovers require omitting nearby stores from the control group, it is important to ensure that the control group remains comparable to the treated stores to the greatest possible extent. We therefore limit the distance at which a store can serve as a control to 60 miles, ensuring that the stores are situated within the same region of the state. Regional comparability is important because Washington state exhibits significant economic and demographic differences across regions, and approximately two-thirds of retailers' wholesale purchases occur within the same region (see Appendix Table \ref{tab:h.1}). Unaffected local markets thus consist of stores located in the concentric ``donut" between 30 and 60 miles from a victimized store.  We conduct several robustness checks to test the sensitivity of our main results to adjusting the boundaries for unaffected local markets (see Appendix \ref{sec:rings_robust}). 


Panel A in Table \ref{tab:sample_summary_stats} compares pre-treatment characteristics of treated and untreated stores in our study. Columns 1 and 2 show that the average unit price at victimized and rival stores is similar prior to treatment. Victimized stores sell more products per month, have higher monthly revenue, and have more product variety compared to rival stores, though the differences are small and never statistically significant. The similarities between victimized and rival stores prior to crimes support our assertion that the location of crime incidents is conditionally quasi-random. Column 3 summarizes the never-treated group and is based on the entire sample period. Note that never-treated stores may or may not serve as controls depending on their location. The mean price at never-treated stores is similar to victimized and rival stores, but the quantity sold, monthly revenue, and product variety are slightly lower than at victimized and rival stores, though the differences are not statistically significant.

\begin{table}[!htbp]
\centering
\caption{Estimation sample summary statistics}
\renewcommand{\tabcolsep}{1pt}{
\def\sym#1{\ifmmode^{#1}\else\(^{#1}\)\fi}
\begin{tabular*}{\hsize}{@{\hskip\tabcolsep\extracolsep\fill}l*{3}{c}}
\toprule
\addlinespace
\addlinespace
\addlinespace

\multicolumn{4}{c}{(A) Pre-treatment characteristics} \\
\addlinespace
\midrule
\addlinespace
& \multicolumn{1}{c}{(1)} &\multicolumn{1}{c}{(2)} &\multicolumn{1}{c}{(3)}  \\

&\multicolumn{1}{c}{\parbox{2cm}{\centering Victimized}}&\multicolumn{1}{c}{\parbox{2cm}{\centering Rivals}} &\multicolumn{1}{c}{\parbox{2cm}{\centering Never-treated}} \\

\addlinespace

\midrule

\parbox{4cm}{Unit price \\ (in dollars)}  & 26.24 & 25.47 & 26.00 \\ 
        & (4.33) & (4.49) & (4.50) \\ 
\addlinespace
\parbox{4cm}{Units sold \\ per month} & 13,260 & 12,304 & 10,908 \\ 
                    & (12,409) & (11,931) & (11,617) \\ 
\addlinespace
\parbox{4cm}{Monthly revenue \\ (in dollars)}& 248,516 & 215,201 & 210,416 \\ 
         & (234,911) & (217,957)  & (235,890) \\
\addlinespace
\parbox{4cm}{Unique products \\ per month} & 505 & 425 & 393 \\ 
                & (388) & (346) & (324) \\ 
\midrule
\addlinespace
\addlinespace
\addlinespace
\multicolumn{4}{c}{(B) Treatment group sizes} \\
\addlinespace
\midrule

\addlinespace
& \multicolumn{1}{c}{(1)} &\multicolumn{1}{c}{(2)} &\multicolumn{1}{c}{(3)}  \\

&\multicolumn{1}{c}{\parbox{2cm}{\centering Victimized}}&\multicolumn{1}{c}{\parbox{2cm}{\centering Rivals}} &\multicolumn{1}{c}{\parbox{2cm}{\centering Never-treated}} \\

\addlinespace

\midrule

Treated stores & 57 & 242 &  \\

\addlinespace

Control stores & 349 & 349 & 152 \\

\addlinespace

Total stores & 372 & 410 & \\

\addlinespace

Total store-months & 30,761 & 32,249 & \\

\addlinespace
\bottomrule
\end{tabular*}
\begin{minipage}[h]{\textwidth}
\medskip
\footnotesize \emph{Notes:} Panel A summarizes store-level variables prior to treatment, with statistics for the never-treated group based on the entire sample period. Standard deviations are in parentheses. The reported variables include mean unit price, mean quantity sold per month, mean revenue per month, and mean number of distinct products sold per month. Columns 1-2 display summary statistics for the two treatment groups: victimized stores and rival stores. Column 3 shows the subset of control stores that are neither victimized nor rival stores. Panel B shows the number of stores in each treatment group.

\end{minipage}
}
\label{tab:sample_summary_stats}
\end{table}

\subsection{Main specification}


Our main specification builds on the stacked DiD estimator. Stacked DiD has been applied in a variety of settings, including investigating the effect of local cost shocks on national retail chains \citep{butters2022}, the effect of application costs on the targeting of disability programs \citep{deshpanade2019}, and the effects of minimum wage hikes on low wage employment \citep{cengiz2019}. Stacked DiD overcomes issues with canonical DiD which can yield biased estimates under staggered treatment adoption and heterogeneous treatment effects (see, e.g., \cite{baker2022,gb2021,sun2020,Callaway2021}). Similar to other DiD estimators, the stacked DiD estimator identifies the causal treatment effect under the assumption of parallel trends and no anticipation. We discuss these identifying assumptions and other threats to identification in more detail in Section \ref{sec:robustness_checks}.

The idea behind stacked DiD is to create a separate dataset (i.e. sub-experiment) for each crime incident comparing the treated group for that incident to a set of ``clean" control stores. 
An identifying variable is generated for each dataset and the datasets are concatenated to create a single ``stacked" dataset. The model is estimated on the stacked data with sub-experiment-specific unit- and time-fixed effects. For each sub-experiment, stacked DiD identifies a group average treatment effect for the treated (ATT), similar to the cohort-specific ATT in \cite{sun2020} and the group-time ATT in \cite{callaway2021a}; see Section \ref{sec:robustness_checks} for further details.

Most studies using stacked DiD define clean controls solely based on treatment timing as this accounts for biases from heterogeneous treatment effects under staggered adoption.\footnote{An exception is \cite{butters2022}, who investigate spillover effects of state excise taxes on chain stores in nearby states. Their approach is similar to ours in that they exclude chain stores whose parent company is affected by a tax hike in another state within the event window.} We add a geographic layer to the clean control criteria to remove additional contamination from treatment effect spillovers. Thus, to be considered a clean control for a given sub-experiment, a store must be (i) located between 30-60 miles of the corresponding crime incident, and ii) a not-yet-treated store outside of its own event window or a never-treated store.  A store can qualify as a clean control for multiple sub-experiments with overlapping event windows. Since the store is included in each of the sub-experiment-specific datasets, the stacked dataset will contain duplicate observations of that store for some calendar months---a common feature of the stacked DiD estimator. Further details about our inclusion criteria for clean controls are discussed in Appendix \ref{sec:inclusion_criteria}.  

The flexibility of the stacked DiD estimator in extending rules for clean controls to geographic criteria is crucial in our setting. In particular, stacked DiD reduces biases from spillovers to untreated stores, maintains geographically comparable treatment and control stores, and ensures a large pool of clean control stores. This feature is the primary reason we choose the stacked DiD estimator over related estimators \citep{Callaway2021, Borusyak2021} which typically impose restrictions on the control group composition for the entire sample and thereby significantly reduce the pool of valid control observations. 

Panel B of Table \ref{tab:sample_summary_stats} summarizes the estimation samples for our victimized and rival specifications. Both specifications have the same control group comprising 349 stores, 152 of which are never-treated. The rival specification (Column 2) has more treated stores than the victimized specification (Column 1), and hence more total stores and more store-months.

We specify a distributed lag model with leads and lags before and after treatment. We define treatment relative to the store's first crime incident. In our main specification, we drop victimized stores with multiple treatments from the sample prior to the second treatment, isolating the effect of the first crime on store prices (we allow for multiple treatments in a robustness check in Section \ref{sec:robustness_checks}). Following the same logic, we assign treatment for rival stores according to a store's first nearby crime incident, so that each rival store is assigned to only one treatment cohort. However, we keep a rival store in the sample in subsequent periods even if additional nearby crimes occur. We do this for two reasons: First, several rival stores have multiple nearby crime incidents during their event window. Therefore, dropping these stores prior to the second nearby crime incident would result in an increasingly unbalanced panel at higher lags. Second, to the extent that multiple nearby crime incidents may have a compounding effect on rival stores' prices, it is important to include these incidents in our analysis. We drop rivals from the sample prior to their second nearby crime incident in a robustness check in Section \ref{sec:robustness_checks}).

Since the store-level price indexes and the treatment adoption indicators are in first-differences, we estimate the following model in first-differences:
\begin{equation}\label{eq:strategy_1}
  \pi_{j,t,d} = \sum_{l =-k+1}^{k} \beta_l \Delta T_{j,d,t-l} + \gamma_{t,d} + \epsilon_{j,d,t}
\end{equation}
Equation \ref{eq:strategy_1} relates the monthly inflation rate at store $j$, $\pi_{j,t,d}$, to leads and lags of the treatment adoption indicator, $\Delta T_{j,d,t-l}$.  $\Delta T_{j,d,t-l}$ is equal to one if store $j$ in sub-experiment $d$ is treated $l$ months after period $t$ (or $l$ months before when $l$ is negative), and equal to zero otherwise. We control for sub-experiment-by-time fixed effects, $\gamma_{t,d}$. As the model is in first differences, store FE are swept out. We cluster standard errors by store to allow for autocorrelation in unobservables within stores, similar to \cite{butters2022,deshpanade2019}.\footnote{\cite{wing2024} show that clustering at the level of the treatment (i.e. stores) leads to valid inference with a stacked DiD estimator.}

The parameter $\beta_l$ measures the change in establishment $j$'s inflation resulting from a robbery or burglary $l$ months after a crime (or $l$ months before when $l$ is negative) compared to the control group.\footnote{Since distributed lag coefficients measure incremental treatment effects, one fewer lead has to be estimated compared to an event study specification. Thus, an 13-month event window requires estimating 12 distributed lag coefficients \citep{renkin2020,siegloch2023}.} While inflation is the dependent variable, we follow previous studies and present the estimates as the effect of treatment on the price level (see e.g. \cite{renkin2020,leung2021}). We thus normalize the effect on the price level to zero in the baseline period one month before treatment and report the cumulative treatment effect as the sum of $\beta_l$ at various lags: $E_L = \sum_{l=0}^{L}\beta_l$. The pre-treatment coefficients are reported in a similar manner with $P_{-L} = -\sum_{l=-1}^{-L+1}\beta_{l}$. This linear transformation is commonly applied in studies with a dependent variable that is in first-differences \citep[][]{renkin2020,leung2021}.\footnote{The linear transformation transfers the statistical properties (consistency and asymptotic normality) of $\hat{\beta}$ to the cumulative $\hat{E}_L$ and $\hat{P}_{-L}$. Standard errors of $\hat{E}_L$ and $\hat{P}_{-L}$ are calculated from the variances and covariances of the vector $\hat{\beta}$ by the standard formula for linear combinations \citep{siegloch2023}.} The cumulative distributed lag coefficients are numerically equivalent to the parameter estimates from an event study design under constant treatment effects outside the effect window \citep{siegloch2023}. 

An important consideration is the number of leads and lags to include in equation \ref{eq:strategy_1}. One limitation is that the establishment panel is not balanced, meaning that changes in the underlying sample may affect estimates when $l$ is large. Therefore, in our baseline estimation, we set $k = 6$. This implicitly assumes constant treatment effects outside of the 13-month event window, i.e. $\beta_l = 0 \; \forall \; l  > | k |$. We show estimation results using a longer event window in Appendix \ref{sec:robust_window}. 

Our main analysis investigates the effects of retail crime on prices at retail cannabis stores. However, we are also interested in the effects of retail crime on other store-level outcomes, as this sheds light on the underlying factors driving retail crime pass-through. For this purpose, we estimate equation \ref{eq:strategy_1} using the store-level quantity indexes and wholesale cost indexes as dependent variables.

\section{Results}\label{sec:results}

\subsection{Main results}

\begin{figure}[!htbp]
\caption{Effect of crime incidents on store prices, quantities sold and wholesale costs}
\label{fig:main_results}
\centering
	\begin{subfigure}{.7\textwidth}
	\centering
	    \includegraphics[width=0.7\linewidth]{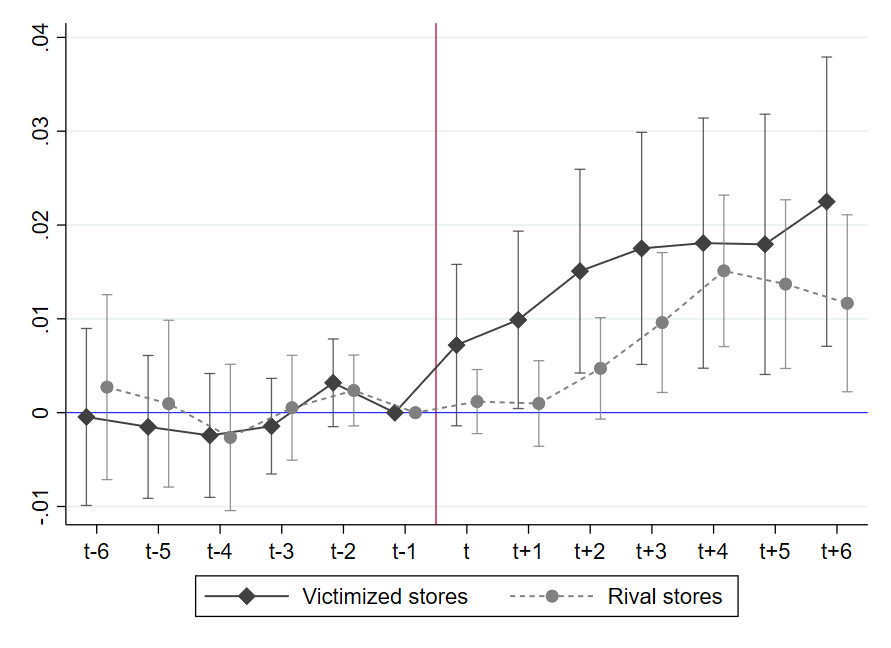}
	\caption{Price level}
	\label{fig:price_main}
    \end{subfigure}\hfil
	\begin{subfigure}{.45\textwidth}
	\centering
	    \includegraphics[width=\linewidth]{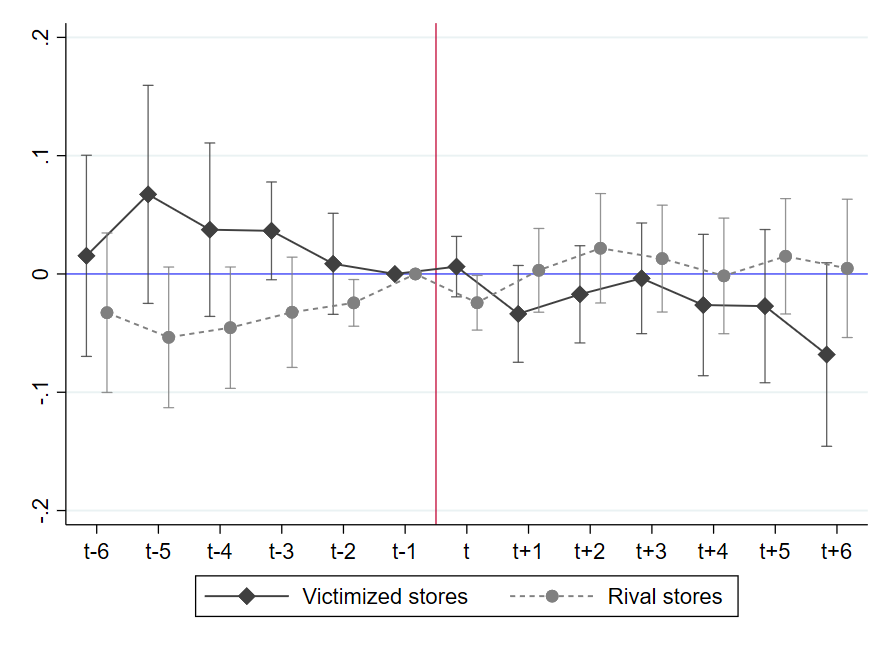}
		\caption{Quantity sold}
	    \label{fig:qty_main}
    \end{subfigure}\hfil
	\begin{subfigure}{.45\textwidth}
	\centering
            \includegraphics[width=\linewidth]{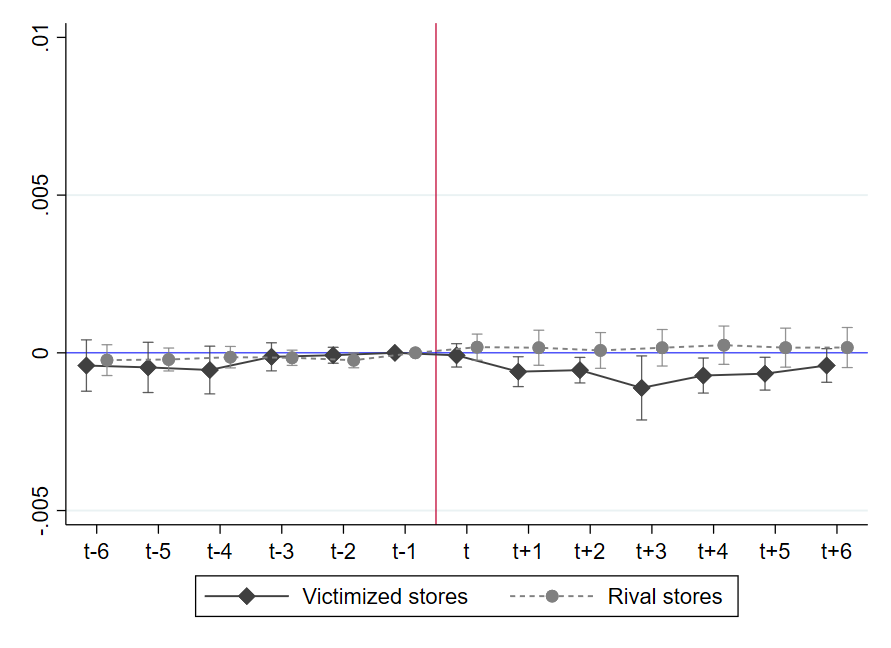}
     \caption{Wholesale cost}
     \label{fig:whole_cost_main}	
	\end{subfigure}\hfil
\par
\rule{\textwidth}{0.4pt}
\begin{minipage}[h]{\textwidth}
\medskip
\footnotesize \emph{Notes:} Each panel shows the cumulative treatment effects ($E_L$) $L$ months after a crime on different outcomes, along with corresponding 90\% confidence intervals based on standard errors clustered at the store level. Coefficients are interpretable as percent increases in outcome levels relative to the month before a crime incident. The black line depicts the cumulative effects of crime on outcomes at victimized stores, while the grey line represents rival stores.  The dependent variables are: store-level log price index (Panel A), store-level log quantity index (Panel B), and store-level log wholesale cost index (Panel C). 
\end{minipage}
\end{figure}

Figure \ref{fig:price_main} shows the estimated price level effects for victimized and rival stores according to our main specification. We report cumulative effects, i.e. the effect on the price level (in \%) relative to the baseline period in $t-1$. 
For both groups, the figure shows no significant pre-treatment effects. In the month of a retail crime incident, however, prices at victimized stores increase and continue to rise for three months. Four months after the crime, prices at victimized stores are 1.8\% higher compared to the month before the crime and stay at this higher price level for the following months. In contrast, rival stores show no price increase in the month that a retail crime hits a nearby store.
Nevertheless, two months after crime incidents at victimized stores, a very similar price increase can be observed at rival stores. Four months after the crime, rival store prices are 1.5\% higher than the month before the crime, and the effect is not statistically significantly different from that at victimized stores. Overall, our estimates reveal that retailers in the Washington state cannabis market increase their prices following retail crime incidents. 

Figure \ref{fig:qty_main} presents estimates from equation \ref{eq:strategy_1} when using the quantity index as the dependent variable.  
Post-incident estimates indicate a 3\% (but insignificant) decrease in quantity sold for victimized stores immediately after a crime incident. However, the quantity sold returns to pre-incident levels within a month. This finding speaks against demand substitution out of victimized and into rival stores due to, e.g., stigmatization or personal safety concerns on the part of consumers. Moreover, our findings suggest that demand in our setting is relatively inelastic in the short run. Nevertheless, when we expand the event window, we find a large and statistically significant negative effect on quantity sold after nine months at victimized stores (see Appendix \ref{sec:robust_window}). This delayed effect is consistent with high consumer search costs and persistently higher prices at affected stores (see Appendix Figure \ref{fig:longer_window}). 
Pre-incident estimates for victimized stores are insignificant at the 10\% level. For rival stores, we observe a slightly negative pre-trend with a statistically significant coefficient at $t-2$. As shown in Appendix \ref{sec:outliers}, winsorizing the quantity index, which exhibits considerably more volatility than the price index, mitigates this pre-trend. This correction suggests that the observed pre-trend is driven by a small number of outliers in the quantity index.

Figure \ref{fig:whole_cost_main} shows estimates with the wholesale cost index as the dependent variable. Estimated monthly treatment effects on wholesale costs are statistically significant but economically small, at approximately 0.1\%. 
These negligible effect sizes imply that retailers do not meaningfully adjust their wholesale expenditures in response to retail crime incidents, and also indicate no major reaction from producer-processors to crime incidents at their client stores. Additionally, these results provide reassurance that changes in COGS are not a confounder driving the observed retail price effects in Figure \ref{fig:price_main}. Pre-incident effects show no discernible divergence in the evolution of wholesale costs between the treatment and control groups. 

\subsection{Heterogeneity analysis}\label{sec:heterogeneity}

In this subsection, we conduct two main heterogeneity analyses to gain more insight into the role of market structure in crime pass-through.

\subsubsection*{Chain vs Independent stores}
We first examine whether crime pass-through differs between chain and independent stores. One concern is that chain size may correlate with profitability and is therefore endogenous. This is less of a concern in our setting for two reasons. First, when the market was initially created, the LCB randomly assigned licenses to applicants, reducing self-selection bias for store types \citet{hollenbeck2022winning}. Second, the number of licenses is capped by the LCB, meaning that profitable firms cannot easily increase their chain size. 

We define chains as stores belonging to firms running three or more stores and consider all other stores as independent. Of the 57 victimized stores, 37 are independent and 20 belong to chains. We sort sub-experiments into two groups depending on whether the corresponding victimized store is a chain store or independent. For each sub-experiment, we preserve the original control group (results are similar if we also subset control groups by chain size). We then estimate our main specification separately for each group. For the rival specification, we sort rival stores into two subgroups depending on whether a rival store is independent or belongs to a chain (of the 242 rival stores, 154 are independent and 88 belong to chains). We then estimate our rival specification separately for each subgroup (using the original control groups).

Table \ref{tab:heterogeneity} reports the results of the subsample analyses. Columns 1 and 2 show that among victimized stores, the pass-through of crime is large at independent stores but smaller and not statistically significantly different from zero at chain stores. Columns 5 and 6 show that the same pattern holds among rival stores. When defining independent as single-store firms and chains as firms with two or more stores (see Appendix Table \ref{tab:heterogeneity_robustness}), we observe a similar difference for victimized stores but not for rival stores.  
Overall, our findings align with studies showing little within-chain price adjustments to local conditions \citep{hitsch2021prices, dellavigna2019uniform} and suggest that owning multiple stores may serve as insurance against negative cost shocks, possibly due to better financing conditions or high returns to scale \citep{hollenbeck2022winning}. 
Alternatively, or as an additional mechanism, the results suggest that chains may already have more effective crime prevention strategies or insurance plans in place that require less adjustment, supporting the notion of firm learning.

\subsubsection*{Market concentration}

Next, we investigate whether effects vary by market concentration. Concentration is endogeneous if profitability affects market entry. However, since the LCB caps the number of cannabis store licenses and distributes them according to population density, profitability does not directly affect concentration in our setting. We consider each victimized store as the focal point of its own market comprising the set of stores (including the focal store) within a 5-mile radius. We calculate the Herfindahl–Hirschman Index (HHI) for each market and sort sub-experiments into two groups depending on whether the victimized store's market HHI is above or below the median HHI for all victimized store markets. A high HHI implies high concentration and thus low market competition. We provide descriptive statistics on the HHI in Appendix Table \ref{tab:hhi_descriptive_statistics}.

We estimate our main specification on each subsample to test whether crime pass-through differs between stores above and below the median HHI. Columns 3 and 4 of Table \ref{tab:heterogeneity} show that among victimized stores, crime causes prices to increase 3.1\% in low-concentration markets while in high-concentration markets the effect is not significantly different from zero. Moreover, the difference in effect sizes is statistically significant. This pattern carries over to rival stores, though the difference in effects sizes between low and high market concentration is less pronounced. Prices in low-concentration markets increase 2.2\% while in high-concentration markets they increase 1.1\%.\footnote{In Appendix \ref{sec:appendix_heterogeneity}, we investigate whether differences between rural and urban markets explain the heterogeneous pass-through rates. Our analysis indicates that once we account for variations in market concentration between rural and urban areas, the disparity in effect sizes between markets with low and high concentration largely diminishes (Appendix Table \ref{tab:heterogeneity_robustness}).}

These results are in line with standard imperfect competition models (e.g. the canonical Cournot model), which predict that when competition increases, prices become more sensitive to marginal costs and pass-through rates rise. However, if marginal cost pass-through rates exceed unity or the shape of the cost function changes, predictions regarding the relationship between competition and pass-through rates can change \citep{weyl2013pass, miller2017, ritz2024does}. Empirical findings on the relationship between competition and cost pass-through rates are similarly mixed. Some studies report similar findings to ours, showing that pass-through rates increase with competition \citep{ cabral2018larger, genakos2022competition}, while other studies find the opposite \citep{doyle20082, miller2017}.

\begin{table}[!htbp] 
\centering
\caption{Heterogeneity of retail crime pass-through}
\renewcommand{\tabcolsep}{1pt}{
\def\sym#1{\ifmmode^{#1}\else\(^{#1}\)\fi}
\begin{tabular*}{\hsize}{@{\hskip\tabcolsep\extracolsep\fill}l*{8}{c}}
\toprule

            &\multicolumn{4}{c}{Victimized} &\multicolumn{4}{c}{Rivals} \\
            \cline{2-5} \cline{6-9} \\
            
            &\multicolumn{1}{c}{(1)}&\multicolumn{1}{c}{(2)}&\multicolumn{1}{c}{(3)}&\multicolumn{1}{c}{(4)}&\multicolumn{1}{c}{(5)}&\multicolumn{1}{c}{(6)} &\multicolumn{1}{c}{(7)} &\multicolumn{1}{c}{(8)} \\

            &\multicolumn{1}{c}{\parbox{1.5cm}{\centering Indep. \\ stores }}&\multicolumn{1}{c}{\parbox{1.5cm}{\centering Chain \\ stores}}&\multicolumn{1}{c}{\parbox{1.5cm}{\centering Low concentration}}&\multicolumn{1}{c}{\parbox{1.5cm}{\centering High concentration}}&\multicolumn{1}{c}{\parbox{1.5cm}{\centering Indep. \\ stores }}&\multicolumn{1}{c}{\parbox{1.5cm}{\centering Chain \\ stores}}&\multicolumn{1}{c}{\parbox{1.5cm}{\centering Low concentration}}&\multicolumn{1}{c}{\parbox{1.5cm}{\centering High concentration}} \\
\midrule
\addlinespace
        $E_0$ & 0.010 & 0.0018 & 0.011 & 0.0041 & 0.00071 & 0.0022 & 0.000077 & 0.0029 \\ 
        ~ & (0.0079) & (0.0026) & (0.010) & (0.0030) & (0.0026) & (0.0030) & (0.0025) & (0.0024) \\ 
\addlinespace
         $E_2$ & 0.021** & 0.0037 & 0.022* & 0.0086 & 0.0077* & -0.00088 & 0.0048 & 0.0046 \\ 
         ~ & (0.0096) & (0.0044) & (0.012) & (0.0056) & (0.0040) & (0.0045) & (0.0044) & (0.0033) \\ 
\addlinespace
         $E_4$ & 0.026** & 0.0035 & 0.031** & 0.0056 & 0.021*** & 0.0052 & 0.017*** & 0.011* \\ 
         ~ & (0.012) & (0.0051) & (0.014) & (0.0077) & (0.0061) & (0.0068) & (0.0065) & (0.0062) \\ 
\addlinespace

        $E_6$ & 0.026** & 0.016 & 0.034** & 0.012 & 0.019*** & -0.0018 & 0.014* & 0.0060 \\ 
        ~ & (0.013) & (0.012) & (0.014) & (0.012) & (0.0071) & (0.0078) & (0.0074) & (0.0080) \\ 
        
        \midrule
$\sum \text{Pre-event}$   
           & 0.00035 & -0.0015 & 0.0040 & -0.0046 & 0.0018 & 0.0041 & -0.0024 & 0.0094 \\ 
        ~ & (0.0079) & (0.0060) & (0.0080) & (0.0077) & (0.0067) & (0.011) & (0.0069) & (0.0070) \\ 
\midrule
\(N\)       & 21,037 & 9,724 & 14,826 & 15,935 & 29,956 & 28,837 & 17,974 & 13,050 \\ 

\bottomrule
\end{tabular*}
\begin{minipage}[h]{\textwidth}
\medskip
\footnotesize \emph{Notes:} Each column shows the cumulative treatment effects on store price levels for different subsamples zero, two, four and six months after a crime, along with the sums of pre-treatment coefficients. Coefficients are interpretable as percent increases in outcome levels relative to the month before a crime incident. The first four columns use victimized stores as the treatment group, and the last four columns consider rival stores. Columns 1 and 5 show effects for independent stores (i.e., owned by firms running one or two stores only), while columns 2 and 6 show effects for stores owned by firms running at least three stores. For the other columns, the Herfindahl-Hirschman Index is calculated for each victimized store's local market market, defined as all stores within a 5-mile radius of the victimized store. The sample is split according to the median market concentration across all victimized markets. Columns 3 and 7 show effects for treated stores in markets with below median concentration, and columns 4 and 8 for treated stores in markets with above median concentration. Standard errors of the sums are clustered at the store level and shown in parentheses. \sym{*} \(p<0.10\), \sym{**} \(p<0.05\), \sym{***} \(p<0.01\). 
\end{minipage}

}
\label{tab:heterogeneity}
\end{table}

\section{Threats to identification, robustness checks and alternative specifications}\label{sec:robustness_checks}

In this section, we address potential endogeneity concerns in our setting. We present various alternative specifications, robustness checks and placebo tests to corroborate the validity of our results.

\subsubsection*{Endogenous treatment timing and parallel trends}

The key identifying assumption of our empirical strategy is the parallel trends assumption. This implies that store-level prices in the control and treatment groups would have followed a common trend in the absence of retail crime incidents. 

In our framework, we account for all time-invariant factors that could influence treatment, such as store location and average revenues, through the inclusion of store-fixed effects. Similarly, we account for all time-varying factors that equally apply to all stores, such as the seasonality of robberies or COVID-19 effects, through month-year fixed effects. However, the parallel trends assumption can be violated if the timing and location of retail crimes are correlated with changes in our outcome variables. An example is if stores that strongly increase or decrease revenues are more likely to be robbed or burglarized. This is less of a concern with rival stores' treatment timing. However, it is also possible that changes in policing in certain areas are correlated with revenues in those areas, which could also threaten the causality for our rival store regression. 

The first clear indication of the validity of the parallel trends assumption in our setting is the lack of significant pre-trends across all main specifications. Figure \ref{fig:main_results} shows no pre-treatment differences for prices, quantity sold and wholesale cost between victimized, rival, and control stores. Furthermore, pre-trends remain insignificant for alternative specifications, for instance, if we expand the event window (Appendix Figure \ref{fig:longer_window}) or when using alternative estimators (Appendix \ref{sec:alt_estimators}). Additionally, the lack of observable pre-trends supports the validity of the second identifying assumption regarding no anticipation. If store owners anticipate robberies, we would expect to see changes in the outcome variables before the actual events.

To further corroborate our findings, we conduct placebo tests in which we run our main regression analysis with a placebo treatment date that occurs 12 months before the actual treatment date. This approach tests for the presence of non-parallel trends or seasonal variations not fully addressed by our fixed effects. The placebo test results, detailed in Appendix Figure \ref{fig:placebo}, reveal no discernible patterns, reinforcing our confidence in the validity of the parallel trends assumption and our empirical model.

\subsubsection*{Alternative estimators}

The stacked DiD estimator offers advantages over related estimators due to its flexibility for applying rules for clean controls based on geographic criteria (for more details, see Appendix \ref{sec:alt_estimators}). Nevertheless, other estimators may have advantages in terms of efficiency and comparability to other studies. To assess whether our findings are sensitive to the choice of estimator, we estimate dynamic treatment effects using three alternative estimators: i) the canonical two-way fixed effect DiD estimator ii) the imputation estimator developed by \citet{Borusyak2021} (BSJ); and iii) the estimator proposed by \citet{Callaway2021} (CS).\footnote{The canonical TWFE estimator is prone to biases under staggered treatment adoption. The BSJ and CS estimators address these biases in a different way than the Stacked DiD estimator.} Following our main approach, we first estimate a distributed lag model (Equation \ref{eq:strategy_1}) for the designated event window and then calculate cumulative treatment effects using the last pre-treatment period as our baseline. We discuss the details of the alternative estimators in Appendix \ref{sec:alt_estimators}.

The dataset used for alternative estimators includes all stores, which means the control group includes stores that are geographically close to victimized stores. Positive price effects at these nearby stores, as suggested by our main results, would thus imply that treatment effects are attenuated in these models. Restricting the sample to stores more than 30 miles away from any treated store reduces the number of control stores from 349 to just 78 stores, reflecting that most stores are within 30 miles of at least one retail crime incident at some point during the sample period. This highlights the benefits of using the stacked DiD framework in our setting.

Appendix Figure \ref{fig:alt_est} displays the cumulative treatment effects for the alternative estimators. As expected, the effects are slightly smaller than in our main specification---about 1.5\% for victimized stores and 1\% for rival stores, except for the CS specification, which yields higher effects for victimized stores. However, in the majority of these alternative specifications, treatment effects are statistically significant and not different from our main specification, demonstrating that our findings are not driven by the choice of our estimator.

\subsubsection*{Alternative definitions of clean control and rival stores}

When defining clean control stores, we balance comparability with treated stores against potential biases from treatment spillovers, as discussed in Section \ref{sec:strategy}. To investigate whether our estimated effects are sensitive to our definition of clean control stores, we estimate our model using alternative definitions unaffected local markets and rival stores. Moreover, we investigate the sensitivity of our results to using more restrictive inclusion criteria.

First, we expand the area of unaffected local markets along two dimensions: i) including all stores between 10 and 60 miles of victimized stores; ii) including all stores that are located more than 30 miles away from victimized stores. Results presented in Appendix \ref{sec:rings_robust} show that treatment effect estimates are very similar and statistically significant when including additional control stores closer to victimized stores. Extending the control group to include more distant stores results in slightly smaller treatment effects---around 1.3\% for victimized stores and 0.8\% for rival stores four months post-crime incident---yet these effects still retain statistical significance for both groups.

Next, we inspect the sensitivity of our results to expanding the inner ring that defines rival stores. We report results in Appendix Figure \ref{fig:robust_innerring}. As expected, the treatment effect size decreases when considering stores further away from victimized stores as rivals, indicating that spillovers decrease with distance from the crime incident. This decline supports our choice of control store as it suggests that treatment spillovers affect stores located further away less strongly. Nonetheless, the treatment effect remains sizeable at around 1\% four months after an incident for stores within 10 miles and only becomes statistically insignificant for those further away. These results imply that spillover effects may extend to a larger set of stores than initially anticipated and considered in our welfare analysis.

Our definition of unaffected local markets and rival stores is constant across locations, but market characteristics may vary depending on whether a store is located in an urban or rural area. For instance, urban stores typically have more nearby rival and control stores, and urban customers may face higher transportation costs when traveling longer distances. To address potential biases related to these differences, we run two alternative specifications, detailed in Appendix \ref{sec:het_spatial_robust}. First, we redefine unaffected local markets by using distance ranks from the victimized store, defining the 150th to 250th closest stores as clean controls and the 20 closest stores as rivals. This approach balances the number of control and rival stores between urban and rural sub-experiments. Second, we tighten the boundaries for defining clean control stores for victimized stores in urban areas, using a 10-30 mile range. The results from these specifications align with our main findings, providing further support for the robustness of our results.

Our main specification excludes victimized and rival stores from serving as clean control stores in another sub-experiment with an overlapping event window. However, untreated stores within 30 miles of a soon-to-be-victimized or recently victimized store can still serve as clean control stores if all inclusion criteria are met. While this permits a large pool of control stores, it introduces potential biases if treatment spillovers affect these untreated stores. In Appendix \ref{sec:cc_criteria_robust}, we discuss this issue in detail and show that our main results remain robust when applying more stringent clean control criteria to account for potential spillover effects.

Finally, some rival stores experience multiple crime incidents within a 5-mile radius in a single month, which can create variations in treatment intensity and lead to a violation of the SUTVA in our rival store specifications. In Appendix \ref{sec:rival_sutva_robust}, we show that our main results are robust to the exclusion of rival stores that encounter more than one nearby crime incident within a given month.

\subsubsection*{Additional Robustness checks}

We conduct a number of additional robustness checks to rule out other factors potentially driving our findings. We present the results of these robustness checks in Table \ref{tab:robustness}. In both panels, Column 1 reproduces our baseline specification from Figure \ref{fig:price_main}. Column 2 shows that our results are similar when we include control variables such as county population, the local house price index (at the three-digit zip code level), and average county wage. These control variables absorb variation in prices stemming from local business cycles, population growth, or fluctuations in the housing market \citep{stroebel2019}. 

Since our price indexes are store-level aggregations of diverse sets of products and product categories, it is important to check whether results hold for a narrower set of homogeneous products. Therefore, in Column 3 we restrict the sample to products belonging to the usable marijuana product category and convert all prices to price per gram.\footnote{All products in the Usable Marijuana product category contain information on the package weight measured in grams. This enables us to convert prices into prices per gram for this product category.} Effect sizes are larger using this price index. In Column 4 we winsorize inflation rates below the bottom 0.5 percentile and above the 99.5 percentile to show that our results are not driven by outliers. We further discuss the winsorized specification, including when the quantity index is used as the dependent variable, in Appendix Section \ref{sec:outliers}.

Our price indexes are constructed with weights based on calendar year revenue shares (see Section \ref{sec:data}). To ensure that our main effects are not an artifact of this weighting scheme, in Column 5 we use indexes constructed with weights based on the fiscal year running from July through June. Estimates are similar to our baseline specification.

Our main analysis only includes a store's first reported crime incident and hence does not capture the effects of subsequent crimes at stores that are victimized more than once (of the 57 victimized stores in our sample period, 11 stores have multiple crime incidents). In Column 6 Panel A, we include all subsequent crimes with non-overlapping event windows, which allows for individual stores to be treated more than once.\footnote{If the first and subsequent crimes for a given store have overlapping event windows, stacked DiD can produce biased estimates if treatment effects are heterogeneous over time. This is similar to the bias that can arise from staggered DiD.} For victimized stores, treatment effects are slightly higher shortly after crime incidents but become attenuated at later lags when allowing for multiple treatments. This is in line with the idea that store owners invest more following the first event but react less to subsequent events. 

In the main rival specification, we retain rival stores in the sample for subsequent periods, even if additional nearby crimes occur. We do this to maintain a balanced panel at higher lags and account for potential compounding effects of nearby crimes on rival stores' prices. In Column 6 of Panel B, we drop rival stores from the sample before their second nearby crime incident to isolate the effect of the first rival treatment. While the effects remain highly statistically significant, dropping rivals before their second incident slightly attenuates the observed price effects at these stores. This attenuation suggests that nearby crime incidents have a compounding effect on rival stores. 


By excluding recently or soon-to-be-treated stores from the control group, stacked DiD identifies a group ATT for each crime incident under the standard DiD assumptions of parallel trends and no anticipation, similar to the cohort-specific ATT in \cite{sun2020} and the group-time ATT in \cite{callaway2021a}. When there is a balance in the number of pre-and post-treatment periods across sub-experiments, the stacked DiD regression recovers an aggregate ATT that is a convex combination of underlying causal effects. However, even when the DiD assumptions hold within each sub-experiment, the aggregated parameter estimate can be biased if the stacked dataset is not balanced in the number of pre- and post-periods across sub-experiments \citep{wing2024}. This is the case if some crime incidents occur near the beginning or end of the period for which data are available so that causal effects are identified for a larger number of event-time periods for some crime incidents than others.\footnote{Since our setting contains several crime incidents near the beginning and end of the sample period, dropping these incidents substantially reduces our treatment sample size. Therefore, in our main specification we allow for compositional imbalance in pre- and post-treatment periods across sub-experiments.} The bias arises because changes in the aggregate parameter may reflect compositional changes rather than treatment effect dynamics. 

To account for this, in Column 7 we estimate our stacked DiD regression using the subset of crime incidents for which the entire event window falls within our sample period.\footnote{In addition, we only include treated stores for which scanner data is reported for at least 75 \% of the event window. We prefer this cutoff because it allows for occasional idiosyncratic store-level gaps in data reporting in the LCB traceability system while at the same time ensuring stability in the composition of treatment and control groups within sub-experiments.} In both panels of Table \ref{tab:robustness}, Column 7 shows that imposing compositional balance in pre- and post-treatment periods substantially reduces our sample size but does not meaningfully change, and even slightly increases, our estimates. Standard errors tend to be larger due to the reduced sample size, but the effects are still significant at the 5\% level four months after the crime.

Stacked DiD implicitly weights treatment and control trends differently across sub-experiments. This can lead to bias in the aggregate parameter estimate if the share of treated and control stores differs across sub-experiments. The bias can arise even if the incident-specific ATTs are unbiased and there is compositional balance in pre- and post-periods across sub-experiments. \cite{wing2024} show that this bias can be corrected by using sample weights that account for relative treatment and control group shares. Using the balanced stacked dataset (from Column 7), we estimate a weighted least squares regression using sample weights based on relative treatment shares in each sub-experiment. We report results in column 8 of Table \ref{tab:robustness}. Effect sizes are comparable and not statistically significantly different from our baseline estimates.

One other concern is that retail crime may cause victimized stores to go out of business. If crime-induced store failure is correlated with unobserved store characteristics (e.g. profitability), then our treatment effect estimates may be biased. However, as detailed in Appendix \ref{sec:closures}, we find no significant difference in failure rates between victimized and non-victimized stores. In fact, 5.3\% of victimized stores close within a year after an incident, which is slightly lower than the 5.5\% average annual closure rate for non-victimized stores. These numbers suggest that crime-induced store closures are an unlikely source of bias.

\begin{table}[!htbp]
\centering
\caption{Price effects at victimized and rival stores - robustness checks}
\renewcommand{\tabcolsep}{1pt}{
\def\sym#1{\ifmmode^{#1}\else\(^{#1}\)\fi}
\begin{tabular*}{\hsize}{@{\hskip\tabcolsep\extracolsep\fill}l*{8}{c}}
\toprule
\addlinespace

\multicolumn{9}{c}{ (A) Victimized stores} \\ \midrule
\addlinespace

            &\multicolumn{1}{c}{(1)}&\multicolumn{1}{c}{(2)}&\multicolumn{1}{c}{(3)}&\multicolumn{1}{c}{(4)}&\multicolumn{1}{c}{(5)}&\multicolumn{1}{c}{(6)}&\multicolumn{1}{c}{(7)}&\multicolumn{1}{c}{(8)} \\ 
            &\multicolumn{1}{c}{\parbox{1.5cm}{\centering Baseline}}&\multicolumn{1}{c}{\parbox{1.5cm}{\centering Controls}}&\multicolumn{1}{c}{\parbox{1.5cm}{\centering Price per gram}}&\multicolumn{1}{c}{\parbox{1.5cm}{\centering Wins.}}&\multicolumn{1}{c}{\parbox{1.5cm}{\centering Alt. weights}}&\multicolumn{1}{c}{\parbox{1.5cm}{\centering Mult. treatments}}&\multicolumn{1}{c}{\parbox{1.5cm}{\centering Bal-anced}}&\multicolumn{1}{c}{\parbox{1.5cm}{\centering Bal-anced (weighted)}} \\
\midrule
\addlinespace

$E_0$ & 0.007 & 0.007 & 0.009 & 0.008 & 0.006 & 0.010 & 0.011 & 0.011 \\ 
        ~ & (0.005) & (0.005) & (0.007) & (0.005) & (0.004) & (0.007) & (0.008) & (0.008) \\ 
\addlinespace
         $E_2$ & 0.015** & 0.015** & 0.024*** & 0.016** & 0.012* & 0.014** & 0.022** & 0.021** \\ 
         ~ & (0.007) & (0.007) & (0.008) & (0.007) & (0.007) & (0.007) & (0.009) & (0.009) \\ 
\addlinespace
$E_4$ & 0.018** & 0.018** & 0.028*** & 0.018** & 0.015* & 0.015 & 0.026** & 0.025** \\ 
         ~ & (0.008) & (0.008) & (0.010) & (0.008) & (0.009) & (0.011) & (0.011) & (0.011) \\ 
\addlinespace

        $E_6$ & 0.022** & 0.022** & 0.033*** & 0.022** & 0.023** & 0.017* & 0.022* & 0.021* \\ 
        ~ & (0.009) & (0.009) & (0.012) & (0.009) & (0.011) & (0.010) & (0.012) & (0.012) \\ 
\midrule
$\sum \text{Pre-event}$   
             & 0.000 & 0.000 & -0.004 & -0.001 & 0.002 & -0.005 & 0.002 & 0.002 \\ 
        ~ & (0.006) & (0.006) & (0.008) & (0.006) & (0.005) & (0.005) & (0.007) & (0.007) \\ 
\midrule
\(N\)       & 30,761 & 30,761 & 30,651 & 30,761 & 30,761 & 31,578 & 20,793 & 20,793 \\ 
\midrule

\addlinespace
\addlinespace

\multicolumn{9}{c}{ (B) Rival stores} \\ \midrule
\addlinespace

            &\multicolumn{1}{c}{(1)}&\multicolumn{1}{c}{(2)}&\multicolumn{1}{c}{(3)}&\multicolumn{1}{c}{(4)}&\multicolumn{1}{c}{(5)}&\multicolumn{1}{c}{(6)}&\multicolumn{1}{c}{(7)}&\multicolumn{1}{c}{(8)} \\ 
            &\multicolumn{1}{c}{\parbox{1.5cm}{\centering Baseline}}&\multicolumn{1}{c}{\parbox{1.5cm}{\centering Controls}}&\multicolumn{1}{c}{\parbox{1.5cm}{\centering Price per gram}}&\multicolumn{1}{c}{\parbox{1.5cm}{\centering Wins.}}&\multicolumn{1}{c}{\parbox{1.5cm}{\centering Alt. weights}}&\multicolumn{1}{c}{\parbox{1.5cm}{\centering Single treatment}}&\multicolumn{1}{c}{\parbox{1.5cm}{\centering Bal-anced}}&\multicolumn{1}{c}{\parbox{1.5cm}{\centering Bal-anced (weighted)}} \\
\midrule

\addlinespace

       $E_0$ & 0.001 & 0.001 & 0.005 & 0.001 & 0.002 & 0.002 & 0.002 & 0.001 \\ 
        ~ & (0.002) & (0.002) & (0.004) & (0.002) & (0.002) & (0.002) & (0.003) & (0.003) \\ 
\addlinespace
        $E_2$ & 0.005 & 0.004 & 0.007 & 0.004 & 0.003 & 0.006* & 0.006 & 0.005 \\ 
        ~ & (0.003) & (0.003) & (0.005) & (0.003) & (0.003) & (0.003) & (0.004) & (0.004) \\ 
\addlinespace
         $E_4$ & 0.015*** & 0.014*** & 0.019*** & 0.013*** & 0.013*** & 0.014*** & 0.015** & 0.013** \\ 
         ~ & (0.005) & (0.005) & (0.006) & (0.004) & (0.005) & (0.005) & (0.006) & (0.006) \\ 
\addlinespace

        $E_6$ & 0.012** & 0.011* & 0.014* & 0.010** & 0.009 & 0.009 & 0.012* & 0.010 \\ 
        ~ & (0.006) & (0.006) & (0.008) & (0.005) & (0.006) & (0.006) & (0.007) & (0.006) \\ 

\midrule
$\sum \text{Pre-event}$   
           &  0.003 & 0.003 & -0.001 & 0.003 & 0.009* & 0.002 & -0.001 & -0.002 \\ 
        ~ & (0.006) & (0.006) & (0.009) & (0.005) & (0.005) & (0.006) & (0.008) & (0.008) \\ 
\midrule
\(N\)       & 31,024 & 31,024 & 30,891 & 31,024 & 31,024 & 25,877 & 17,913 & 17,913 \\ 

\bottomrule
\end{tabular*}
\begin{minipage}[h]{\textwidth}
\medskip
\footnotesize \emph{Notes:} Each column shows the cumulative treatment effects on store price levels for different specifications zero, two, four and six months after a crime, along with the sums of pre-treatment coefficients. Coefficients are interpretable as percent increases in outcome levels relative to the month before a crime incident. Panel A uses victimized stores as the treatment group, and Panel B considers rival stores. Column 1 in each panel shows the cumulative treatment effects from the main specification. Column 2 includes additional time-varying control variables: county population, local house price index (at the three-digit zip code level), and average county wage. Column 3 restricts the sample to Usable Marijuana products and converts all prices to price per gram. Column 4 uses store-level price indexes winsorized at the 0.5 and 99.5 percentiles as the dependent variable. Column 5 uses price indexes weighted based on the fiscal year running from July through June as the dependent variable. Column 6 (Panel A) considers subsequent store-level crimes as additional treatment events. Column 6 (Panel B) drops rival stores from the sample the month before their second rival treatment. Column 7 shows effects for events with a completely observable and balanced event window. Column 8 applies the sample weights proposed by \citet{wing2024} to the balanced stacked data set. Standard errors of the sums are clustered at the store level and shown in parentheses. \sym{*} \(p<0.10\), \sym{**} \(p<0.05\), \sym{***} \(p<0.01\).
\end{minipage}
}

\label{tab:robustness}
\end{table}

\section{Policy analysis---Retail crime as a hidden tax}\label{sec:policy}

We model retail crime as a marginal cost shock that functions like a hidden tax---an analogy first introduced by \cite{jackson2020}. 
The marginal cost shock could arise from multiple sources. On the one hand, the shock may represent direct marginal costs proportional to the quantity sold. Industry reports suggest that retailers respond to crime by hiring additional security personnel, expanding surveillance, or increasing theft insurance coverage, all of which scale with store size, operating hours, and product quantities \citep{uscc2023,rila2021,nrf_retail_security_2022,hs2022}. Alternatively, the shock could arise from changes in retailers' subjective perceptions of marginal costs. For example, retailers may use average costs as a proxy for marginal costs, meaning fixed costs such as property loss or one-time security expenditures could indirectly influence subjective marginal costs and, in turn, pricing strategies. In our welfare analysis, we remain agnostic about the exact sources of the marginal cost shocks, as the welfare implications generally hold irrespective of their underlying source.

In Section \ref{sec:welfare}, we adopt the symmetric oligopoly model of \citet{weyl2013pass}---which nests common imperfect competition models as well as the monopoly and perfect competition cases---to derive the general welfare implications of a hidden crime tax. Section \ref{sec:cost_pass} examines how retailers adjust their prices in response to changes in marginal cost, utilizing methodologies from the extensive literature on cost pass-through \citep[e.g.][]{hollenbeck2021,muehlegger2022pass,conlon2020,miller2017}. Our objective in estimating the marginal cost pass-through rate is twofold. First, assuming that crime-related cost shocks are passed through at a similar rate as other marginal cost shocks, the marginal cost pass-through rate enables us to quantify the hidden crime tax. Second, our analysis of marginal cost pass-through encompasses not just a retailer's own marginal costs but also the average marginal costs of it's competitors, which allows us to examine strategic complementarity in pricing in the cannabis industry. The degree of strategic price complementarity is informative about the mechanisms driving rivals' price response to crime and has important implications for our welfare analysis. In Section \ref{sec:welfare_quant}, we combine the estimated marginal cost and retail crime pass-through rates with the theoretical insights from \cite{weyl2013pass} to derive and quantify the welfare effects of the hidden crime tax in our context. 

\subsection{Welfare implications of a hidden crime tax}\label{sec:welfare}

We consider a market with $N$ firms. Firm $j$ maximizes profits by setting a unidimensional strategic variable, $r_j$, that can be price, $p_j$, or quantity, $q_j$. Each firm produces a single good with marginal costs equal to $mc_j=c'(q_j)+\tau$, where $c'(q_j)$ is the first derivative of the cost function, $c(q_j)$, which is identical for all firms. $\tau$ is the unit cost shock stemming from a retail crime incident, that is, the hidden crime tax. We make the simplifying assumption that the hidden crime tax applies equally to all $N$ firms in the affected local market (i.e. stores within a 5-mile radius of the crime).\footnote{This assumption is supported by our our main results and our marginal cost pass-through estimates (see Section \ref{sec:cost_pass}). The former show similar treatment effects between victimized and rival stores, while the latter rules out strategic complementarity in prices as a main mechanism driving the treatment effects.} In this case, \citet{weyl2013pass} show that under symmetric imperfect competition the pass-through rate (in dollars) for a small unit tax is:
\begin{equation}\label{eq:pass}
    \rho=\dfrac{dp}{d\tau}=\dfrac{1}{1+\dfrac{\epsilon_D-\theta}{\epsilon_S}+\dfrac{\theta}{\epsilon_{ms}}+\dfrac{\theta}{\epsilon_{\theta}}},
\end{equation}
where $\epsilon_S$ is the elasticity of the supply function, i.e., of the inverse marginal cost function, and $\epsilon_D=-\frac{p}{qp'(q)}$ is the elasticity of market demand. $\theta$ is a conduct parameter summarizing the degree of market competition and can be understood as the ratio of the markup in a market to the (fictional) monopoly markup. Consequently, $\theta$ is zero for perfect competition and one for the monopoly case. $\epsilon_{\theta}$ is the elasticity of the conduct parameter with respect to quantity. $\epsilon_{ms}$ is the elasticity of the inverse marginal surplus function equal to $\epsilon_{ms}=\frac{ms}{ms'q}$, which describes the curvature of the demand function. The marginal effect of the unit tax on consumer surplus (CS) and producer surplus (PS) is given by:

\begin{equation}
    \dfrac{dCS}{d\tau}=-\rho q
\end{equation}

\begin{equation}\label{eq:ps_welfare}
    \dfrac{dPS}{d\tau}=-\left[1-\rho (1-\theta)\right] q
\end{equation}
Accordingly, the incidence of the unit tax, that is the ratio of consumer to producer harm from an infinitesimal unit tax increase, is given by:

\begin{equation}
    I=\dfrac{\rho}{1-\rho (1-\theta)}
\end{equation}

These results offer important insights regarding the welfare implications of a hidden crime tax. Specifically, they suggest that the pass-through of retail crime is influenced by four factors: the demand elasticity, the supply elasticity, the curvature of the demand function, and the conduct parameter (together with its elasticity with respect to $q$ which is often zero in common models, such as the Cournot model). Under perfect competition ($\theta=0$), $\rho$ is only determined by the ratio of supply and demand elasticities, where the more inelastic side of the market bears more of the tax burden---a familiar result of the tax literature. 

Even if we assume that the crime-induced increase in marginal costs is fully redistributed to other players in the economy (e.g., security service providers) or raises retailers' profits in the case of subjective marginal cost increases, crime-induced price hikes still lead to an excess burden in the case of imperfect competition. This loss results from price distortions caused by the market power of firms. To see this, consider the monopoly case where $\theta=1$. In this case, the monopolist fully pays the hidden tax out of its profits ($\frac{dPS}{d\tau}=-q$). Yet, consumers still bear $\frac{dCS}{d\tau}=-\rho q$, implying that the tax is more than fully shared by market participants. This excess burden is zero for the perfect competition case, $\theta \to 0$. 

The results of the model also show that the pass-through rate serves as a sufficient statistic (together with $\theta$) for deriving the welfare effects of a unit tax, its incidence, and the excess burden. This has the advantage that one need not impose restrictive assumptions about the underlying market structure. 

\subsection{The marginal cost pass-through rate} \label{sec:cost_pass}

Next, we estimate the marginal cost pass-through rate which we can directly relate to the sufficient statistic approach from the tax pass-through literature \citep{weyl2013pass}. 
To estimate the marginal cost pass-through rate, we follow the industrial organization literature that measures the pass-through of cost shocks and taxes \citep{gana2020,hollenbeck2021,miller2017,muehlegger2022pass}. A major advantage of our setting is that, because we observe wholesale unit prices, we can directly measure how changes in marginal cost are passed through to prices.\footnote{Wholesale costs are typically estimated from supply-side first order conditions as such granular information on vertical relations is rarely observable by researchers \citep{dubois2022use}.}

We estimate the marginal cost pass-through rate, that is, the increase in retail unit prices (in dollars) stemming from a \$1 increase in wholesale unit prices. We specify a model at the store-product-month level that relates a store-product’s retail price to (i) that store-product’s wholesale price and (ii) the average wholesale price paid by rival stores for that same product (with rivals defined as all stores within a 5-mile radius of the focal store). We include rivals' cost changes to capture the effect of rivals' cost-induced price changes, i.e. strategic complementarity in prices \citep{muehlegger2022pass}. We estimate the following model for stores in affected markets (i.e. victimized and rival stores):
\begin{equation}\label{eq:pt_reg_2}
    \Delta p_{i,j,t} = \rho \Delta w_{i,j,t} + \beta \Delta w_{i,r(j),t} + \gamma_t + \Delta \varepsilon_{i,j,t},
\end{equation}
where $p_{i,j,t}$ is the average price (in dollars) of product $i$ sold at store $j$ in month $t$, $w_{i,j,t}$ is the average wholesale price that retailer $j$ pays for product $i$ in month $t$, $w_{i,r(j),t}$ is the average wholesale price that store $j$'s rivals pay for product $i$ in month $t$, and $\gamma_t$ is the year-month FE.\footnote{Since cannabis transaction data is publicly available, stores have full information on competitors' unit costs and prices updated on an almost weekly basis. Therefore, we focus on contemporaneous changes in costs and prices. This is in line with pass-through literature from other industries \citep[see e.g.][]{hollenbeck2021,muehlegger2022pass,conlon2020,miller2017}.} Since the model is in first differences, product and retailer FE are swept out. The pass-through rate, $\rho$, is the dollar increase in price at store $j$ from a $\$1$ increase in store $j$'s marginal cost. $\beta$ measures the effect of marginal cost changes at rival stores on store $j$'s prices. In Appendix \ref{sec:pt_figure}, we include the costs of stores located further away to gain insight into the geographic scope of sensitivity to costs and to guide our choice of the clean control group discussed in Section \ref{sec:strategy}.

Table \ref{tab:pt_elasticities} reports the results from the pass-through regression. We find that a \$1 increase in unit wholesale cost corresponds to a retail unit price increase of \$1.67 (Column 1). Such over-shifting of costs onto consumers is in line with the findings of \citet{hollenbeck2021}, and indicates that cannabis retailers exercise substantial market power.\footnote{Pass-through rates greater than one have been found in a number of empirical studies estimating cost pass-through in other industries \cite[see e.g.][]{pless2019pass,conlon2020}.}  According to equation \ref{eq:pass}, for pass-through to exceed one it is sufficient that firms have market power ($\theta>0$), marginal costs are constant, and demand is log-convex ($\epsilon_{ms}<0$) \citep{weyl2013pass}. 

Column 2 shows that the own-cost pass-through estimate is robust to including county-level average wages, the county-level unemployment rate, and the home price index (at the three-digit zip code level) to absorb variation in retail cannabis prices due to local business cycles or changes in house prices \citep[see][]{stroebel2019}. Similarly, estimates are unchanged when we include region-time FE to account for other spatially correlated shocks that may covary with wholesale and retail cannabis prices (Column 3). We present further robustness checks in Appendix \ref{sec:pt_figure}, including estimates in levels, in log-log terms, and using store-level price and cost indexes.

We also find economically small but significant effects of rival stores' costs on a store's own retail prices. For a given product sold at store $j$, a \$1 increase in the unit cost at rival stores corresponds to a \$0.02 increase in store $j$'s unit price (Table \ref{tab:pt_elasticities} Column 1). This aligns with \citet{hollenbeck2021} who find that cannabis retailers in Washington behave like local monopolists. Consequently, it is unlikely that the price increase at rival stores after a crime incident reflects a strategic response to increasing prices at victimized stores. Instead, rivals' price increase appears consistent with an own-cost shock e.g. from precautionary security expenditures or higher commercial crime insurance premia. This conclusion remains unchanged across all of the specifications in Table \ref{tab:pt_elasticities} and Appendix \ref{sec:pt_figure}. 

In our welfare analysis, we therefore abstract from strategic complementarity in pricing and assume that the effect of crime on prices at victimized and rival stores runs entirely through the own-cost channel. 

\begin{table}[!htbp] 
\centering
\caption{Unit cost pass-through rates}
\renewcommand{\tabcolsep}{1pt}{
\def\sym#1{\ifmmode^{#1}\else\(^{#1}\)\fi}
\begin{tabular*}{\hsize}{@{\hskip\tabcolsep\extracolsep\fill}l*{3}{c}}
\toprule
            \addlinespace

            &\multicolumn{1}{c}{(1)}&\multicolumn{1}{c}{(2)}&\multicolumn{1}{c}{(3)} \\ 

            &\multicolumn{1}{c}{\parbox{1.5cm}{\centering Baseline}} &\multicolumn{1}{c}{\parbox{1.5cm}{\centering Controls}} &\multicolumn{1}{c}{\parbox{1.5cm}{\centering Reg. $\times$ time FE}} \\ 
\midrule
\addlinespace
Own wholesale cost       & 1.67*** & 1.67*** & 1.67*** \\ 
                         & (0.027) & (0.027) & (0.027) \\ 
\addlinespace
Rivals' wholesale      & 0.016*** & 0.017*** & 0.016*** \\ 
cost (0-5 miles)            & (0.006) & (0.006) & (0.06) \\ 
\midrule
\(N\)               & 5,260,269 & 5,245,497 & 5,260,269 \\ 

\bottomrule
\end{tabular*}
\begin{minipage}[h]{\textwidth}
\medskip
\footnotesize \emph{Notes:} The table reports the estimates of wholesale unit cost pass-through rates from equation \ref{eq:pt_reg_2}. We report estimates for own wholesale cost changes and for average changes in wholesale costs at rival stores located within 5 miles of the respective store. All specifications control for month-year fixed effects. Coefficients for Columns 1-4 are interpretable as pass-through rates in dollars. 
Standard errors are clustered at the store level and shown in parentheses. \sym{*} \(p<0.10\), \sym{**} \(p<0.05\), \sym{***} \(p<0.01\).
\end{minipage}

}
\label{tab:pt_elasticities}
\end{table}

\subsection{Quantifying the welfare effects of retail crime pass-through in cannabis} \label{sec:welfare_quant} 

To quantify the hidden tax of crime and the corresponding welfare effects in the Washington state cannabis industry, we combine the sufficient statistics approach outlined in Section \ref{sec:welfare}, our DiD estimates from Section \ref{sec:results}, and our marginal cost pass-through estimates from Section \ref{sec:cost_pass}. Our goal is not to measure the overall welfare effects of retail crime in the Washington state cannabis market, as these include other direct and indirect costs of crime that we do not observe. Instead, we seek to quantify the welfare effects specifically stemming from the pass-through of retail crime onto stores' prices.

As a first step, we quantify the hidden unit tax from crime. In Section \ref{sec:results}, we found that stores in affected markets (i.e., victimized and rival stores) raise prices by 1.8\% four months after a crime incident. 
Multiplying this semi-elasticity by the mean unit price of \$27.93 in affected markets, we find that crime causes an increase in the average unit price of around \$0.50. We then compare this to the marginal cost pass-through rate obtained in Section \ref{sec:cost_pass}. 
Assuming that crime-related changes in marginal cost are passed through at a similar rate, we can calculate the implied hidden unit tax of crime by dividing our treatment effect in dollar terms by the pass-through rate.\footnote{There is no scope for factor substitution in cannabis retail, meaning the marginal cost pass-through rate should be independent of the source of the marginal cost change \citep{gana2020}.} This yields a unit tax per crime incident of about \$0.30---or approximately 1\% at the average unit price. 
Next, we calculate the fictional annual tax revenue. Depending on the source of the marginal cost shock, this fictional revenue could represent increased expenditures on security or insurance. If the marginal cost shock is subjective, the fictional tax revenue would instead reflect unanticipated future profits for retailers.
 The fictional tax revenue can be approximated by multiplying the unit tax per crime (\$0.30) by the average annual units sold in affected local markets. The average annual quantity sold in affected markets is 45,520,552 units, implying a fictional tax revenue of around \$13.7 million per year.

As a next step, we calculate the effect on consumer surplus by multiplying the crime-induced increase in unit prices (\$0.50) by the average annual units sold in affected markets. This yields a negative annual effect on consumer surplus of around \$22.8 million. To quantify the effect on producer surplus and the incidence of the hidden tax, we rely on the conduct parameter estimate by \citet{hollenbeck2021}, who study the same industry using comparable data. They estimate an average $\theta$ of 0.89, which aligns well with our results of substantial market power by retailers. Together with our marginal cost pass-through estimate and the average annual quantity sold, equation \ref{eq:ps_welfare} implies that a \$1 unit tax reduces producer surplus by around \$37.2 million, or a decrease of \$11.1 million for the estimated hidden crime tax.\footnote{Note that the reduction in producer surplus entails a decrease in quantity demanded. While our main results reveal no change in quantity sold following a retail crime incident, we find evidence of a delayed reduction in quantity sold at higher lags (see Appendix \ref{sec:appendix_robustness}).} Accordingly, the hidden crime tax incidence falls about 67\% on consumers and 33\% on producers. The incidence of the tax on consumers is equal to around 62.5\% in the monopoly case ($\theta=1$) and increases if retailers' market power decreases.

The combined annual welfare effect for market participants is about \$33.9 million. Assuming that the fictional tax revenue stays within the Washington state economy, we find an excess burden from retail crime pass-through of around \$20.2 million per year (\$33.9 million - \$13.7 million). Note that these welfare calculations assume that only stores within a 5-mile radius of an incident are affected. Further analysis presented in Appendix \ref{sec:rings_robust} indicates that while the effect of retail crime on prices decreases for stores located further away, it remains significant. This finding suggests that our welfare analysis is conservative, indicating that the actual welfare impacts on Washington state's economy could be higher than estimated. 

\section{Discussion and conclusion}\label{sec:discussion}

In this paper, we present the first causal evidence that retailers pass costs related to organized retail crime through to consumer prices, uncovering an overlooked aspect of retail crimes' social costs. We show this using novel matched administrative datasets from the Washington state cannabis industry which provides an ideal setting to estimate causal relationships and disentangle the channels driving the pass-through of retail crime. 

While the cannabis industry is distinct, our findings have significant welfare implications beyond our context. 
The cannabis industry shares several important characteristics with other retail sectors. These include a comparable variable cost structure (see Appendix \ref{sec:industry}) and similar demand elasticities, both of which are important determinants of cost pass-through \citep{hollenbeck2021}. The lack of a competing black market \citep{hollenbeck2021} underscores the generalizability of our results. Further support for the broad applicability of our findings comes from a variety of news reports, a recent survey by the \citet{nrf_retail_security_2022}, and from findings of a study by \citet{jackson2020}, who show that a decreased likelihood of a felony conviction is connected to increased prices for automobiles and computers. These sources highlight that increased crime-related costs are a widespread concern for retailers in numerous sectors, suggesting that these costs are passed on to consumers in a variety of settings.

We estimate an annual excess burden for the Washington state cannabis industry of around \$20.2 million. In 2020, the state's total cannabis sales amounted to \$1.4 billion, while total U.S. retail sales reached \$5,572 billion \citep{census2022}. When we scale the excess burden of retail crime pass-through to all U.S. retailers based on the relative shares of sales, our estimates imply an additional annual welfare cost of about \$80 billion nationwide. This naive extrapolation requires careful interpretation since the impact of retail crime on prices likely differs across sectors due to different market structures and crime rates. Nevertheless, even if the pass-through rate of retail crime in other sectors is substantially lower, our analysis underscores the importance of considering pass-through effects when evaluating the social costs of retail crime. Furthermore, with state and local governments allocating approximately \$126 billion to police protection in 2020 \citep{census2022expend}, our findings provide a compelling argument for increasing public expenditure on crime reduction.

Our welfare analysis makes some restrictive assumptions on the implications of retail crime pass-through. For instance, we abstract from the potential negative external effects of cannabis consumption. While the cannabis sales tax aims to internalize these external effects, the identified price hikes may still have beneficial corrective effects for some consumers.  Moreover, we neglect the distributional implications of our results. The price effects may be regressive in the sense that they fall disproportionally on low-income and younger individuals, aligning with the demographics of cannabis consumers (see Appendix \ref{sec:industry}). A detailed analysis of these additional welfare implications goes beyond the scope of this paper and is left for future research.

\clearpage
\appendix

\begin{center}
\LARGE{\textbf{ Appendix}}
\end{center}

\renewcommand{\thefigure}{A.\arabic{figure}}
\setcounter{figure}{0}

\renewcommand{\thetable}{A.\arabic{table}}
\setcounter{table}{0}
\setcounter{page}{1}


\section{Cannabis industry background}

\subsection{Cannabis consumption and supply in Washington state}\label{sec:industry}

\paragraph{Cannabis consumers}

\begin{figure}[!htbp]
\caption{Demographic characteristics of regular cannabis consumers}
\centering
	\begin{subfigure}{.4\textwidth}
	\centering
	    \includegraphics[width=\linewidth]{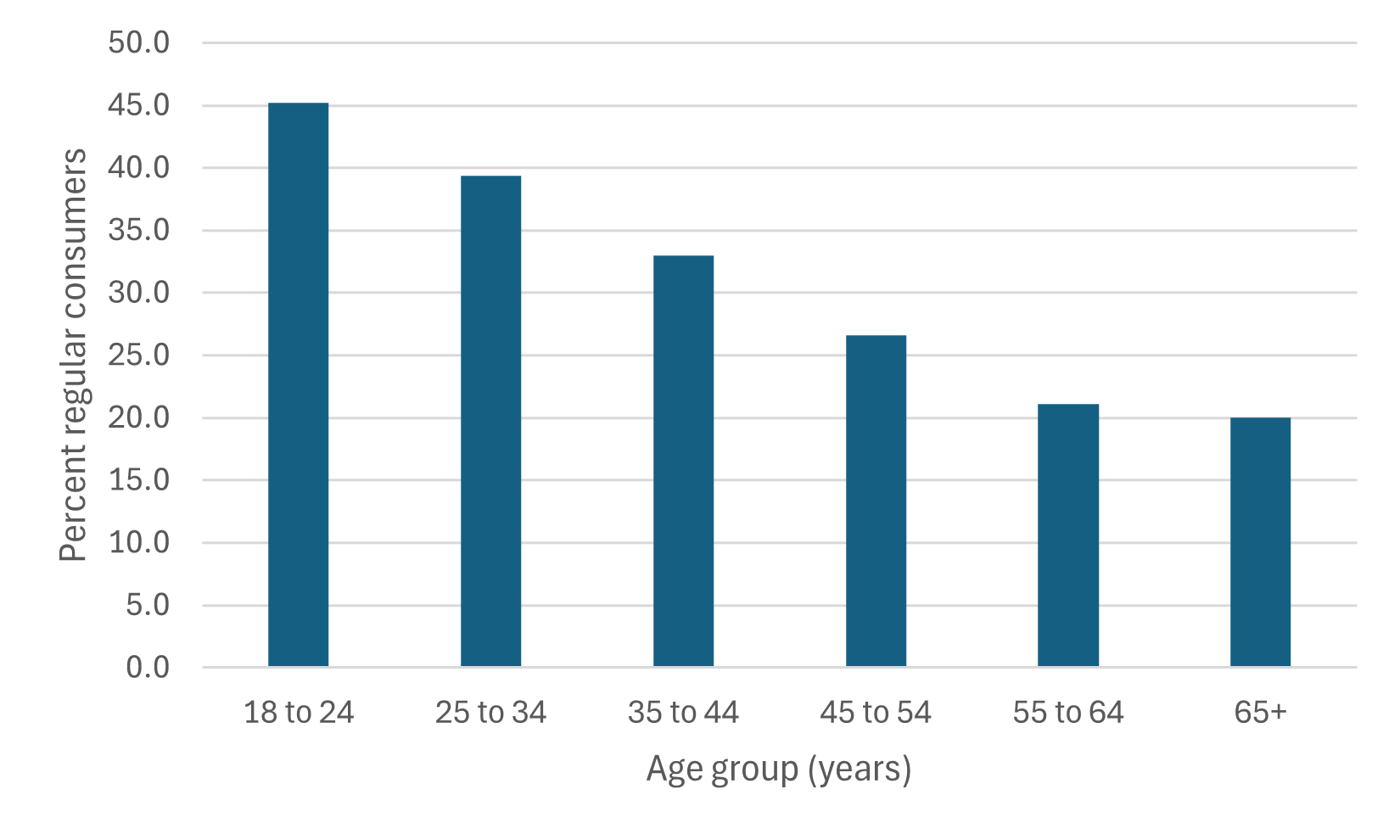}
		\caption{Age}
	\end{subfigure}\hfil
	\begin{subfigure}{.4\textwidth}
	\centering
	    \includegraphics[width=\linewidth]{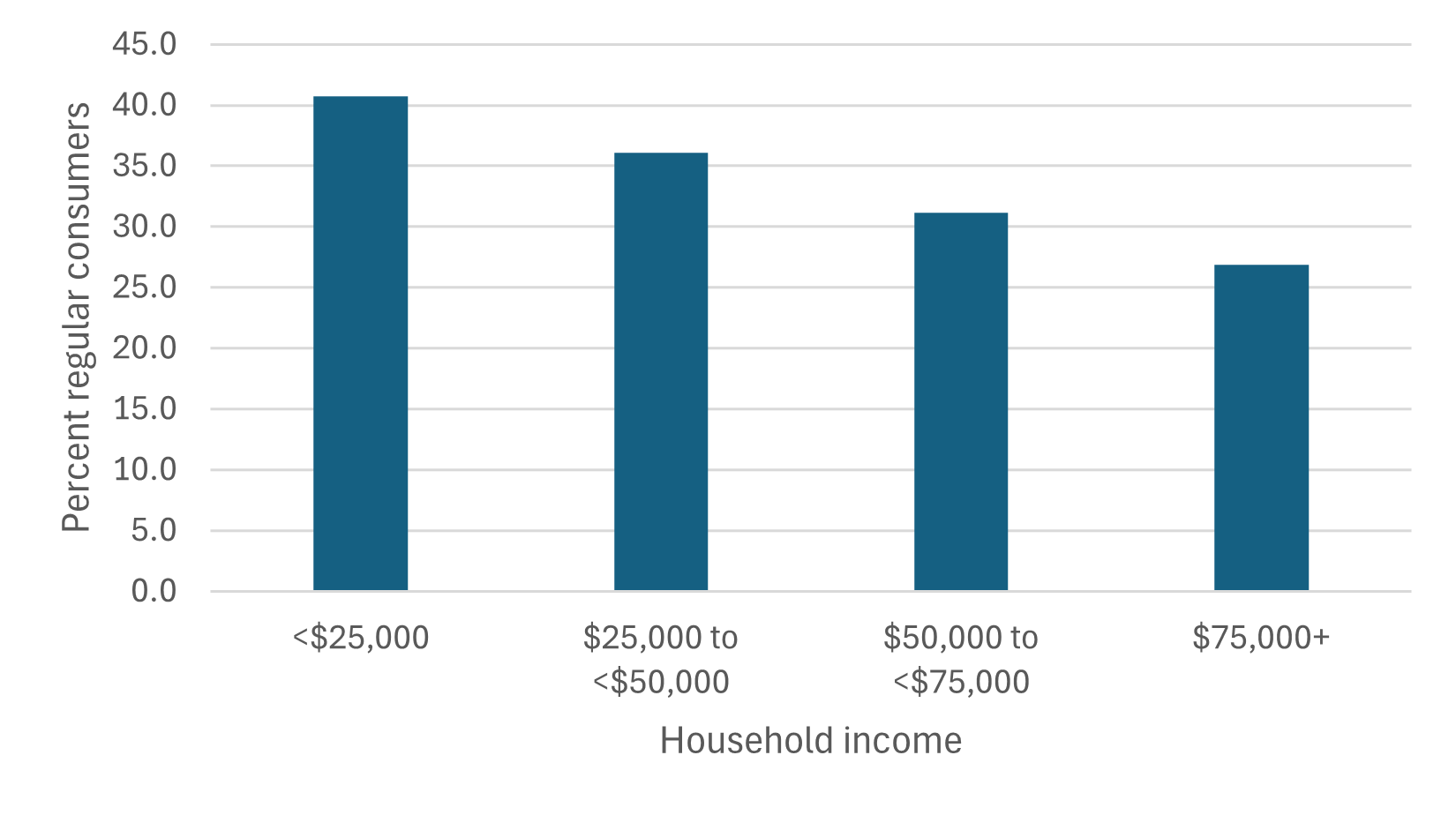}
		\caption{Household income}
	\end{subfigure}\hfil
 \begin{subfigure}{.4\textwidth}
	\centering
	    \includegraphics[width=\linewidth]{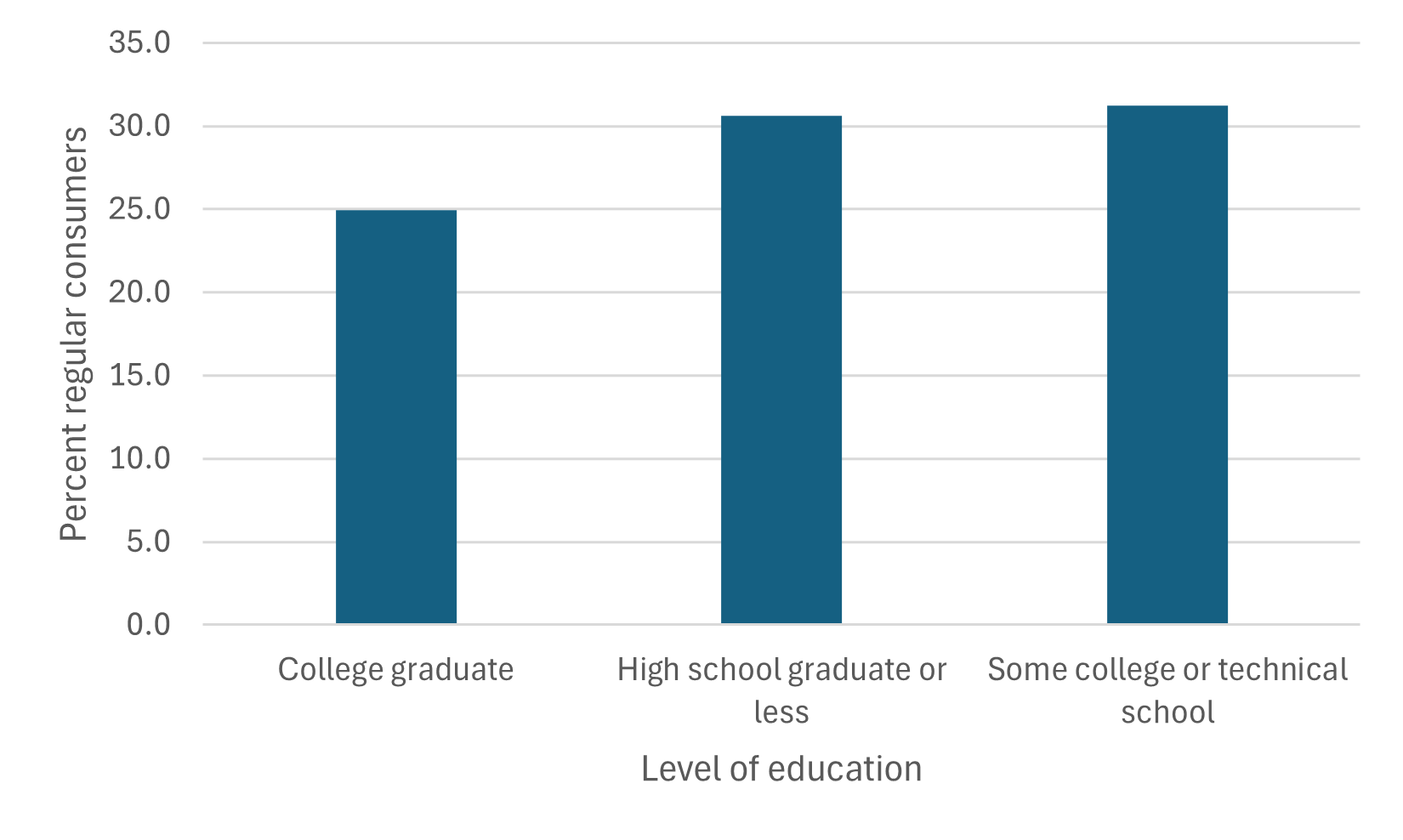}
		\caption{Education}
	\end{subfigure}\hfil
	\begin{subfigure}{.4\textwidth}
	\centering
	    \includegraphics[width=\linewidth]{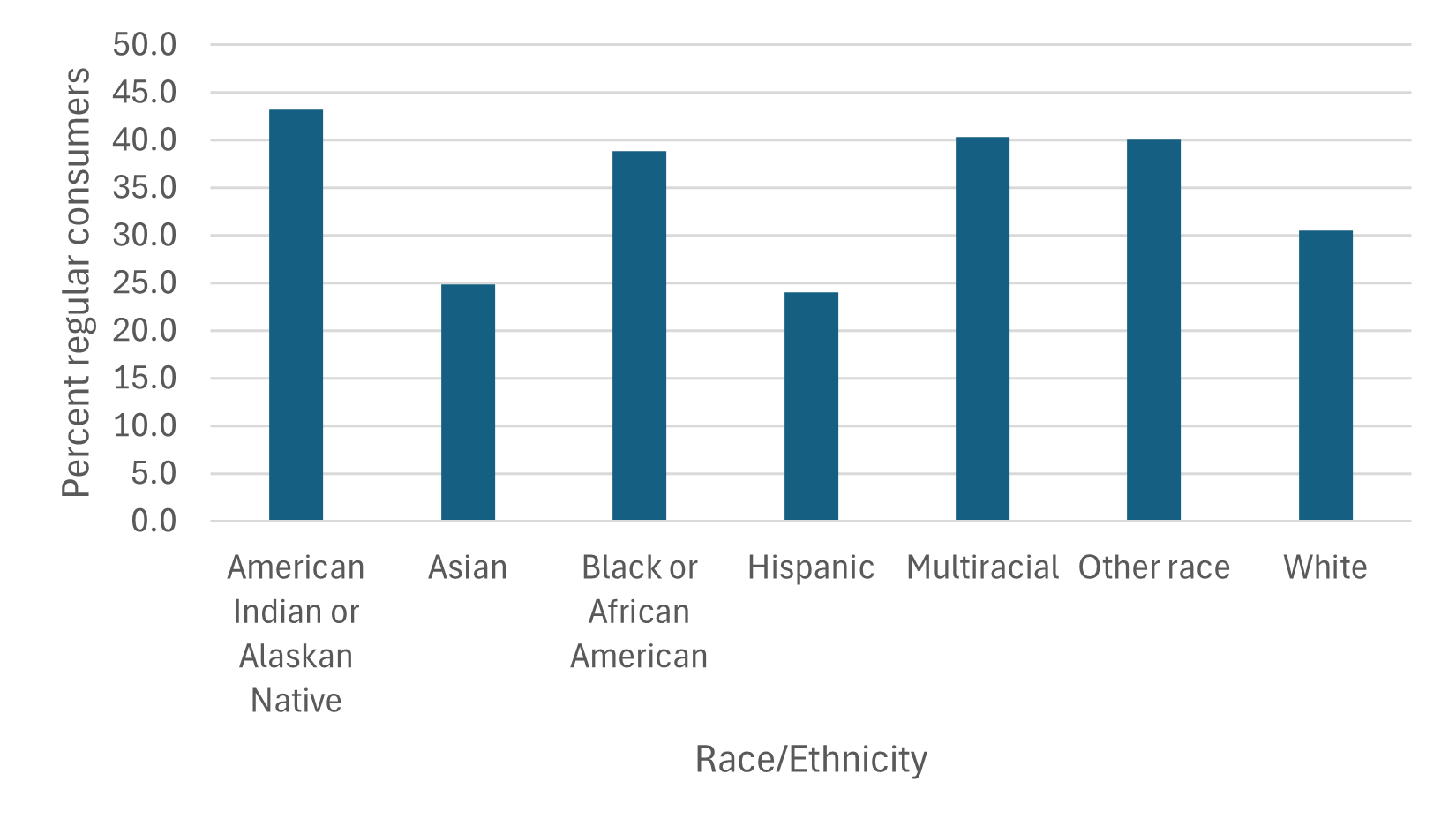}
		\caption{Race/Ethnicity}
	\end{subfigure}\hfil
	\begin{subfigure}{.4\textwidth}
	\centering
	    \includegraphics[width=\linewidth]{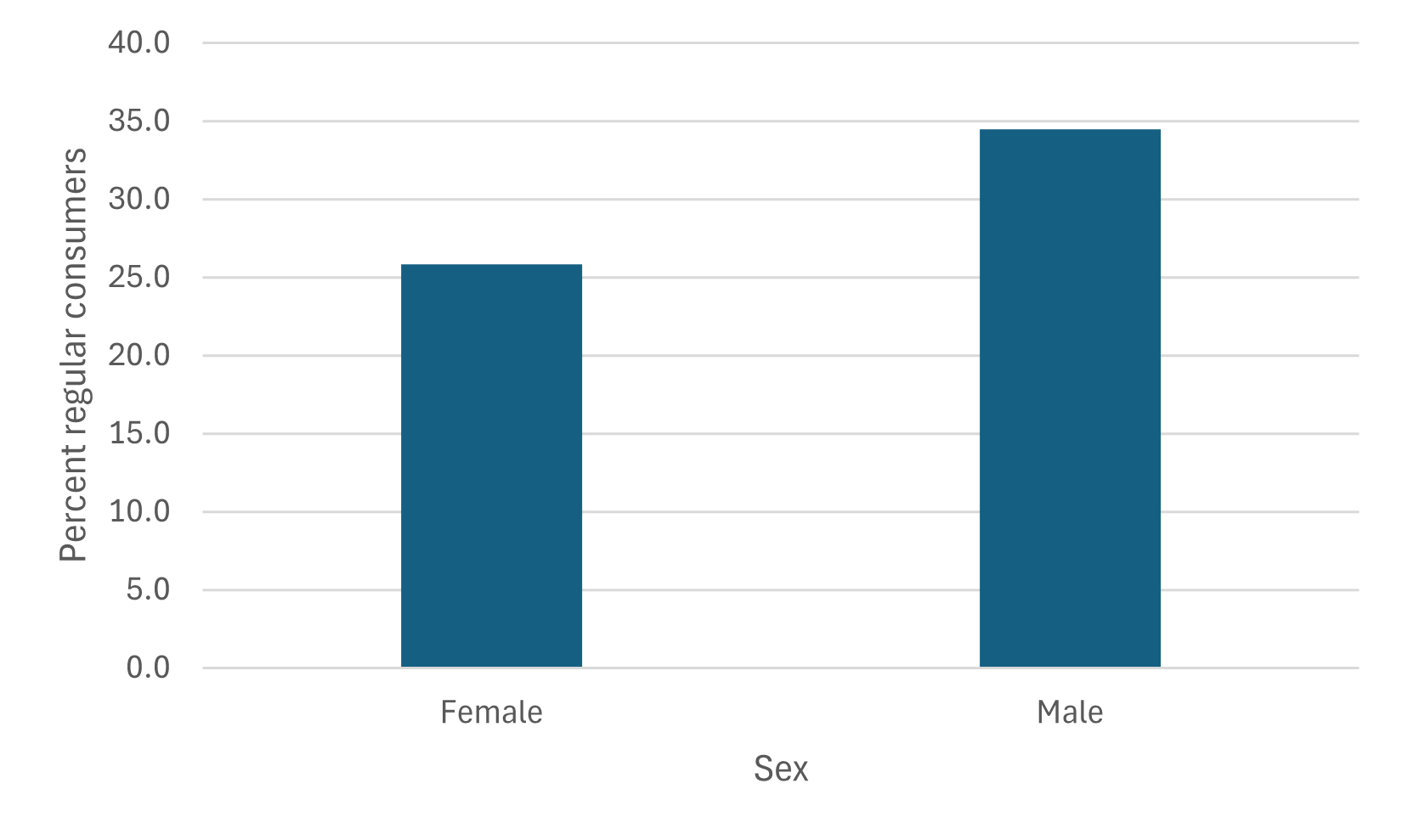}
		\caption{Sex}
	\end{subfigure}
\label{fig:demographics}
\par
\rule{\textwidth}{0.4pt}
\begin{minipage}[h]{\textwidth}
\medskip
\footnotesize \emph{Notes:} This figure presents the distribution of regular cannabis consumers in Washington state, broken down by various demographic characteristics: (A) age, (B) household income, (C) education level, (D) race/ethnicity, and (E) gender. Each bar represents the proportion of regular cannabis consumers within the respective subgroup. The data is from the 2021 Behavioral Risk Factor Surveillance System by the Washington State Department of Health, Center for Health Statistics.
\end{minipage}
\end{figure}

Cannabis use is widespread in Washington state. Approximately 30\% of Washington adults consume cannabis on a monthly basis \citep{wadoh}. For context, about 10\% of adults in Washington consume cigarettes. Figure \ref{fig:demographics} shows cannabis consumption in Washington state along various demographic lines. The data come from the Behavioral Risk Factor Surveillance System (BRFSS) survey, an annual survey conducted by the Department of Health in partnership with the Centers for Disease Control and Prevention (CDC). The purpose of the survey is to measure changes in the health behaviors of people in Washington state \citep{wadoh}.

The figures show the percent of each demographic group that consumes cannabis at least once a month. Panel A illustrates that over 40\% of adults age 18 to 24 use cannabis regularly; the same holds for adults age 25 to 34. 33\% of adults age 35 to 44 consume regularly, while only one in five adults aged 55+ consume regularly. Panel B shows that regular cannabis consumption decreases monotonically with household income. There are no major differences between levels of education (Panel C). Panel D shows that consumption is highest among American indian, black, multiracial, and other (approx 40\%), while consumption is lowest for asian and hispanic adults (approx. 25\%). Panel E shows that males consume more than females (35\% vs 25\%).


\begin{figure}[!htbp]
\caption{The cannabis supply chain}
\centering
	\includegraphics[width=0.6\textwidth]{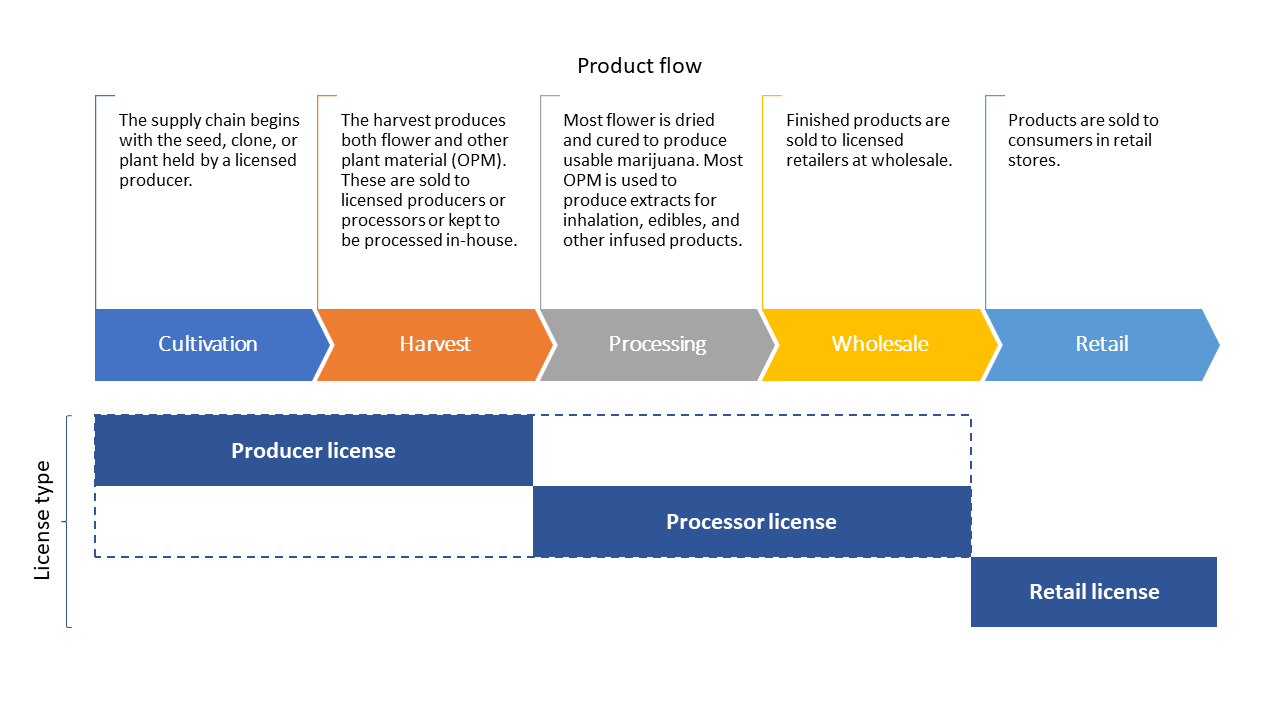}
\label{fig:supplychain}
\par
\rule{\textwidth}{0.8pt}
\begin{minipage}[h]{\textwidth}
\medskip
\footnotesize \emph{Notes:} This figure depicts the flow of cannabis products, from left to right, as they move through the supply chain. Only licensed producers are permitted to cultivate and harvest cannabis plants; producers can only sell to licensed processors, who in turn are permitted to process products; only processors can sell finished products at wholesale to retailers; licensed retailers can sell finished products to end consumers. An establishment can jointly hold producer and processor licenses, so the overwhelming majority of upstream establishments hold both licenses (i.e. producer-processors). Retailers may not hold a producer or a processor license and vice versa. As a result, production and retail activities are legally separated.
\end{minipage}
\end{figure}

\paragraph{The cannabis supply chain}
Figure \ref{fig:supplychain} shows the stages of the cannabis supply chain. Most cannabis in Washington is grown in indoor facilities ranging in size from 2,000 to 30,000 square feet of plant canopy. When plants reach a mature stage, their buds are harvested, dried, and cured. The majority of cannabis is consumed in this unprocessed form (called "Usable marijuana") while the rest is processed into derivative subproducts like edibles and concentrates.

\newpage

\subsection{Cannabis retail stores}

\paragraph{Store characteristics}\label{sec:sales_across_stores}

\begin{figure}[!htbp]

\caption{Distribution of monthly averages across stores}
\centering
	\begin{subfigure}{.4\textwidth}
	\centering
	    \includegraphics[width=\linewidth]{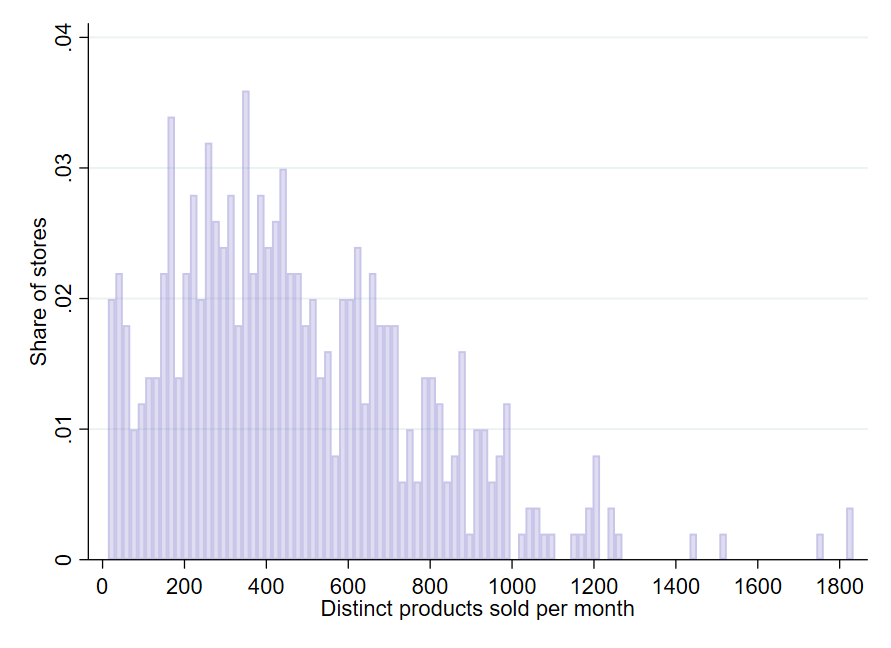}
		\caption{Product variety}
	\end{subfigure}
	\begin{subfigure}{.4\textwidth}
	\centering
	    \includegraphics[width=\linewidth]{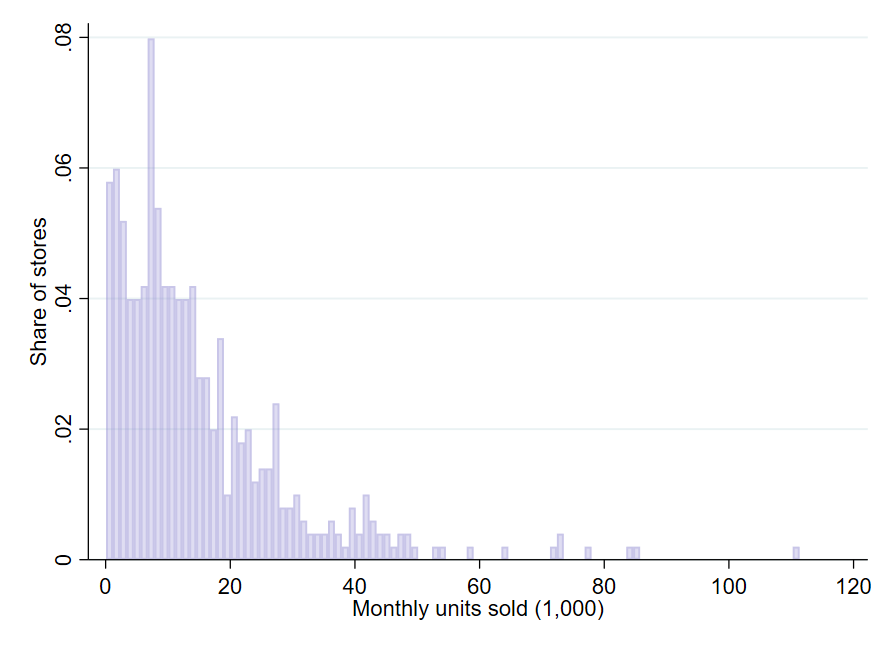}
		\caption{Units sold}
	\end{subfigure}
 \begin{subfigure}{.4\textwidth}
	\centering
	    \includegraphics[width=\linewidth]{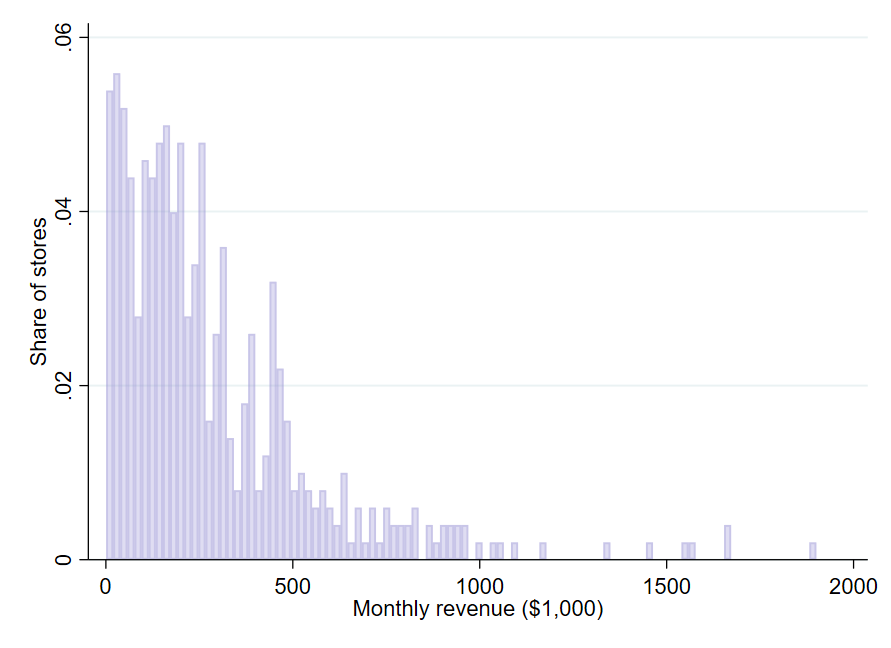}
		\caption{Revenue (tax-inclusive)}
	\end{subfigure}
\label{fig:descriptive_distributions}
\par
\rule{\textwidth}{0.4pt}
\begin{minipage}[h]{\textwidth}
\medskip
\footnotesize \emph{Notes:} The figures show the distribution of store-level average statistics across all stores in our sample. Panel A presents the distribution of the average monthly number of unique products sold. Panel B displays the distribution of the monthly average units sold. Panel C illustrates the distribution of average monthly sales revenue across stores. 
\end{minipage}
\end{figure}

Figure \ref{fig:descriptive_distributions} shows the distribution of store-level monthly averages for various store characteristics. Panel A shows the average number of distinct products sold per month across stores. A 1.0 gram package and a 2.0 gram package of Sunset Sherbert usable marijuana (i.e. unprocessed dried flower) produced by Northwest Harvesting Co are examples of distinct products in our data. The average store in our sample sells 473 distinct products per month (median: 419). However, Panel A reveals substantial variation across stores in our sample, with values ranging from as low as 13 to a maximum of 1,833 products per month. Panel B reports the average units sold per month across stores (in thousands). The average store in our sample sells 14,836 units per month (median: 10,905). As is the case with product variety, there is large variation in units sold across stores. Stores at the 1st percentile sell 287 units per month, while those at the 99th percentile sell 72,826 units per month. Panel C shows the distribution of tax-inclusive monthly revenue across stores. The average store generates \$285,320 revenue per month (median: \$205,377). Again, revenue varies across stores: stores at the 1st percentile generate \$3,765 while those at the 99th percentile generate \$1,447,000 per month. 

\newpage 
\paragraph{Variable cost structure for cannabis retailers}\label{sec:variable_costs}

To ascertain the variable cost structure for cannabis retailers, we use aggregate payroll data on cannabis retailers from the Washington State Employment Security Department (ESD) and High Peak Strategy. The ESD collects data on employment and wages in industries covered by unemployment insurance (95\% of U.S. jobs). The data spans the years 2018-2020. Table \ref{tab:e1} illustrates that cannabis retailers have a similar variable cost structure as other retail industries studied in the literature. \cite{renkin2020}, for example, find that for U.S. grocery stores, COGS accounts for 83\% of variable costs. Note that in most retail settings, cost of goods sold (COGS) and labor cost together account for 99\% of variable cost while other expenditures like packaging and transport typically make up less than 1\% of variable cost \citep{renkin2020}.
 
\begin{table}[!htbp] 
\centering
\caption{COGS and the labor share of costs for cannabis retailers}
\label{tab:e1}
\renewcommand{\tabcolsep}{.5pt}{
\def\sym#1{\ifmmode^{#1}\else\(^{#1}\)\fi}
\begin{tabular*}{\hsize}{@{\hskip\tabcolsep\extracolsep\fill}l*{5}{c}}
\toprule 

& \multicolumn{2}{c}{Average expenditure} & & \multicolumn{2}{c}{Variable cost share} \\
\cline{2-3} \cline{5-6} \\

\multicolumn{1}{c}{Year} & \multicolumn{1}{c}{Labor} & \multicolumn{1}{c}{COGS} & & \multicolumn{1}{c}{Labor} & \multicolumn{1}{c}{COGS} \\

\midrule

2018  &  \$324,582 & \$702,358 & & 0.32 & 0.68 \\
\addlinespace

2019 & \$370,897 & \$1,187,462 & & 0.24 & 0.76 \\
\addlinespace

2020 & \$407,273 & \$1,584,301 & & 0.20 & 0.80 \\
\bottomrule

\end{tabular*}
\begin{minipage}[h]{\textwidth}
\medskip
\footnotesize \emph{Notes:} This table compares average annual labor expenditure and COGS expenditure for cannabis retail stores in Washington state for the years 2018-2020. Aggregate payroll data on cannabis retailers is from the Washington State Employment Security Department and High Peak Strategy (2018-2020). Labor expenditure equals total wages divided by the number of active stores. Stores with missing UI data are excluded from total wages and establishment counts. COGS is the average annual wholesale expenditure for cannabis retailers in the estimation sample. Wholesale purchases from processor-only licenses are included. 
\end{minipage}

}
\end{table}

\newpage
\paragraph{The geography of wholesale costs}

Table \ref{tab:h.1} shows the percentage of retailers' wholesale costs in relation to a producer's geographic location. Column 1 shows that only 5.22\% of retailers' wholesale expenditures go to producers located in the same city as the retailer. Column 2 shows that less than 15\% goes to producers in the same county as the retailer. For Column 3, we sort counties into their respective 3-digit zip codes (retailers are located in 14 3-digit zip codes compared to 37 counties). Column 3 shows that less than 16\% of wholesale cost goes to producers located in the same 3-digit zip code. Next, we sort counties into three regions (west, central, east), defined by well-established topographic and economic boundaries. Column 4 shows that 62\% of wholesale sales go to retailers in a different region than the producer. Column 5 looks at the subset of establishments located in the west and east regions of the state, thus dropping producers in the central region. The east and west regions are non-contiguous and are located on opposite sides of the state. For establishments located in these two regions, 23.9\% of wholesale sales go to retailers located in the other region, that is to say, retailers on the opposite side of the state. Because the majority of retail establishments are located in the west and east regions, this share amounts to 21.4\% of all of wholesale expenditures in the industry. Taken together, the results from Table \ref{tab:h.1} illustrate that there is no home bias in wholesale cannabis purchases.

\begin{table}[!htbp] 
\centering
\caption{Share of retailers' wholesale costs by geographic proximity}
\label{tab:h.1}
\renewcommand{\tabcolsep}{1pt}{
\def\sym#1{\ifmmode^{#1}\else\(^{#1}\)\fi}
\begin{small}\begin{tabular*}{\hsize}{@{\hskip\tabcolsep\extracolsep\fill}l*{6}{c}}
 \toprule
 \addlinespace
            &\multicolumn{1}{c}{(1)} &\multicolumn{1}{c}{(2)} & \multicolumn{1}{c}{(3)} &\multicolumn{1}{c}{(4)} &\multicolumn{1}{c}{(5)}  &\multicolumn{1}{c}{(6)}\\
\addlinespace
            &\multicolumn{1}{c}{\parbox{1cm}{Same city}}&\multicolumn{1}{c}{\parbox{1cm}{Same county}} &\multicolumn{1}{c}{\parbox{1cm}{Same 3-digit zip code}} &\multicolumn{1}{c}{\parbox{1cm}{Same region}} &\multicolumn{1}{c}{\parbox{1cm}{Non-contiguous region}}  &\multicolumn{1}{c}{\parbox{1cm}{Same state}}\\
\addlinespace
\midrule
\multicolumn{1}{l}{\parbox{2cm}{Percent of \\ wholesale expenditure}}   & 5.22\% & 14.67\% & 15.59\% & 62.08\% & 23.90\%  & 100\%\\
\bottomrule
\end{tabular*}\end{small}
\begin{minipage}[h]{\textwidth}
\medskip
\footnotesize \emph{Notes:} This table shows the share of retailers' wholesale expenditure according to wholesalers' geographic proximity. The shares are based on 5.92 million unique wholesaler-retailer-product-month observations from August 2018 through July 2021. Retailers are located in 14 3-digit zip codes and 35 counties. Region groups counties into three categories: west, central, or east. Data from Top Shelf Data.
\end{minipage}
}

\end{table}

\subsection{Seasonality of crimes}\label{sec:seasonality}

\begin{figure}[!htbp]
\caption{Reported crimes at cannabis retailers by calendar month}
\centering
	\includegraphics[width=0.6\textwidth]{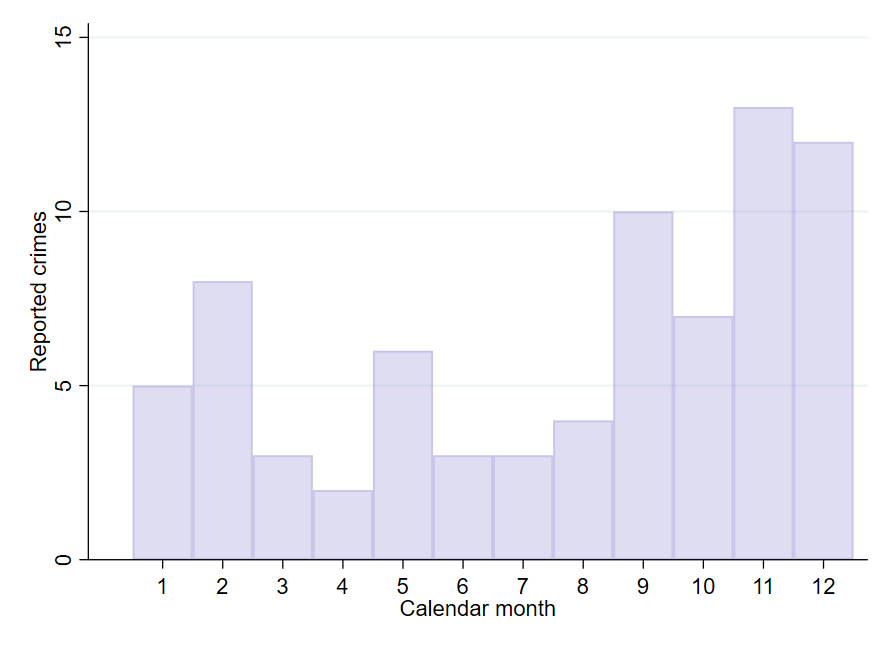}
\label{fig:crime_seasonality}
\par 
\rule{\textwidth}{0.8pt}
\begin{minipage}[h]{\textwidth}
\medskip
\footnotesize \emph{Notes:} This figure illustrates the seasonality in treatment timing within our sample. Each bar represents the number of crimes that occurred in the respective calendar month, with 1 indicating January and 12 indicating December.
\end{minipage}
\end{figure}

Table \ref{fig:crime_seasonality} illustrates the seasonality in crime incidents within our sample. Each bar represents the number of crimes that occurred in the respective calendar month, with 1 indicating January and 12 indicating December. The figure shows that crime occurs throughout all months but that rates tend to be higher towards the end of the year.

\newpage
\section{Establishment-level indexes}\label{sec:indexes}

This section provides more detail on the store-level indexes used in our analysis.

\subsection{Price indexes}
Our empirical analysis uses traceability data provided by the data analytic firm Top Shelf Data (TSD), which ingests the raw tracking data from the Washington state Liquor and Cannabis Board (LCB) and matches it with additional product information. Note that the raw tracking data from the LCB includes each product's Stock Keeping Unit (SKU), but TSD does not report this. Instead, each product is identified by a unique combination of five elements: retailer-producer-category-product name-unit weight. For products with no unit weight (such as liquid edibles), the first four elements identify the product. TSD then calculates the average price of product $i$ at retail store $j$ in month $t$ as
\begin{equation}
    P_{i,j,t} = \frac{TR_{i,j,t}}{TQ_{i,j,t}}.
\end{equation}
 where $TR_{i,j,t}$ is the revenue from product $i$ at retailer $j$ in month $t$, and $TQ_{i,j,t}$ is total quantity.
 
 To construct store-level price indexes, we employ a two-step process similar to that used by \cite{renkin2020}. In the first step, we use $P_{i,j,t}$ to construct a geometric mean of month-over-month changes for product subcategory $c$ at store $j$:
\begin{equation}\label{eq:a.1}
    I_{c,j,t} = \prod_i \left( \frac{P_{i,j,t}}{P_{i,j,t-1}}\right)^{\omega_{i,c,j,y(t)}}
\end{equation}
where each subcategory is a unique category-unit weight combination.\footnote{Since unit weight is a major component of cannabis product differentiation (akin to volume in beverage sales), the majority of sales contain information on unit weight. Therefore, in the first step of the establishment index, we choose to aggregate at category-unit weight level rather than the category level.} For example, usable marijuana is a category, whereas 1.0 gram usable marijuana and 2.0 gram usable marijuana are separate subcategories. Following \cite{renkin2020}, the weight $\omega_{i,c,j,y(t)}$ is the share of product $i$ in total revenue of subcategory $c$ in establishment $j$ during the calendar year of month $t$.\footnote{As pointed out by\cite{renkin2020}, price indexes are sometimes constructed using lagged quantity weights. Since product turnover can be high in retail settings---and cannabis is no exception---lagged weights would limit the number of products used in constructing the price indexes. Thus, contemporaneous weights are used.}

In the second step, we aggregate across subcategories to get the price index for store $j$ in month $t$:
\begin{equation}
    I_{j,t} = \prod_c I_{c,j,t}^{\omega_{c,j,y(t)}}.
\end{equation}
Similar to the first step, the weight $\omega_{c,j,y(t)}$ is the share of subcategory $c$ in total revenue in store $j$ during the calendar year of month $t$. The store-level inflation rate is then simply the natural logarithm of the index
\begin{equation}
    \pi_{j,t} = \ln I_{j,t}
\end{equation}

\subsection{Quantity indexes}
The quantity indexes are constructed the same way as the price indexes. The only difference is in the first step
\begin{equation}
    I_{c,j,t} = \prod_i \left( \frac{Q_{i,j,t}}{Q_{i,j,t-1}}\right)^{\omega_{i,c,j,y(t)}}
\end{equation}
where $Q_{i,j,t}$ is the quantity sold of product $i$ at store $j$ in month $t$. The index weights are otherwise identical to those from the price index.

\subsection{Wholesale cost indexes}
The wholesale cost indexes are constructed similar to the price indexes. TSD calculates the average monthly wholesale price at the SKU-level as
\begin{equation}
    W_{i,j,t} = \frac{TE_{i,j,t}}{TQ_{i,j,t}}.
\end{equation}
where $TE_{i,j,t}$ is the total expenditure on product $i$ at retailer $j$ in month $t$, and $TQ_{i,j,t}$ is total quantity purchased at wholesale. In the first step of the establishment-level cost index, we construct a geometric mean of the month-over-month changes in wholesale price for product subcategory $c$ at store $j$:
\begin{equation}\label{eq:whole_cost_index}
    I_{c,j,t} = \prod_i \left( \frac{W_{i,j,t}}{W_{i,j,t-1}}\right)^{\omega_{i,c,j,y(t)}}
\end{equation}
where the subcategories are the same as before. In contrast to the price indexes, $\omega_{i,c,j,y(t)}$ is an expenditure weight equal to the share of product $i$ in the wholesale expenditure of subcategory $c$ at store $j$ in the calendar year of month $t$. Note that by calculating this weight based on annual (rather than monthly) expenditure shares, we may not capture short-term wholesale substitution patterns on the part of retailers. To check this, we also construct the index using monthly expenditure shares. This increases the variance of the indexes but does not change our main estimation results. Therefore, we use the annual expenditure weights for our preferred specification.\footnote{Results using monthly weights are available on request.}

In the second step, we aggregate across subcategories to get the wholesale cost index for store $j$ in month $t$:
\begin{equation}
    I_{j,t} = \prod_c I_{c,j,t}^{\omega_{c,j,y(t)}}.
\end{equation}
Similar to the first step, the weight $\omega_{c,j,y(t)}$ is the share of subcategory $c$ in total wholesale expenditure in store $j$ during the calendar year of month $t$. The store-level wholesale cost index is then simply the natural logarithm of the index, $\ln I_{j,t}$.

\begin{figure}[!htbp]
\caption{Distributions for store-level indexes}
\label{fig:index_distributions}
\centering
	\begin{subfigure}{.4\textwidth}
	\centering
	    \includegraphics[width=\linewidth]{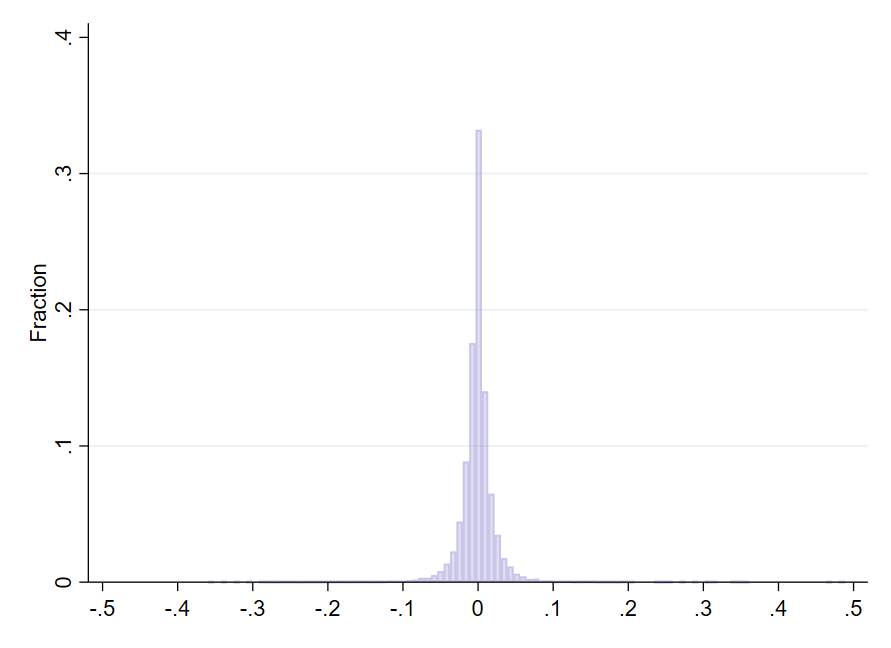}
	\caption{Price}
    \end{subfigure}\hfil
	\begin{subfigure}{.4\textwidth}
	\centering
	    \includegraphics[width=\linewidth]{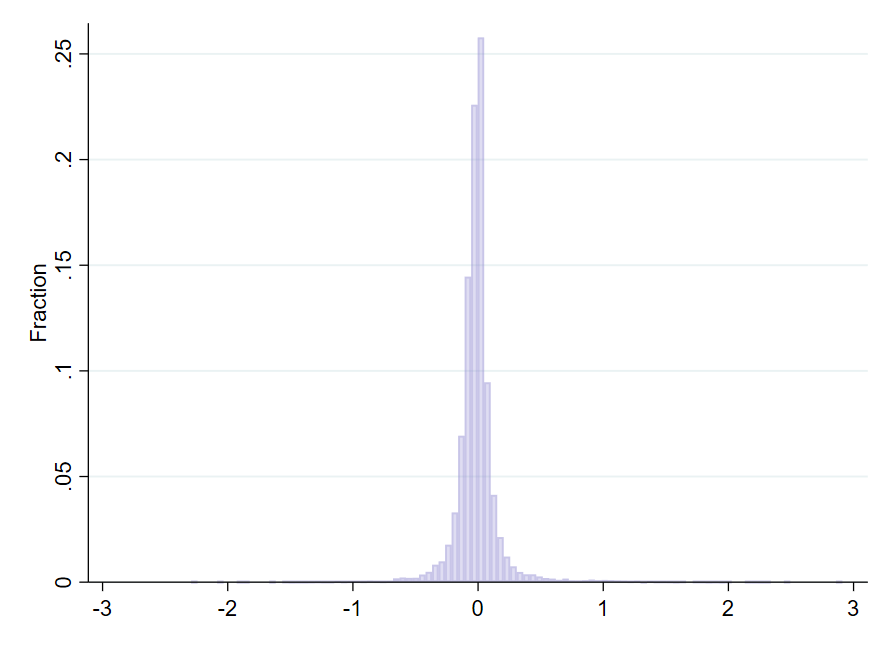}
		\caption{Quantity sold}
    \end{subfigure}\hfil
	\begin{subfigure}{.4\textwidth}
	\centering
	    \includegraphics[width=\linewidth]{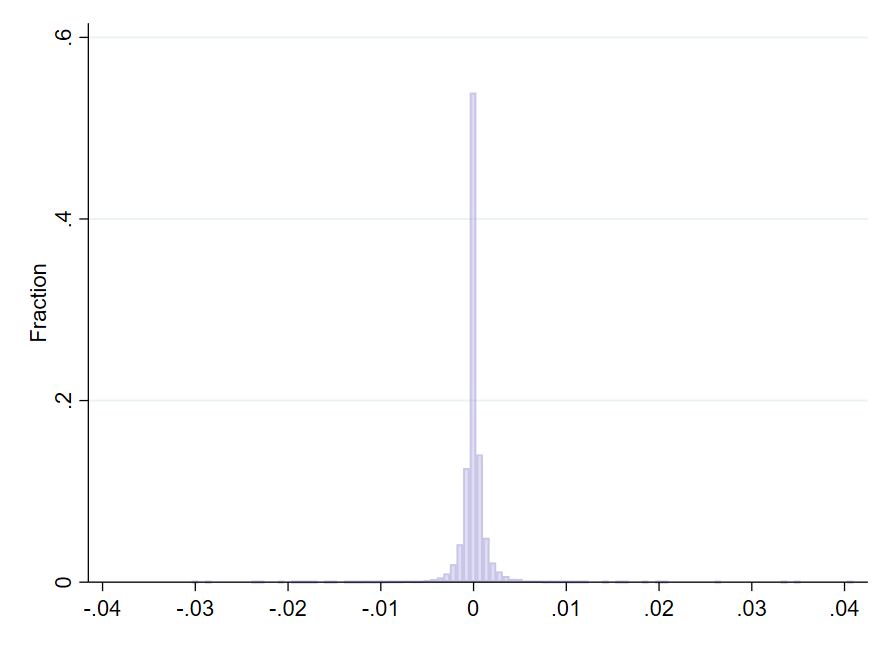}
     \caption{Wholesale cost}
	\end{subfigure}\hfil
\par
\rule{\textwidth}{0.4pt}
\begin{minipage}[h]{\textwidth}
\medskip
\footnotesize \emph{Notes:} The figures show the distribution of the main dependent variables over the sample period. Panel A depicts the monthly store-level price index. Panel B illustrates the monthly store-level quantity index. Panel C displays the monthly store-level wholesale cost index. For visual clarity, extreme outliers are omitted from the figures, although they are included in our regressions.
\end{minipage}
\end{figure}

\newpage
\section{Additional heterogeneity analysis and descriptive statistics}\label{sec:appendix_heterogeneity}

\subsection{Independent versus chain stores}

\begin{table}[!htbp] 
\centering
\caption{Cannabis chain store characteristics}
\renewcommand{\tabcolsep}{1pt}{
\def\sym#1{\ifmmode^{#1}\else\(^{#1}\)\fi}
\begin{tabular*}{\hsize}{@{\hskip\tabcolsep\extracolsep\fill}l*{2}{c}}
\toprule

 &\multicolumn{1}{c}{(1)}&\multicolumn{1}{c}{(2)} \\
 
&\multicolumn{1}{c}{Independent} &\multicolumn{1}{c}{Chains}  \\
\addlinespace
\midrule
\addlinespace
\parbox{4cm}{Unit price \\ (in dollars)} & 26.74 & 26.78 \\
& (4.29) & (4.58) \\ 
\addlinespace
\parbox{4cm}{Units sold \\ per month} & 13,421 & 16,161 \\ 
            & (13,068) & (15,399) \\ 
\addlinespace
\parbox{4cm}{Monthly revenue \\ (in dollars)} & 257,473 & 312,154 \\
&  (252,444) & (301,285) \\ 
\parbox{4cm}{Unique products \\ per month} & 460 & 553 \\ 
                & (379) & (391) \\ 
\addlinespace
$N$ & 330 & 181 \\
\addlinespace
\bottomrule
\addlinespace
\end{tabular*}
\begin{minipage}[h]{\textwidth}
\footnotesize \emph{Notes:} The table reports descriptive statistics for chains and independent stores using the sample of active cannabis retail stores as defined in the main paper. Standard deviations are in parentheses. The variables reported are: unit price, average store quantity sold per month, average store revenue per month, and average number of distinct products sold per month.
\end{minipage}

}
\label{tab:chains_vs_independent_balance}
\end{table}

In Section \ref{sec:heterogeneity}, we distinguished between small and large chains, with the former defined as stores belonging to firms with 1-2 stores and the latter corresponding to three or more stores. Table \ref{tab:chains_vs_independent_balance} reports descriptive statistics for the different store types, based on all 511 active cannabis retailers.

In Table \ref{tab:heterogeneity_robustness} we report alternative specifications for our heterogeneity analyses. In this specification we define independent stores as stores belonging to a single-store firm, and chain stores as stores at firms with two or more stores. Column 1 and 5 show that effects at independent stores tend to be smaller than the effects at small chains found in Section \ref{sec:heterogeneity}. Moreover, Columns 2 and 6 show a larger the effect at chains than in those found in Section \ref{sec:heterogeneity}. This redistribution of treatment effects reflects that stores in small chains (i.e. two-store chains) have a large treatment effect that is similar in magnitude to independent stores. This points to small chain stores facing similar competitive pressures as independent stores, and validates our definition of independent and chain stores used in our main heterogeneity analysis in Section \ref{sec:heterogeneity}.

\subsection{Market concentration} 

Table \ref{tab:hhi_descriptive_statistics} shows the descriptive statistics for our store-level market concentration measure, as defined in our main text.

\begin{table}[!htbp] 
\centering
\caption{Market concentration descriptive statistics}
\renewcommand{\tabcolsep}{1pt}{
\def\sym#1{\ifmmode^{#1}\else\(^{#1}\)\fi}
\begin{tabular*}{\hsize}{@{\hskip\tabcolsep\extracolsep\fill}l*{3}{c}}
\toprule

 &\multicolumn{1}{c}{(1)}&\multicolumn{1}{c}{(2)}&\multicolumn{1}{c}{(3)} \\
 
&\multicolumn{1}{c}{Mean} &\multicolumn{1}{c}{Median} &\multicolumn{1}{c}{Std} \\
\addlinespace
\midrule
\addlinespace
Victimized HHI & 0.15 & 0.11 & 0.19 \\
All HHI & 0.27 & 0.15 & 0.28 \\

\bottomrule
\addlinespace
\end{tabular*}
\begin{minipage}[h]{\textwidth}
\footnotesize \emph{Notes:} The table shows descriptive statistics for our market concentration measure, calculated on the store-level. The first row only considers victimized stores. the second row considers all stores withing the Washington State cannabis industry.
\end{minipage}

}
\label{tab:hhi_descriptive_statistics}

\end{table}

In Section \ref{sec:heterogeneity}, we found large effects at stores in markets with low concentration, and no effect at markets with high concentration. One possible reason for this finding is that low concentration markets are generally located in urban areas, where labor costs (and hence the security cost shock) may be higher compared to rural areas. To test this, we use an alternative definition of low and high concentration markets that accounts for urban-rural heterogeneity. We proceed as follows. First, we categorize as urban stores that are located in the four largest and most densely populated cities (Seattle, Tacoma, Bellevue, and Spokane), and consider all other stores as rural. Next, we calculate the median HHI for stores in the urban subsample and the median HHI for stores in the rural subsample. We split the urban subsample into high concentration and low concentration parts, and do the same for the rural subsample. This leaves us with four subsamples: urban low concentration, urban high concentration, rural low concentration, and rural high concentration. Finally, we pool the high concentration subsamples into a single subsample, and the low concentration subsamples into another subsample. Our low concentration subsample thus contains stores in markets considered low concentration in an urban setting, but also stores in markets considered low concentration in a rural setting. Finally, we estimate our main regression equation separately for the high concentration and the low concentration subsamples. Columns 3-4 show that at victimized stores, six months after a crime the difference in effect sizes between low and high concentration markets is 0.007 while the difference in effect sizes in Section \ref{sec:heterogeneity} is 0.022. Thus, accounting for the urban-rural divide appears to explain the comparatively large effect size for low concentration markets found in Section \ref{sec:heterogeneity}. Similarly, Columns 7-8 show an equalization of the effect sizes for rivals in low- and high-concentration markets. Without accounting for the urban-rural divide, the difference in effect sizes at $t+6$ is 0.008 (Table \ref{tab:heterogeneity}, while accounting for the rural-urban divide results in a slightly higher effect in high concentration markets.

To summarize, for both victimized and rival stores stores alike, urban-rural heterogeneity appears to explain much of the heterogeneous effects for low and high concentration markets, particularly at higher lags.

\begin{table}[!htbp] 
\centering
\caption{Heterogeneity analysis robustness checks}
\renewcommand{\tabcolsep}{1pt}{
\def\sym#1{\ifmmode^{#1}\else\(^{#1}\)\fi}
\begin{tabular*}{\hsize}{@{\hskip\tabcolsep\extracolsep\fill}l*{8}{c}}
\toprule

            &\multicolumn{4}{c}{Victimized} &\multicolumn{4}{c}{Rivals} \\
            \cline{2-5} \cline{6-9} \\
            
            &\multicolumn{1}{c}{(1)}&\multicolumn{1}{c}{(2)}&\multicolumn{1}{c}{(3)}&\multicolumn{1}{c}{(4)}&\multicolumn{1}{c}{(5)}&\multicolumn{1}{c}{(6)} &\multicolumn{1}{c}{(7)} &\multicolumn{1}{c}{(8)} \\

            &\multicolumn{1}{c}{\parbox{1.5cm}{\centering Indep. \\ stores }}&\multicolumn{1}{c}{\parbox{1.5cm}{\centering Chain \\ stores}}&\multicolumn{1}{c}{\parbox{1.5cm}{\centering Low concentration}}&\multicolumn{1}{c}{\parbox{1.5cm}{\centering High concentration}}&\multicolumn{1}{c}{\parbox{1.5cm}{\centering Indep. \\ stores }}&\multicolumn{1}{c}{\parbox{1.5cm}{\centering Chain \\ stores}}&\multicolumn{1}{c}{\parbox{1.5cm}{\centering Low concentration}}&\multicolumn{1}{c}{\parbox{1.5cm}{\centering High concentration}} \\
\midrule
\addlinespace
 $E_0$ & 0.017 & 0.0014 & 0.013 & 0.0017 & 0.0013 & 0.00097 & 0.00066 & 0.0018 \\ 
        ~ & (0.012) & (0.0032) & (0.010) & (0.0027) & (0.0044) & (0.0019) & (0.0025) & (0.0031) \\ 
\addlinespace
         $E_2$ & 0.029** & 0.0060 & 0.025** & 0.0067 & 0.0025 & 0.0052 & 0.0019 & 0.0081 \\ 
         ~ & (0.014) & (0.0051) & (0.012) & (0.0051) & (0.0050) & (0.0037) & (0.0037) & (0.0053) \\ 
\addlinespace
        $E_4$ & 0.035** & 0.0074 & 0.031** & 0.0068 & 0.0094* & 0.017*** & 0.010* & 0.021*** \\ 
         ~ & (0.016) & (0.0076) & (0.014) & (0.0074) & (0.0053) & (0.0062) & (0.0063) & (0.0072) \\ 
\addlinespace

        $E_6$ & 0.023 & 0.023*** & 0.026 & 0.019** & 0.011* & 0.011 & 0.010 & 0.013* \\ 
        ~ & (0.020) & (0.0082) & (0.017) & (0.0086) & (0.0067) & (0.0072) & (0.0079) & (0.0073) \\ 
        \midrule
$\sum \text{Pre-event}$   
            & 0.010 & -0.0075 & -0.0055 & 0.0044 & -0.00011 & 0.0040 & -0.0034 & 0.0096 \\ 
        ~ & (0.0089) & (0.0068) & (0.0092) & (0.0062) & (0.011) & (0.0064) & (0.0070) & (0.0079) \\ 
\midrule
\(N\)       & 11,968 & 18,793 & 15,004 & 15,757 & 28,889 & 29,904 & 16,023 & 15,001 \\ 

\bottomrule
\end{tabular*}
\begin{minipage}[h]{\textwidth}
\medskip
\footnotesize \emph{Notes:} Each column shows the cumulative treatment effects on store price levels for different subsamples zero, two, four and six months after a crime, along with the sums of pre-treatment coefficients. Coefficients are interpretable as percentage increases in outcome levels relative to the month before a crime incident. The first four columns use victimized stores as the treatment group, and the last four columns consider rival stores. Columns 1 and 5 show effects for independent stores (owned by firms running one store only), while columns 2 and 6 show effects for stores owned by firms running at least two stores. For the other columns, the Herfindahl-Hirschman Index, conditional on a store being located in a rural or urban area, is calculated for the market around each store (including non-treated stores) within a 5-mile radius. The sample is then split according to the median market concentration in rural and urban areas, respectively. Columns 3 and 7 show effects for treated stores in rural and urban markets with below median concentration, and columns 4 and 8 for treated stores in markets with above median concentration. Standard errors of the sums are clustered at the store level and shown in parentheses. \sym{*} \(p<0.10\), \sym{**} \(p<0.05\), \sym{***} \(p<0.01\). 
\end{minipage}

}
\label{tab:heterogeneity_robustness}
\end{table}

\newpage 
\section{Strategic complementarity in pricing}\label{sec:pt_figure}

In this appendix section, we discuss the implications and potential issues arising from strategic complementarity in pricing in our setting. Furthermore, we present estimation results measuring the extent of strategic pricing in the Washington state cannabis industry.

\subsection{Theoretical framework}

We follow the framework of \cite{muehlegger2022pass} and consider the pass-through of a tax (or input cost shock) $\tau$ onto the price of firm $j$. Firm $j$ sets the profit-maximizing price $p_j$ and faces tax-inclusive marginal costs $\alpha_j$. Each firm in the market can have a different exposure to the tax, with $\frac{\partial \alpha_j}{\partial \tau}$ capturing the marginal unit tax rate faced by firm $j$. In oligopolistic markets, the price a firm sets is a function of not just its own costs, but also those of its rivals.  The pass-through of the tax onto firm $j$'s price can thus be decomposed as a direct (own-cost) and an indirect (competitors' cost) effect: 

\begin{equation}
    \frac{\partial p_{j}}{\partial \tau} = \frac{\partial p_{j}}{\partial \alpha_{j}}\frac{\partial \alpha_{j}}{\partial \tau} + \sum_{i \neq j} \frac{\partial p_j}{\partial p_i}\frac{\partial p_{i}}{\partial \alpha_{i}}\frac{\partial \alpha_{i}}{\partial \tau}
\end{equation}
where $\frac{\partial p_j}{\partial p_i}$ is firm $j$'s best response to a change in firm $i$'s price.\footnote{For ease of exposition we consider competition in prices. \cite{muehlegger2022pass} show that this framework extends to a broad class of oligopolistic settings.} Consequently, in the presence of imperfect competition, the strategic response of (untreated) competitors may disqualify them as a valid control group. It is therefore important to quantify the size and geographic scope of strategic complementarity in prices.

\subsection{Quantifying strategic complementarity in prices}

To identify the scope of strategic complementary in prices, we follow the industrial organization literature that measures the pass-through of cost shocks and taxes. In particular, we build on the approach of \cite{hollenbeck2021}, who use similar data to evaluate the optimal cannabis sales tax. A major advantage of this approach is that, because we observe wholesale unit prices, we can directly measure how changes in unit cost are passed through to prices.\footnote{Wholesale costs are typically estimated from supply-side first order conditions. For similar approaches, see, for instance, \cite{muehlegger2022pass,gana2020} who use variation in energy input costs to estimate the price pass-through of a hypothetical carbon tax or \cite{miller2017} who estimate the pass-through of carbon pricing in the portland cement industry.} In addition to stores' own wholesale unit costs, we also observe the wholesale unit costs of their competitors. By relating stores' prices to competitors' cost changes, we can measure the effect of competitors' cost-induced price changes, i.e. strategic complementarity in prices. Moreover, we can test whether this effect is a function of the geographic distance between stores. We use the results of this analysis to define unaffected local markets (a clean control inclusion criterion in Section \ref{sec:strategy}).

To investigate the geographic scope of strategic complementarity of prices, we sort competitors into 5-mile bins and calculate average wholesale unit price for each store-product-month-bin. We specify a model at the store-product-month level that relates a store-product’s retail price to (i) the wholesale unit price and (ii) the average wholesale unit price paid by stores in each distance bin. By including both own costs and competitors' costs, we capture the total effect (i.e. own-cost and strategic price response) of an aggregate unit cost shock on stores' prices. Since cannabis transaction data is publicly available, stores have full information on competitors' unit costs and prices updated on an almost weekly basis. Therefore, we focus on contemporaneous changes in costs and prices. This is in line with the pass-through literature from other industries \citep[see e.g.][]{hollenbeck2021,muehlegger2022pass,conlon2020,miller2017}. We estimate the following model in first-differences:
\begin{equation}\label{eq:pt_reg_2_appendix}
    \Delta p_{i,j,t} = \rho \Delta w_{i,j,t} + \sum_{r = 1}^{R} \beta_r \Delta w_{i,r(j),t} + \Delta \gamma_t + \Delta \varepsilon_{i,j,t},
\end{equation}
where $p_{i,j,t}$ is the average price (in dollars) of product $i$ sold at store $j$ in month $t$, $w_{i,j,t}$ is the average wholesale price that retailer $j$ pays for product $i$ in month $t$, $w_{i,r(j),t}$ is the average wholesale price that competitors pay for product $i$ in month $t$, and $\gamma_t$ is the year-month FE. In our baseline specification, we set $R=9$ ($R>9$ does not meaningfully affect estimates but changes the sample size and standard errors). We cluster standard errors at the store level to allow for autocorrelation in unobservables within stores.

The effect of an aggregate change in unit costs on store $j$'s prices comprises two parts. The first is the own cost pass-through rate, $\rho$, i.e. the increase in retail unit price at store $j$ from the increase in store $j$'s own wholesale unit cost. The second part is $\beta_r$ which measures the pass-through of wholesale unit costs at competing stores in bin $r$ to unit prices at store $j$. This is equivalent to the strategic price response between store $j$ and competing stores in bin $r$. 

\begin{figure}[!htbp]
\caption{The pass-through of competitors' wholesale unit costs to own unit prices}
\label{fig:pt_rate_cumulative}
\centering
    	\begin{subfigure}{.45\textwidth}
	\centering
	    \includegraphics[width=\linewidth]{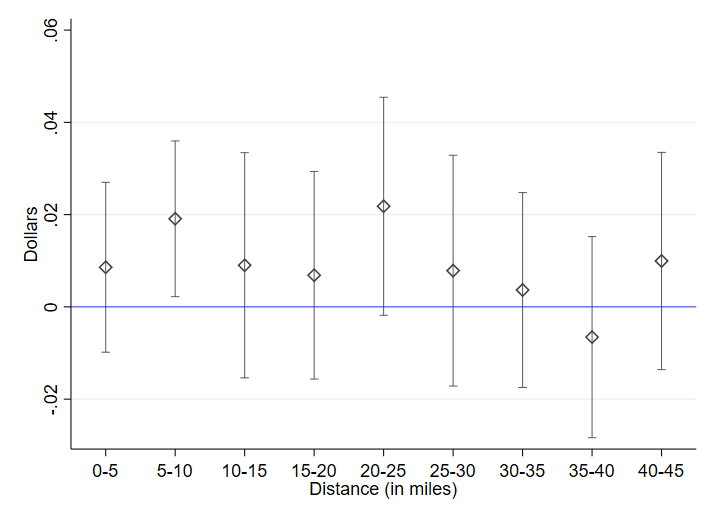}
	\caption{Marginal effects}
    \end{subfigure}
	\begin{subfigure}{.45\textwidth}
	\centering
	    \includegraphics[width=\linewidth]{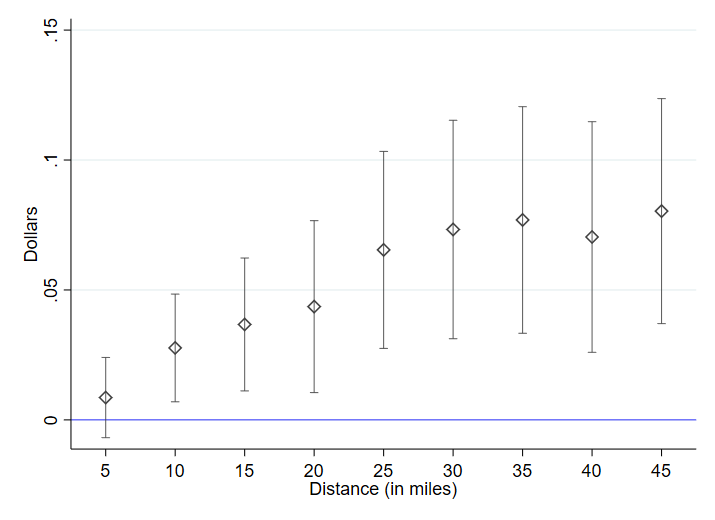}
		\caption{Cumulative effects}
    \end{subfigure}
    \label{fig:strategic_pricing}
\par 
\rule{\textwidth}{0.4pt}
\begin{minipage}[h]{\textwidth}
\medskip
\footnotesize \emph{Notes:} Panel A shows estimated coefficients $\beta_r$ for $r \in [1,9]$ obtained from the pass-through regression (equation \ref{eq:pt_reg_2_appendix}). Coefficients in Panel A are interpretable as the effect (in dollars) on store $j$'s retail unit price from a \$1 increase in wholesale unit costs affecting all stores in distance bin $r$. Panel B shows cumulative sums of coefficients $\sum_{r=1}^{R} \beta_r$ for $R \in [1,9]$. Coefficients in Panel B are interpretable as the effect on store $j$'s retail unit price of a \$1 increase in wholesale unit cost affecting all stores up to $r$ miles away. The figure shows 90\% confidence intervals of the sums based on SE clustered at the store level. Data: Top Shelf Data, March 2018 through December 2021.
\end{minipage}
\end{figure}

In Figure \ref{fig:strategic_pricing} Panel A, we report estimated pass-through rates of competitors' unit costs, $\beta_r$, from our baseline specification. The estimates differ across bins with the largest effects in the 5-10 mile and 20-25 mile bins. The fact that the effect fluctuates with distance could reflect commuting patterns, with the average daily distance travelled in Washington state ranging from less than 20 miles in some counties to more than 70 in others \citep{axios2024}. Nevertheless, at the 25-30 mile bin, effects shrink and remain close to zero for three consecutive bins. 

The $\beta_r$ estimates in Panel A can be interpreted as marginal effects in that they measure the additional effect on store $j$'s prices of increasing the geographic scope of an aggregate change in costs by another 5 miles. While this is informative about the geographic scope of strategic complementarity in cannabis prices, quantifying the actual effect on prices of an aggregate change in costs requires summing the marginal effects $\sum_{r = 1}^{R} \beta_r$ up to a given distance bin $R$. The sum can be interpreted as the effect on store $j$'s prices of an aggregate change in costs that affects all stores up to a given distance (while holding store $j$'s costs constant). We report these sums at increasing distances in Panel B of Figure \ref{fig:strategic_pricing}. Panel B further highlights that sensitivity to competitors' costs increases up to the 30-mile mark before plateauing thereafter. This aligns with a growing literature showing that the scope of cost shocks matters and that aggregate (i.e market-wide) cost shocks elicit a larger strategic price response than idiosyncratic or highly localized shocks \citep{muehlegger2022pass}. 

In addition to our main specification, we report several variants of our pass-through regression in Table \ref{tab:cum_pt_rate}. In Column 2, we add controls (average wages, the county-level unemployment rate, and the home price index at the three-digit zip code level) to absorb variation in retail cannabis prices due to local business cycles or changes in house prices \citep[see][]{stroebel2019}. In Column 3, we include region-time FE to account for other spatially correlated shocks that may covary with wholesale and retail cannabis prices. We also estimate equation \ref{eq:pt_reg_2_appendix} in levels rather than first-differences, with store-product FE (Column 4). In Column 5, we specify equation \ref{eq:pt_reg_2_appendix} using the first-difference of logs. This minimizes the influence of outliers and delivers pass-through elasticities instead of pass-through rates. 

Columns 1-5 in Table \ref{tab:cum_pt_rate} show that stores' sensitivity to competitors' costs plateaus at the 30-mile mark across across all specifications. Moreover, the results indicate that an aggregate cost shock with sufficient geographic scope has non-negligible strategic price effects. When estimated in first-differences (Column 1), a \$1 increase in wholesale unit costs at all stores within a 30-mile radius corresponds to a \$0.07 increase in retail prices solely due to strategic complementarities. When estimated in levels (Column 4), prices increase \$0.48 from a \$1 increase in wholesale cost.

\begin{table}[!htbp] 
\centering
\caption{Cumulative pass-through of competitors' unit costs}
\renewcommand{\tabcolsep}{1pt}{
\def\sym#1{\ifmmode^{#1}\else\(^{#1}\)\fi}
\begin{tabular*}{\hsize}{@{\hskip\tabcolsep\extracolsep\fill}l*{6}{c}}
\toprule
            &\multicolumn{3}{c}{Dollars (FD)} &\multicolumn{3}{c}{} \\
            \cline{2-4} 
            \addlinespace

            &\multicolumn{1}{c}{(1)}&\multicolumn{1}{c}{(2)}&\multicolumn{1}{c}{(3)}&\multicolumn{1}{c}{(4)} &\multicolumn{1}{c}{(5)} &\multicolumn{1}{c}{(6)} \\ 

            &\multicolumn{1}{c}{\parbox{1.5cm}{\centering Baseline}} &\multicolumn{1}{c}{\parbox{1.5cm}{\centering Controls}} &\multicolumn{1}{c}{\parbox{1.5cm}{\centering Reg. $\times$ time FE}} &\multicolumn{1}{c}{\parbox{1.5cm}{\centering Dollars (levels)}} &\multicolumn{1}{c}{\parbox{1.5cm}{\centering Logs (FD)}} &\multicolumn{1}{c}{\parbox{1.5cm}{\centering Store-level index}} \\
\midrule
\addlinespace
Own wholesale cost       &  1.735*** & 1.735*** & 1.735*** & 1.217*** & 0.736*** & 1.247*** \\ 
                         & (0.0515) & (0.0515) & (0.0515) & (0.426) & (0.00991) & (0.213) \\ 
\addlinespace
\multicolumn{2}{l}{Competitors' wholesale cost} &&&&& \\
\midrule

$< 5$ miles & 0.00860 & 0.00864 & 0.00880 & 0.0682* & 0.000241 & -0.0167 \\ 
   & (0.00935) & (0.00936) & (0.00939) & (0.0349) & (0.00239) & (0.0617) \\

\addlinespace
       $< 10$ miles & 0.0277** & 0.0278** & 0.0280** & 0.231*** & 0.00356 & -0.00656 \\ 
        ~ & (0.0126) & (0.0126) & (0.0126) & (0.0641) & (0.00323) & (0.0844) \\ 
\addlinespace
        $< 15$ miles & 0.0367** & 0.0368** & 0.0379** & 0.250*** & 0.00283 & -0.0224 \\ 
        ~ & (0.0155) & (0.0155) & (0.0155) & (0.0818) & (0.00381) & (0.110) \\ 
\addlinespace
        $< 20$ miles & 0.0436** & 0.0437** & 0.0445** & 0.321*** & 0.00618 & -0.0346 \\ 
        ~ & (0.0200) & (0.0201) & (0.0201) & (0.0946) & (0.00479) & (0.135) \\ 
\addlinespace
        $< 25$ miles & 0.0654*** & 0.0656*** & 0.0664*** & 0.346*** & 0.00964* & -0.0615 \\ 
        ~ & (0.0230) & (0.0230) & (0.0230) & (0.104) & (0.00559) & (0.151) \\ 
\addlinespace
        $< 30$ miles & 0.0733*** & 0.0733*** & 0.0734*** & 0.481*** & 0.0162*** & -0.0581 \\ 
        ~ & (0.0255) & (0.0255) & (0.0255) & (0.129) & (0.00611) & (0.167) \\ 
\addlinespace
        $< 35$ miles & 0.0769*** & 0.0768*** & 0.0766*** & 0.511*** & 0.0157*** & -0.0852 \\ 
        ~ & (0.0264) & (0.0264) & (0.0264) & (0.144) & (0.00597) & (0.187) \\ 
\addlinespace
        $< 40$ miles & 0.0704*** & 0.0705*** & 0.0702*** & 0.491*** & 0.0152*** & -0.0291 \\ 
        ~ & (0.0269) & (0.0269) & (0.0269) & (0.149) & (0.00532) & (0.214) \\ 
\addlinespace
        $< 45$ miles & 0.0803*** & 0.0805*** & 0.0799*** & 0.446*** & 0.0203** & 0.00962 \\ 
        ~ & (0.0262) & (0.0263) & (0.0262) & (0.145) & (0.00856) & (0.229) \\ 
\midrule
\(N\)               & 2,290,818 & 2,290,726 & 2,290,818 & 3,683,684 & 2,290,818 & 9,729 \\         

\bottomrule
\end{tabular*}
\begin{minipage}[h]{\textwidth}
\medskip
\footnotesize \emph{Notes:} The table reports the estimates of wholesale cost pass-through rates from equation \ref{eq:pt_reg_2}. We report estimates for own wholesale cost changes and for average changes in wholesale costs at competitor stores located within 5 miles of the respective store. All specifications control for month-year fixed effects. Coefficients for Columns 1-4 are interpretable as pass-through rates in dollars. Columns 5-6 are interpretable as pass-through elasticities. Standard errors are clustered at the store level and shown in parentheses. \sym{*} \(p<0.10\), \sym{**} \(p<0.05\), \sym{***} \(p<0.01\).
\end{minipage}

}\label{tab:cum_pt_rate}
\end{table}

Overall, the results from Figure \ref{fig:strategic_pricing} and Table \ref{tab:cum_pt_rate} provide suggestive evidence of strategic complementarity in prices for cannabis stores within 30 miles of each other. Increasing the geographic scope of an aggregate cost shock appears to have little additional effect on store prices beyond the 30-mile mark. This suggests that stores located more than 30 miles from a victimized store will not have a strategic price response to the crime-induced cost shock at victimized and rival stores. We therefore view Figure \ref{fig:strategic_pricing} and Table \ref{tab:cum_pt_rate} as providing supportive evidence for our definition of unaffected local markets from Section \ref{sec:strategy}.

\subsubsection*{Wholesale unit cost pass-through with store-level price and cost indexes}
It is worth noting that equation \ref{eq:pt_reg_2_appendix} is specified at the store-product-month level of aggregation. There are two reasons for using this disaggregated level in our baseline specification. First, our target parameter is the pass-through of \emph{unit} (i.e. per product) cost to \emph{unit} price. The disaggregation allows us to estimate this parameter by directly relating these two values. In contrast, a regression based on store-level indexes first aggregates products within their respective subcategory and then aggregates across subcategories within a store. While this is preferable when estimating treatment effects at the store level, the two-step aggregation necessarily breaks the direct vertical link between unit cost and unit price. Second, and more importantly, this link is further cleaved by the fact that rivals' cost indexes contain cost changes for products and subcategories that store $j$ may not actually sell. In the extreme case, one might relate costs and prices for adjacent stores that sell non-overlapping baskets of goods and hence do not compete in prices at the product level. As a result, the estimated coefficient for rivals' cost changes---and by extension the own-cost pass-through rate---may be biased when using store-level indexes. Nevertheless, we estimate the pass-through regression using store-level price and cost indexes in Column 6 of Table \ref{tab:cum_pt_rate}. The own-cost pass-through elasticity is 1.25 aligns with our baseline finding that costs are more than fully passed through. However, the coefficients on competitors' costs fluctuate and are not statistically significant, in line with the idea that competitors' cost indexes may not capture strategic complementarity in prices for the reasons outlined above. 

\newpage

\section{Spatial treatment and alternative definitions of unaffected local markets and rival stores}\label{sec:spatial_treat}

This section elaborates on the inclusion criteria for our control group, with a special emphasis on the geographical criteria. Subsequent subsections conduct several robustness checks to assess the sensitivity of our estimated effects to these geographical criteria and our definitions of rival stores.

\subsection{Stacked DiD with spatial treatment}\label{sec:inclusion_criteria}

\begin{figure}[!htbp]
\caption{Clean controls}
\centering
	\includegraphics[width=.5\textwidth]{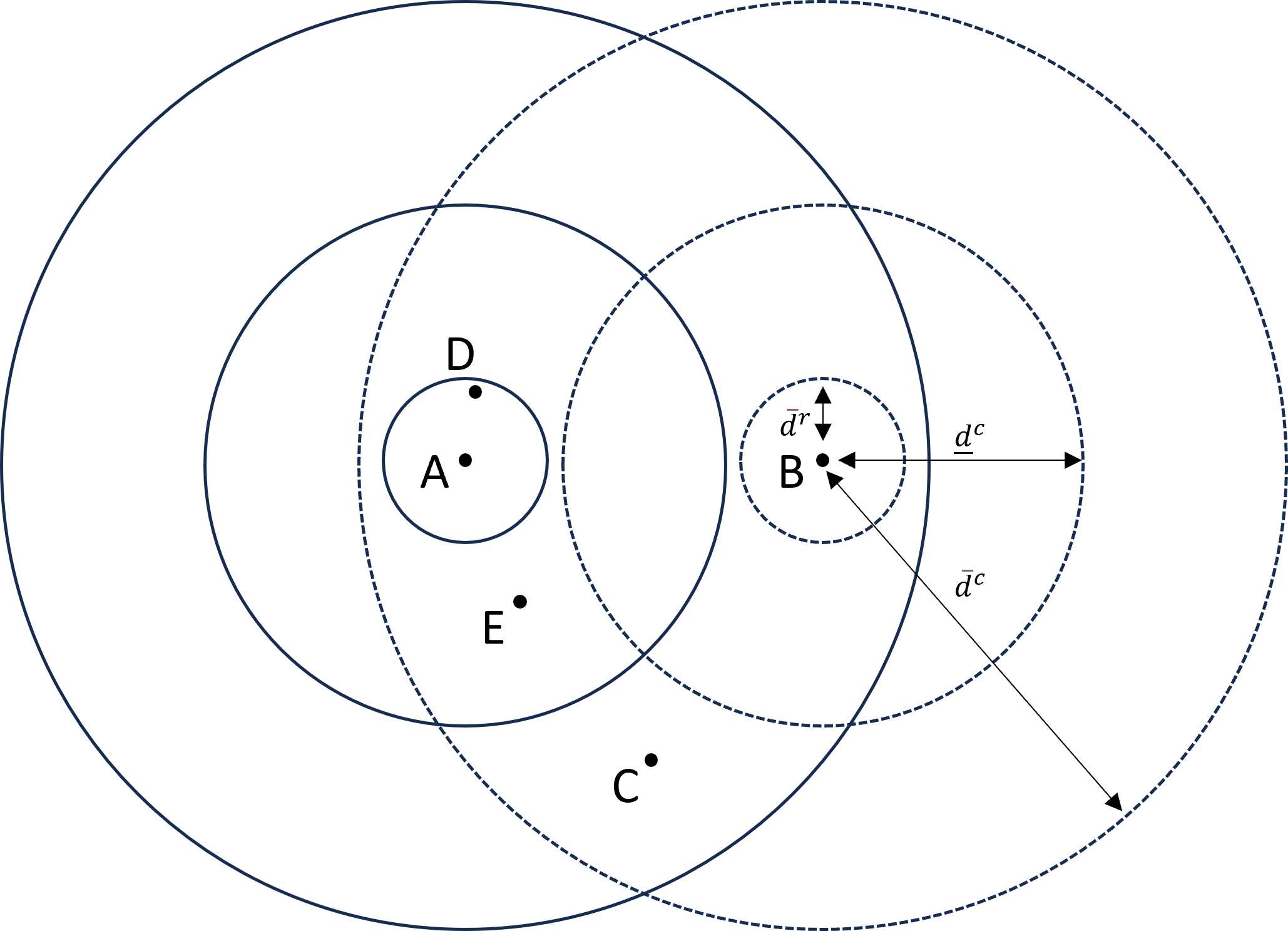}
 \label{fig:strategy_1}
\par \bigskip
\rule{\textwidth}{0.5pt}
\begin{minipage}[h]{\textwidth}
\medskip
\footnotesize \emph{Notes:} The figure depicts the locations of four stores (A, B, C, D) in relation to each other. Stores A and B are victimized while stores C and D are not. Concentric rings of diameter $\underline{d}$ and $\overline{d}$ denote the boundaries defining clean control stores in our setting.
\end{minipage}
\end{figure}

To provide a clearer intuition behind our inclusion criteria for defining clean control stores, we provide an illustrative example in Figure \ref{fig:strategy_1}.
The figure depicts five stores: A and B are both victimized while stores C, D, and E are not victimized. Three concentric rings surround each victimized store: the innermost ring denotes the boundary for rival stores and has a radius of $\overline{d}^r$. The other two rings, with radius $\underline{d}^c$ and $\overline{d}^c$, denote the boundaries for control candidates. 

Store C is a control candidate for both A and B while store E is a control candidate for B only. Store D is at the same time a control candidate for B and a rival of A. If the event windows for stores A and B overlap, D is disqualified from the control group for B. However, store D qualifies as a clean control for store B if the event windows for stores A and B do not overlap. This reflects our assumption that store D's contamination from store A's crime incident is zero outside the treatment window. Similarly, an earlier-treated store (e.g. A) would qualify as a control store for a later-treated store (e.g. B) if the earlier-treated store is a control candidate and the event windows do not overlap. 

Using the intuition from Figure \ref{fig:strategy_1}, we can formally express the inclusion criteria for clean controls in our stacked DiD design. Let $H$ denote the set of all stores in our sample and $J$ the set of all victimized stores, chronologically labeled $j = 1, \ldots, n$. Define $R_j = \{h \neq j \in H : d_{h,j} < \overline{d}^{r}\}$ as the set of rival stores related to victimized store $j$, with $R = \bigcup_{j=1}^{n} R_j$ representing the set of all unique rival stores. $t_j$ denotes the month a store $j$ is treated (i.e., the month of the crime incident in sub-experiment $j$). $w$ is the size of the event window (in months), $d_{h,j}$ is the geodesic distance between stores $j$ and $h$ (in miles). 

Store $i$ qualifies as a clean control and is included in the sub-experiment corresponding to the crime incident at store $j$ if the following conditions jointly hold:
\begin{enumerate}
    \item []\textbf{IC.1}: $i \in O_j$ where $O_j = \{h \in H : \underline{d}^c \leq d_{h,j} \leq \overline{d}^c \}$
    \item []\textbf{IC.2}: $i \notin T_{j^-}$, where $T_{j^-}= \{h \in J : (t_h + w > t_j) \lor (t_h - w < t_j) \}$
    \item []\textbf{IC.3}: $i \notin D_{{j^-}}$, where $D_{{j^-}} = \{h \in R: (t_h + w > t_j) \lor (t_h - w < t_j) \}$
\end{enumerate}

 Condition IC.1 requires that store $i$ be located between $\underline{d}^c $ and $\overline{d}^c $ miles of store $j$. This geographic criterion designates candidate control stores for each treated store, e.g., stores A, C, D, and E for store B in Figure \ref{fig:strategy_1}. Condition IC.2 ensures that store $i$ is either never victimized or, if it is victimized, its treatment event window does not overlap with that of victimized store $j$. This is the standard stacked DiD inclusion criterion that prevents unclean comparisons between early- and late-treated stores. In our example, this criterion excludes store A from the set of clean control stores for B and vice versa if the event window of A and B overlaps.  Condition IC.3 requires that store $i$ must not be a rival store of any victimized store whose event window overlaps with that of store $j$.\footnote{Note that rival treatment is only assigned according to a store’s first nearby crime incident, implying that each rival store is assigned to only one treatment cohort and that multiple treated rival stores can serve as clean control stores for subsequent crimes even if the event window overlaps with store $j$. This cannot happen to victimized stores, as we exclude multiple victimized stores before the second treatment.} This condition would disqualify store D as a clean control for store B if the treatment windows of stores A and B overlap. However, IC.2 and IC.3 ensure that stores D and A qualify as a clean control candidate for store B if store B's event window does not overlap with that of store A. 

 Note that store $i$ may satisfy IC.1-IC.3 for some crime incidents but not for others. As long as store $i$ satisfies all three criteria for a given crime incident, then it will be included in that sub-experiment. Accordingly, store $i$ may qualify as a clean control for multiple sub-experiments that overlap in calendar-time. This is a common feature of the stacked DiD estimator and it implies that certain store-month observations may be recycled and appear in multiple sub-experiments \citep{wing2024}. For example, in Figure \ref{fig:strategy_1}, if stores A and B are both victimized in the same period, then store C serves as a clean control for both sub-experiments (assuming the other conditions are satisfied). Store C is then included in each of the sub-experiment-specific datasets, meaning the stacked dataset will contain duplicate observations for store C for the calendar months corresponding to that particular event window.

\subsection{Alternative inner and outer ring sizes}\label{sec:rings_robust}

\begin{figure}[!htbp]
\caption{Price effects with alternative inner and outer rings}\label{fig:robust_ring}
\centering
	\begin{subfigure}{.4\textwidth}
	\centering
	    \includegraphics[width=\linewidth]{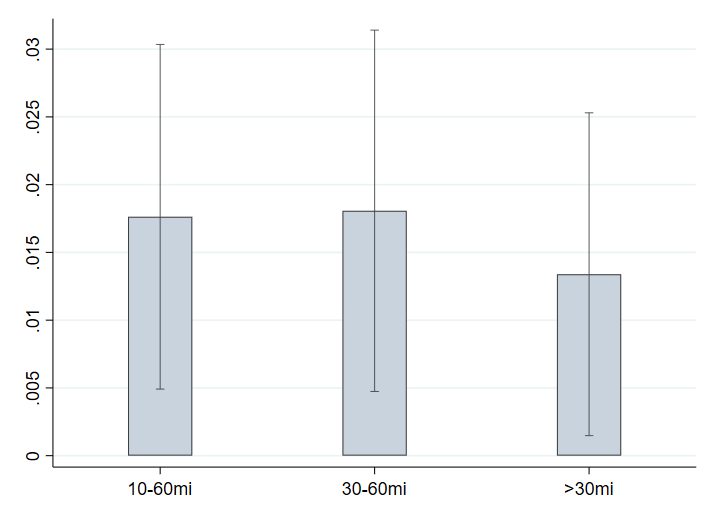}
		\caption{Victimized stores, outer ring}
  \label{fig:robust_outerring1}
	\end{subfigure}\hfil
	\begin{subfigure}{.4\textwidth}
	\centering
	    \includegraphics[width=\linewidth]{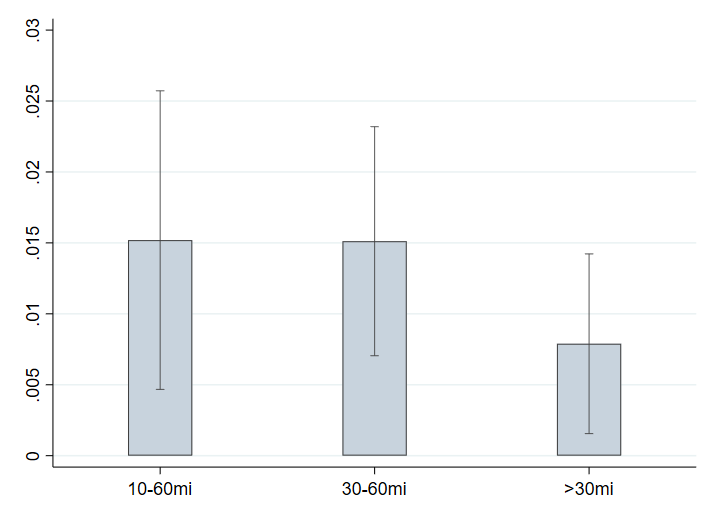}
		\caption{Rival stores, outer ring}
  \label{fig:robust_outerring2}
	\end{subfigure}\hfil
	\begin{subfigure}{.4\textwidth}
	\centering
	    \includegraphics[width=\linewidth]{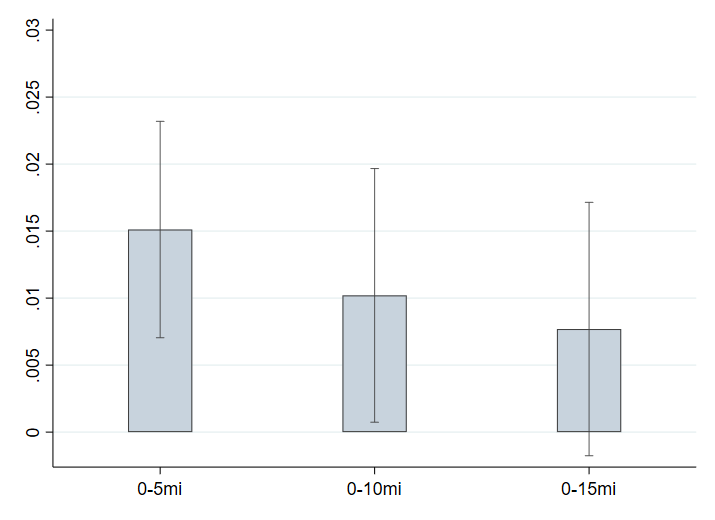}
		\caption{Rival stores, inner ring}
  \label{fig:robust_innerring}
	\end{subfigure}\hfil
\par 
\rule{\textwidth}{0.4pt}
\begin{minipage}[h]{\textwidth}
\medskip
\footnotesize \emph{Notes:} Panels A and B show the cumulative treatment effects four months after a crime incident for victimized stores and for rival stores, respectively, along with corresponding 90\% confidence intervals based on standard errors clustered at the store
level. The first bar in both panels shows the effects when considering all stores within 10 and 60 miles as potential control stores. The middle bar refers to our main specification. The third bar shows the effects when considering all stores located more than 30 miles away from victimized stores as clean control stores. Panel C illustrates estimated effects as we expand the definition for rival stores. The first bar reflects cumulative effects in our main specification; the second bar shows effects when considering all stores within 10 miles as rival stores, and the third bar shows effects when considering all stores within 15 miles as rivals.
\end{minipage}
\end{figure}

In our main specification, we set $\overline{d}^{r}=5$, $\underline{d}^c=30$ and $\overline{d}^c=60$. To investigate whether our estimated effects hinge on our definition of clean control stores, we estimate our model using alternative spatial definitions of clean control and rival stores.  Figures \ref{fig:robust_outerring1} and \ref{fig:robust_outerring2} show the cumulative treatment effects four months after a crime incident for victimized stores and for rival stores, respectively. The middle bar refers to our main specification. The first bar in both figures shows the effects when expanding the inner boundary ($\underline{d}^c $) to include all stores within 10 and 60 miles as potential control stores, and the third bar shows the effects of expanding the outer boundary ($\overline{d}^c $) to include all stores that are located more than 30 miles away from victimized stores. 

Similarly, Figure \ref{fig:robust_innerring} illustrates estimated effects as we expand the definition of the inner ring defining rival stores ($\overline{d}^{r}$). The first bar reflects cumulative effects in our main specification; the second shows effects for stores within 10 miles, and the third for stores within 15 miles of victimized stores. 

\subsection{Heterogeneous spatial treatment and inclusion criteria}\label{sec:het_spatial_robust}

\begin{figure}[!htbp]
\caption{Price effects with spatial heterogeneity}\label{fig:spatial_heterogeneity}
\centering
	\begin{subfigure}{.4\textwidth}
	\centering
	    \includegraphics[width=\linewidth]{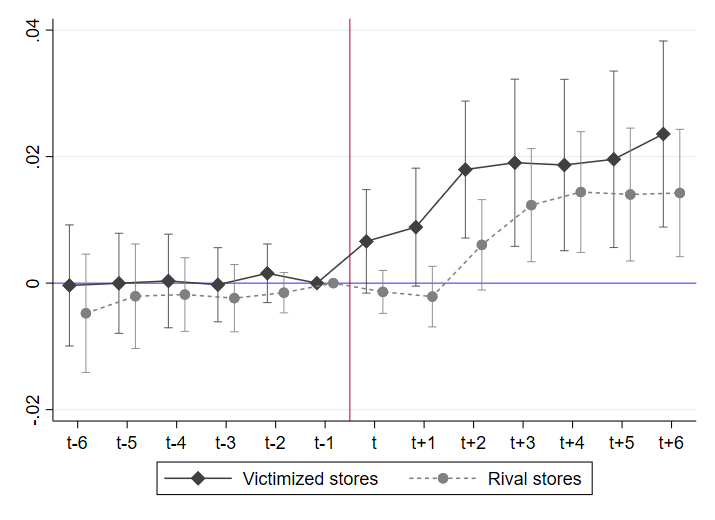}
		\caption{Rank-based criteria}
  \label{fig:}
	\end{subfigure}\hfil
	\begin{subfigure}{.4\textwidth}
	\centering
	    \includegraphics[width=\linewidth]{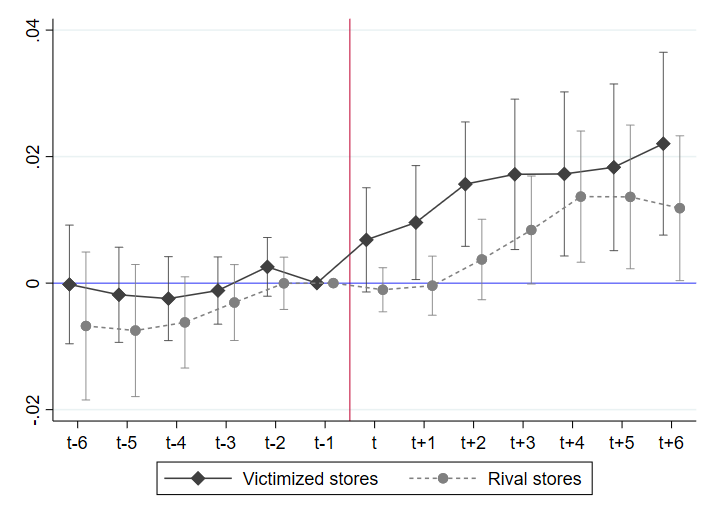}
		\caption{Distance-based criteria}
  \label{fig:}
	\end{subfigure}\hfil
\par 
\rule{\textwidth}{0.4pt}
\begin{minipage}[h]{\textwidth}
\medskip
\footnotesize \emph{Notes:} Panels A and B show cumulative treatment effects for victimized stores and for rival stores, respectively, along with corresponding 90\% confidence intervals based on standard errors clustered at the store level. In Panel A, control candidate stores are the 150th to 250th closest stores. Rivals are the 10 closest stores. In Panel B, control candidate stores are 10-30 miles if the victimized store is in an urban area, and 30-60 miles if the victimized store is in a rural area. Rivals are based on a 5-mile radius.
\end{minipage}
\end{figure}

The LCB distributes retail cannabis licenses to counties according to population density, meaning urban areas have a higher concentration of stores than rural areas. Our baseline specification defines the control group in a way that does not account for such heterogeneity. In this subsection, we conduct two robustness checks that allow for heterogeneity in the geospatial density of stores across local markets. First, we sort stores by their proximity to each victimized store and define clean control candidates as the 150th to 250th closest stores (plus the other clean control criteria discussed in Section \ref{sec:inclusion_criteria}). This definition roughly coincides with the average number of candidate controls in our baseline specification, but allows the distances of the inner and outer boundary to vary according to the local market density. We define rivals as the 20 closest stores, which after applying the stacked DiD treatment timing criteria aligns with the average number of rivals in our baseline specification. Panel A in Figure \ref{fig:spatial_heterogeneity} shows that the rank-based criteria produce similar treatment effect sizes as in our baseline specification.

Second, we revert to distance-based criteria but allow for different inner/outer ring boundaries for stores in urban and rural areas. This allows for heterogeneity across local markets in a restrictive manner.
Using data from the Washington State Office of Financial Management, we define urban stores as those located in the largest municipalities in the state: Seattle, Spokane, Tacoma, Vancouver, Bellevue, Kent, Everett, Renton, Spokane Valley \citep{waofm}. For victimized stores in these cities, we set the inner/outer boundary for clean control candidates at 10-30 miles. For all other victimized stores (which we categorize as ``rural" for ease of exposition), the inner/outer boundary is 30-60 miles as in the baseline specification. For both urban and rural stores, the definition of rival stores remains 5 miles (results are similar if we reduce the radius for urban rivals). Panel B of Figure \ref{fig:spatial_heterogeneity} shows results for this specification. Estimated treatment effects for victimized stores are very similar to our baseline specification: four months after a crime, prices at victimized stores are 1.7\% higher than the month before the crime. The price effect at rival stores is 1\%, a slight attenuation compared to the baseline specification. Nevertheless, the timing and path of treatment effects are identical to the baseline specification. Taken together, the results from Figure \ref{fig:spatial_heterogeneity} show that our main results hold when we account for geospatial heterogeneities across urban and rural areas.

\subsection{Treatment spillovers from other crime incidents}\label{sec:cc_criteria_robust}

As discussed above,  a store must be at least 30 miles from a victimized store to qualify as a clean control in a sub-experiment. However, unlike victimized and rival stores, this distance restriction does not apply to stores between $\overline{d}^r$ and $\underline{d}^c$ from other victimized stores with overlapping event windows. Thus, stores within 30 miles of a soon-to-be-victimized or recently victimized store can still serve as clean controls if they meet all other criteria. For instance, in Figure \ref{fig:strategy_1}, store E is a clean control store for store B, even if the event windows of store A and B overlap. In this section, we investigate the sensitivity of our results to applying more restrictive clean control criteria to address potential spillover effects on these stores.

Like our definition of rival stores, let $G_j = {g \neq j \in H : d_{g,j} < \underline{d}^{c}}$ denote the set of contaminated stores near victimized store $j$, with $G = \bigcup_{j=1}^{n} G_j$ representing all unique contaminated stores. In this section, we maintain IC.1 and IC.2, but modify IC.3 to include all contaminated stores: 

\begin{enumerate}
        \item []\textbf{IC.3}: $i \notin D_{{j^-}}$, where $D_{{j^-}} = \{h \in G: (t_h + w > t_j) \lor (t_h - w < t_j) \}$
\end{enumerate}

\begin{figure}[!htbp]
\caption{Price effects with more restrictive clean control criteria}
\label{fig:alt_cc_criteria}
\centering
    	\begin{subfigure}{.45\textwidth}
	\centering
	    \includegraphics[width=\linewidth]{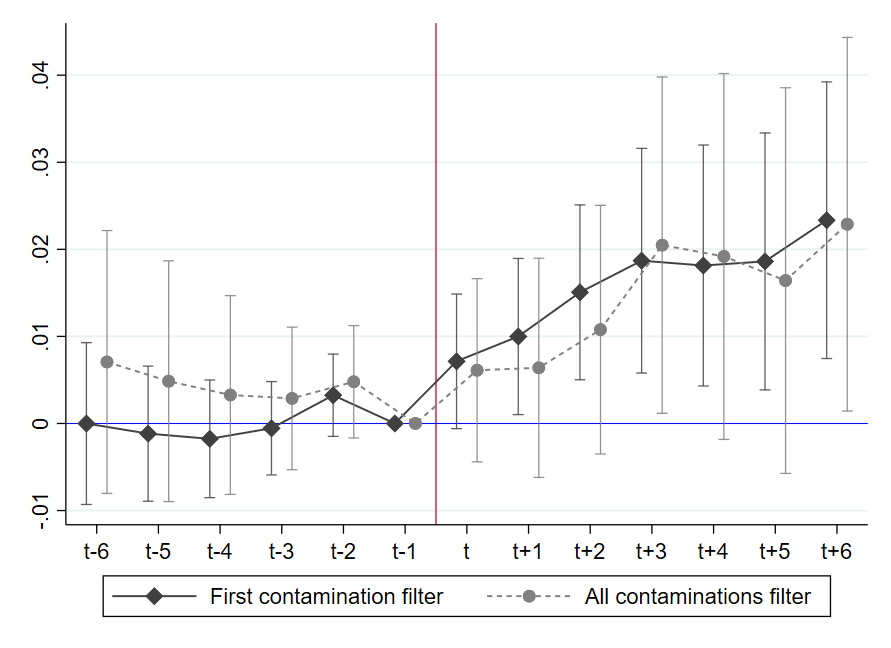}
	\caption{Victimized stores}
    \end{subfigure}
	\begin{subfigure}{.45\textwidth}
	\centering
	    \includegraphics[width=\linewidth]{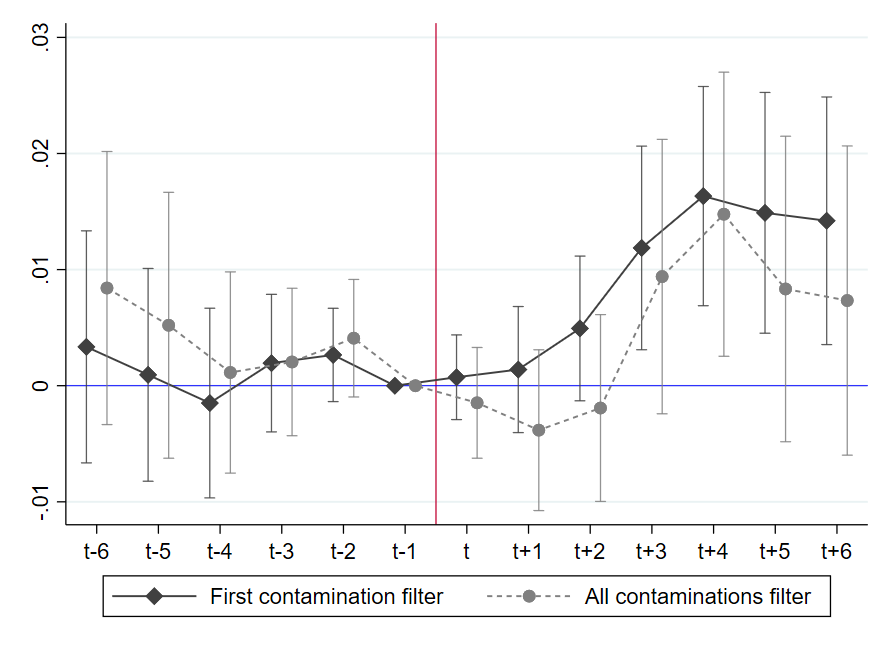}
		\caption{Rival stores}
    \end{subfigure}
\par 
\rule{\textwidth}{0.4pt}
\begin{minipage}[h]{\textwidth}
\medskip
\footnotesize \emph{Notes:} Panels A and B show cumulative treatment effects, ($E_L$), $L$ months after a crime for victimized stores and for rival stores, respectively, along with corresponding 90\% confidence intervals based on standard errors clustered at the store
 level. The black lines show estimates when contamination status is based on a store’s first nearby crime incident. The grey lines show estimates when contamination status is based on any nearby crime incident.
 \end{minipage}
\end{figure}

Figure \ref{fig:alt_cc_criteria} shows our treatment effect estimates for victimized (Panel A) and rival stores (Panel B) using the revised IC.3 condition. The black lines represent estimates under the condition that contamination status is assigned based on a store’s first nearby crime incident, as with rival treatment. Consequently, each contaminated store is assigned to only one treatment cohort, allowing multiply-contaminated stores to serve as clean controls for subsequent crimes, even if their event windows overlap with store $j$. Our results show minimal differences from the main specification but exhibit slightly larger standard errors. If we consider contamination from any nearby crime incident with overlapping event windows, the standard errors increase further due to fewer clean controls, yet the effect sizes remain similar (grey lines in Figure \ref{fig:alt_cc_criteria}). Overall, this reinforces that our main findings are not substantially biased by spillover effects from non-rival stores located within 30 miles of other treated stores.

\subsection{Variation in rival treatment intensity}\label{sec:rival_sutva_robust}

\begin{figure}[!htbp]
\caption{Rival price effects without multiple contemporaneous rival treatments}
\label{fig:sutva}
\centering
	    \includegraphics[width=.4\linewidth]{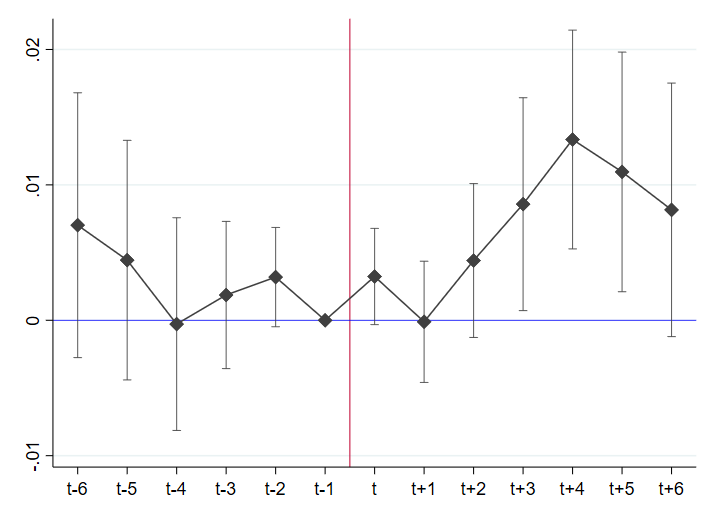}
\par
\rule{\textwidth}{0.4pt}
\begin{minipage}[h]{\textwidth}
\medskip
\footnotesize \emph{Notes:} The figure shows the cumulative effect of crime incidents on rival store prices, ($E_L$), $L$ months after a crime, along with corresponding 90\% confidence intervals based on standard errors clustered at the store level. Compared to our main specification, we exclude rival stores with more than one nearby crime incident in a given month.
\end{minipage}
\end{figure}

Some stores experience multiple nearby crime incidents within a 5-mile radius in a single month. Such underlying differences in treatment intensity violate the SUTVA and may lead to biased estimates in our rival specification. Therefore, in this section we estimate a variant of our rival specification that excludes rival stores with more than one nearby crime in a given month. The number of rival stores falls from 242 to 225 and three sub-experiments drop out of the estimation sample. However, Figure \ref{fig:sutva} shows that the treatment effects, while more volatile, are similar to those from our main specification in Figure \ref{fig:main_results}. Four months after a retail crime incident, prices at rival stores are 1.3\% higher than the month before the incident, and the estimate is significant at the 1\% level. This suggests that underlying differences in rival treatment intensity are not an important source of bias in our main analysis.

\newpage

\section{Alternative estimators}\label{sec:alt_estimators}

One advantage of the stacked DiD estimator, compared to other estimators that address biases from staggered treatment adoption \citep[e.g.][]{Callaway2021, Borusyak2021}, is that rules defining clean controls readily extend to geographic criteria. This feature allows us to mitigate biases from spillovers to untreated stores and/or differences between stores, while at the same time ensuring a large number of clean control stores. In contrast, with alternative estimators (including canonical TWFE) restrictions on the control group composition must be imposed for the entire sample period, which considerably decreases the number of valid control observations. However, other estimators may be more efficient and better comparable to other studies.

To assess whether our findings are sensitive to the choice of estimator, we use three alternative approaches: i) the canonical two-way fixed effect Difference-in-Differences estimator (TWFE) prone to biases under staggered treatment adoption; ii) the imputation estimator developed by \citet{Borusyak2021} (BSJ); and iii) the estimator proposed by \citet{Callaway2021} (CS). We estimate all models separately for both of our treatment groups (victimized and rival stores). Similar to our main specification, we first estimate a distributed lag model for the same event window (equation \ref{eq:strategy_1}) and then calculate the cumulative treatment effects using the period before treatment as our reference period. 

The underlying dataset for the alternative estimators includes all stores, which means the control group may comprise stores within 30 miles of victimized stores. Positive price effects at these stores due to spillovers (see Section \ref{sec:groups}) would imply that we underestimate the treatment effects in these models.\footnote{Supporting this line of argument, including rivals in the victimized store specification slightly decreases the treatment effects, as expected, that remain statistically significant at the 5\% level for all alternative specifications four months after treatment (results not shown).} To limit the bias from spillovers, we exclude all rival stores in the victimized store specifications and all victimized stores in the rival store specifications. If we restrict the sample to all stores that are located more than 30 miles from any treated stores, the number of clean control stores would fall from 329 to 78 stores, demonstrating the advantages of the stacked DiD framework in our setting.

For the BSJ estimator, we use the accompanying Stata package "did\_imputation" from \citet{Borusyak2021}. Standard errors allow for clustering at the store level. We calculate standard errors (detailed in \citet{Borusyak2021}) using the treatment effect averages across treatment cohorts (time since treatment) excluding the own unit.\footnote{Some observations (one victimized store and a few rival stores) are dropped through the "autosample" option. We cannot impute non-treated potential outcomes for those stores, implying that the BSJ estimator cannot estimate unbiased treatment effects for these stores.} For the CS estimator, we employ the Stata package "csdid" with clustered, asymptotic standard errors and stabilized inverse probability weighting. For both estimators, we consider all never-treated and not-yet-treated stores as potential control observations. 

One notable difference among the estimators is the benchmark against which the average distributed lag coefficients are calculated. The canonical TWFE, similar to our main specification, calculates the distributed lag coefficients relative to the last period before the event window. The CS estimator uses the last pre-treatment period as a reference, while BJS bases its comparison on the average of all pre-treatment periods \citep{roth2023s}. If all benchmarks were zero, these differences would be irrelevant. However, since the last pre-treatment period coefficient is slightly negative, the cumulative coefficients from the CS estimator turn out to be higher than those from other estimators. For the CS estimators, we employ "long" differences, yet the interpretation of pre-treatment coefficients still varies slightly between BSJ and other estimators. Importantly, all distributed lag coefficients are insignificant, and pre-tests included in the packages reveal no concerning pre-trends.

\begin{figure}
    \centering
    \caption{Effect of crime on stores price levels using alternative estimators}
    \begin{subfigure}{.45\textwidth}
           \includegraphics[width=\linewidth]{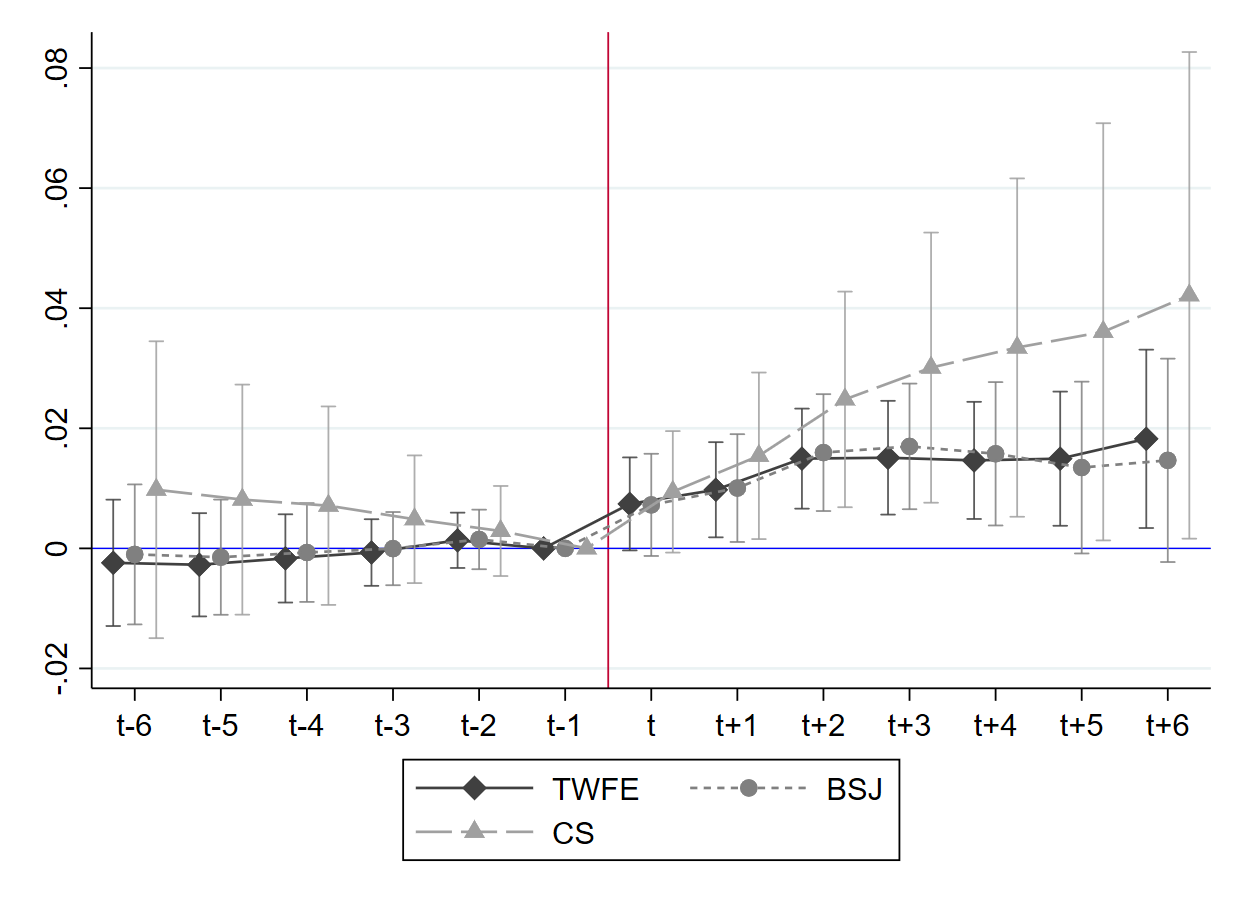}
           \caption{Victimized stores}
    \end{subfigure}\hfil
    \begin{subfigure}{.45\textwidth}
        \includegraphics[width=\linewidth]{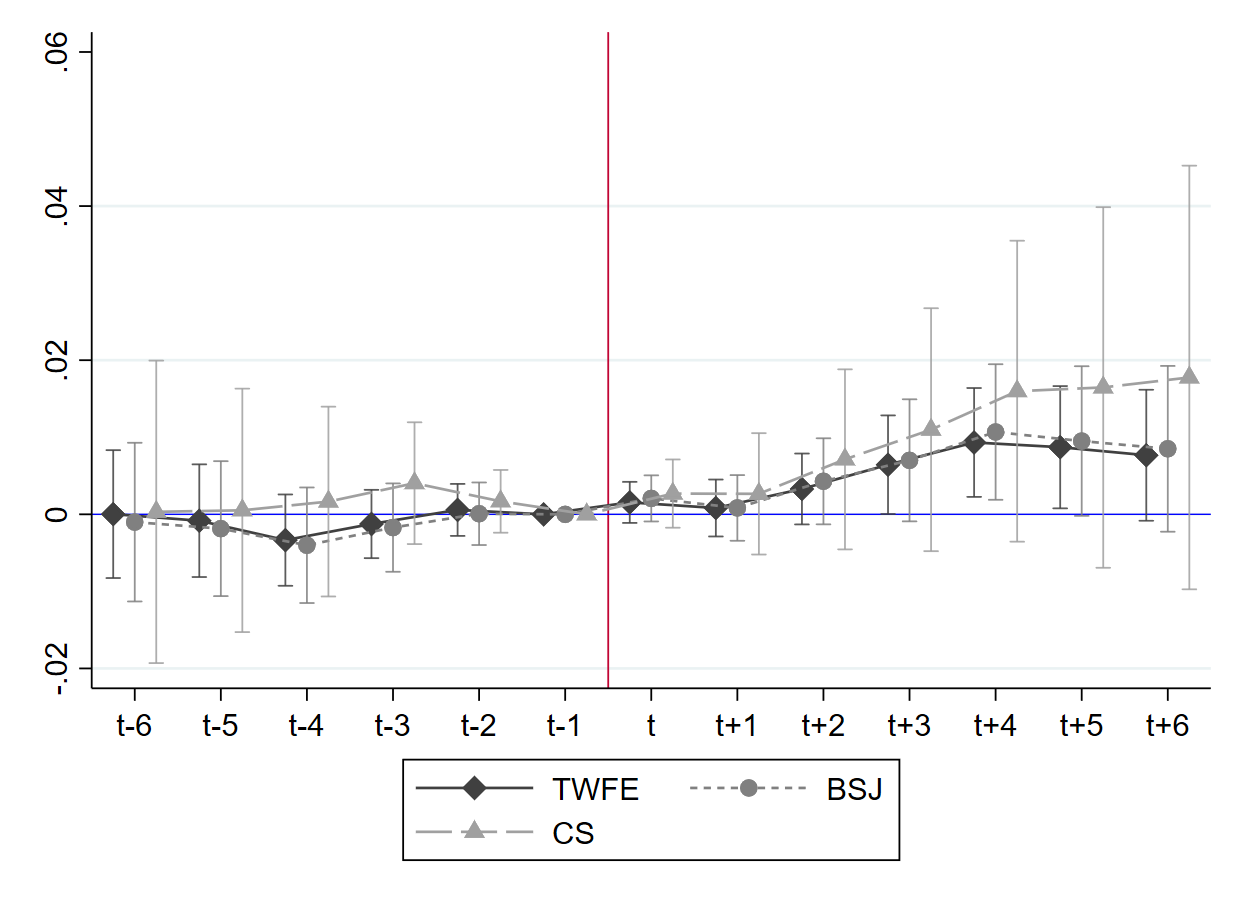}
        \caption{Rival stores}
    \end{subfigure}
\label{fig:alt_est}
\par 
\rule{\textwidth}{0.8pt}
\begin{minipage}[h]{\textwidth}
\medskip
\footnotesize \emph{Notes:} This figure displays the cumulative treatment effects ($E_L$) $L$ months after a crime on store price levels, using different alternative estimators. The effects are shown with corresponding 90\% confidence intervals based on standard errors, which allow for clustering at the store level. Coefficients are interpretable as percentage increases in outcome levels relative to the month before a crime incident. Panel A presents results for victimized stores, while Panel B focuses on rival stores.  The black line (TWFE) represents cumulative effects estimated with the canonical two-way fixed effects estimator. The dotted line (BSJ) shows estimates from the imputation estimator proposed by \citet{Borusyak2021}, and the light grey line (CS) uses the estimator developed by \citet{Callaway2021}.

\end{minipage}
\end{figure}

Figure \ref{fig:alt_est} presents the cumulative treatment effects for the alternative estimators. For victimized stores, the effect size three months after the incident is approximately 1.5\% using the TWFE estimator and 1.7\% with the BSJ estimator (Panel A, Figure \ref{fig:alt_est}). Both coefficients are statistically significant at the 1\% level and remain at this elevated level six months after the incident, which aligns with the estimates in our main specification. As expected, we estimate larger effects in the CS specification, with prices approximately 3.0\% higher three months after the crime, followed by a slight upward trend over time. Although the standard errors are larger for the CS estimator, its statistical significance is comparable to that of the other estimators. 

As anticipated, the treatment effects are slightly smaller than those in our main specification for rival stores. Similar to the main results, effects materialize with some delay, becoming statistically significant at the 5\%-level after four months in the TWFE and BSJ specifications, with an effect size of around 1.0\%.  Again, estimates are slightly higher in the CS specification but statistically insignificant due to higher standard errors. We find no statistically significant pre-trend coefficients in either the victimized or rival store specifications. Overall, the evidence from the alternative estimators supports our conclusion that the effects of crime on prices are robust and not dependent on the choice of estimator.

\newpage

\section{Additional robustness checks}\label{sec:appendix_robustness}

\subsection{Placebo tests}
We conduct placebo tests to check for a potential violation of the parallel trends assumption. We shift the treatment assignment forward by 12 months and estimate our main empirical equation (equation \ref{eq:strategy_1}) with leads and lags based on this placebo treatment assignment. 
If the placebo treatment timing were to produce large and and statistically significant treatment effects, then this would cast doubt on our parallel trends assumption.

\begin{figure}[!htbp]
\caption{Placebo tests}
\centering
	\includegraphics[width=.5\textwidth]{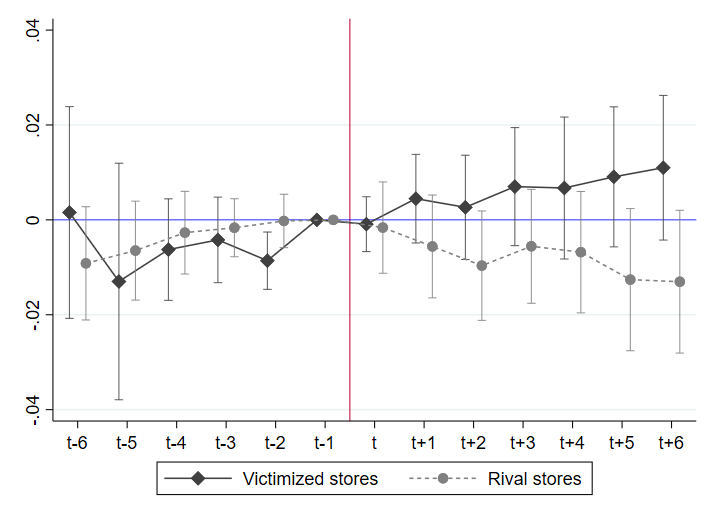}
 \label{fig:placebo}
\par 
\rule{\textwidth}{0.5pt}
\begin{minipage}[h]{\textwidth}
\medskip
\footnotesize \emph{Notes:} The figure shows the cumulative effect of crime incidents on prices, ($E_L$), $L$ months after a crime, along with corresponding 90\% confidence intervals based on standard errors clustered at the store level. Compared to our main specification, we assign a placebo treatment 12 months prior to a store's actual treatment date. The black line depicts the cumulative effects of crime on prices at victimized stores, while the grey line represents rival stores. 
\end{minipage}
\end{figure}

We report our placebo test estimates in Figure \ref{fig:placebo}. The figure indicates that when the treatment timing is moved forward by 12 months, the treatment effects on both victimized and rival stores are negligible and not statistically significant.
Notably, victimized stores exhibit a minor positive effect at higher lags, whereas rival stores show a slight negative post-trend, but all coefficients remain statistically insignificant at the 10\% level. Overall, the placebo test results support the validity of the parallel trends assumption.

\subsection{Longer event window}\label{sec:robust_window}

\begin{figure}[!htbp]
\caption{Effect of crime on store outcomes with extended event window}
\label{fig:longer_window}
\centering
	\begin{subfigure}{.7\textwidth}
	\centering
	    \includegraphics[width=.7\linewidth]{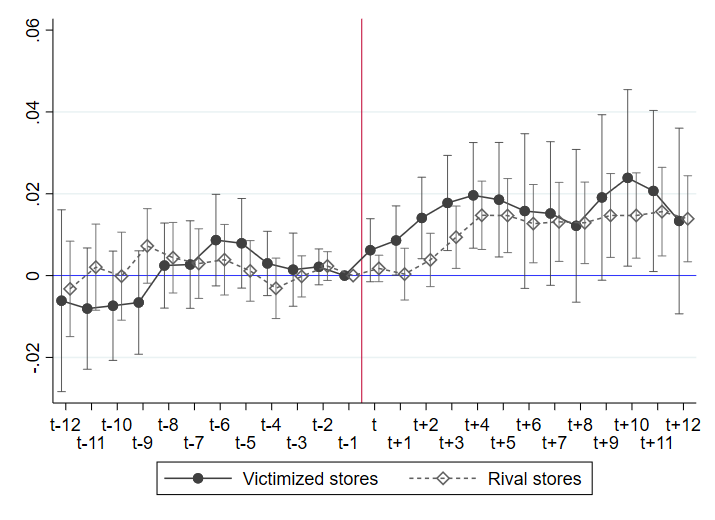}
	\caption{Price level}
    \end{subfigure}\hfil
	\begin{subfigure}{.45\textwidth}
	\centering
	    \includegraphics[width=\linewidth]{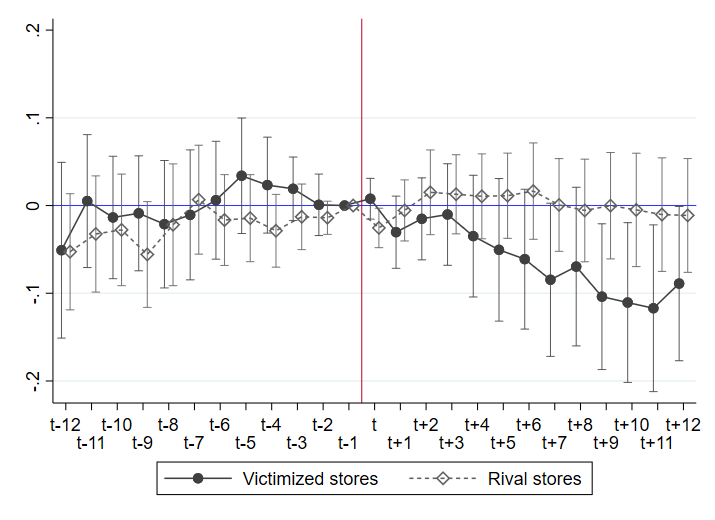}
		\caption{Quantity sold}
    \end{subfigure}\hfil
	\begin{subfigure}{.45\textwidth}
	\centering
	    \includegraphics[width=\linewidth]{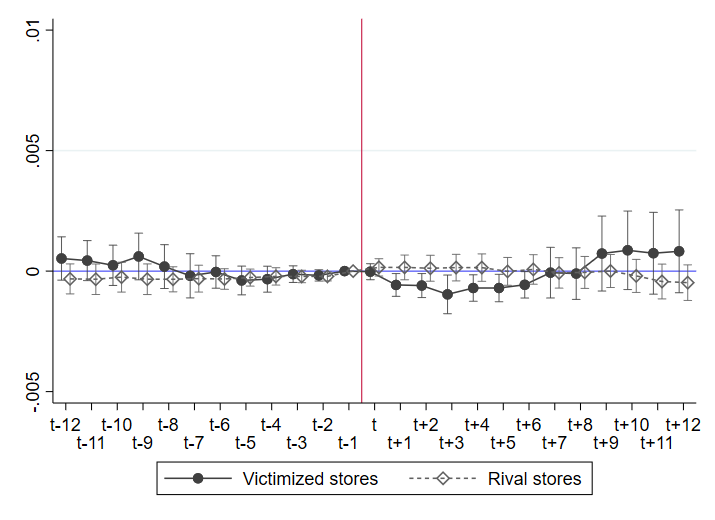}
     \caption{Wholesale cost}
	\end{subfigure}\hfil
\par 
\rule{\textwidth}{0.4pt}
\begin{minipage}[h]{\textwidth}
\medskip
\footnotesize \emph{Notes:} Each panel shows the cumulative treatment effects ($E_L$) $L$ months after a crime on different outcomes, along with corresponding 90\% confidence intervals based on standard errors clustered at the store level. Compared to the main specification, we extend the event window from 11 to 19 months.  Coefficients are interpretable as percentage increases in outcome levels relative to the month before a crime incident. The black line depicts the cumulative effects of crime on outcomes at victimized stores, while the grey line represents rival stores.  The dependent variables are: store-level price index (Panel A), store-level quantity index (Panel B), and store-level wholesale cost index (Panel C).
\end{minipage}
\end{figure}

Figure \ref{fig:longer_window} illustrates that our main results carry over when we extend the event window from 12 months to 18 months. Prices at victimized stores rise for three months following a crime incident and maintain a higher level for nine months. Rival stores experience similar price increases, albeit with a two-month delay. The positive treatment effects remain intact beyond the original 12-month event window, reinforcing our empirical strategy's assumption of constant treatment effects outside this period. 

The effects on quantity sold (Panel B) are more volatile than those for price, reflecting the much larger variance for the quantity indexes compared to the other indexes. Pre-treatment, the effects at both victimized and rival stores are insignificant and show no significant differences. Interestingly, several months post-crime, we notice a decrease in quantities sold at victimized stores. This reduction is likely due to consumers shifting away from these stores in response to the higher prices. Initially, demand at these stores appears inelastic, but as consumers adjust to the increased prices, demand decreases as expected.

 Panel C shows little effect on wholesale prices, as in the main section. Taken together, the results from Figure \ref{fig:longer_window} illustrate that treatment effects materialize and stabilize well before $t+5$, which validates our choice of an 11-month event window for our main analysis in Section \ref{sec:results}.

\subsection{Sensitivity to outliers}\label{sec:outliers}

\begin{figure}[!htbp]
\caption{Effect of crime incident on prices and quantity sold, winsorized}
\label{fig:wins}
\centering
    	\begin{subfigure}{.45\textwidth}
	\centering
	    \includegraphics[width=\linewidth]{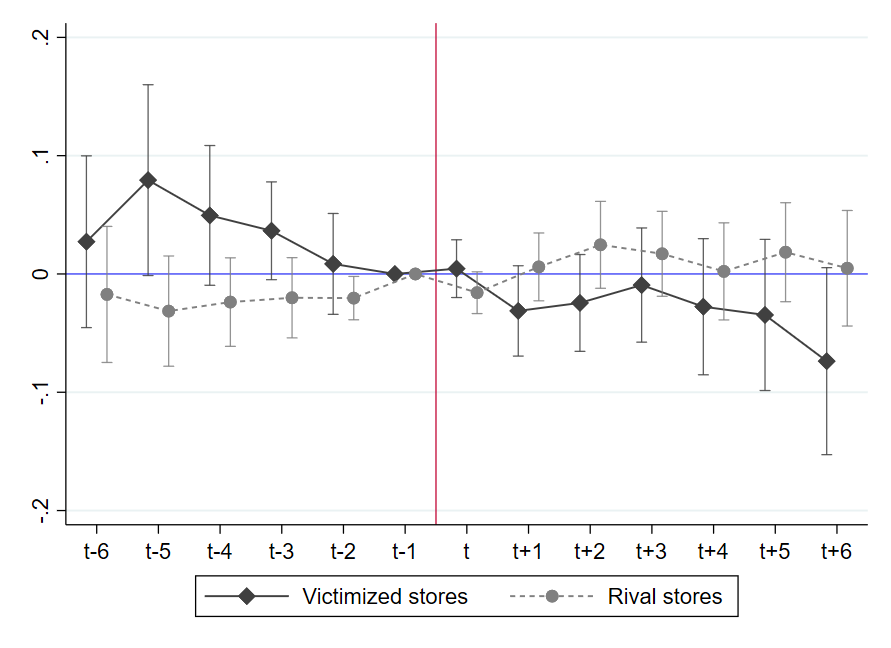}
	\caption{Quantity sold}
    \end{subfigure}
	\begin{subfigure}{.45\textwidth}
	\centering
	    \includegraphics[width=\linewidth]{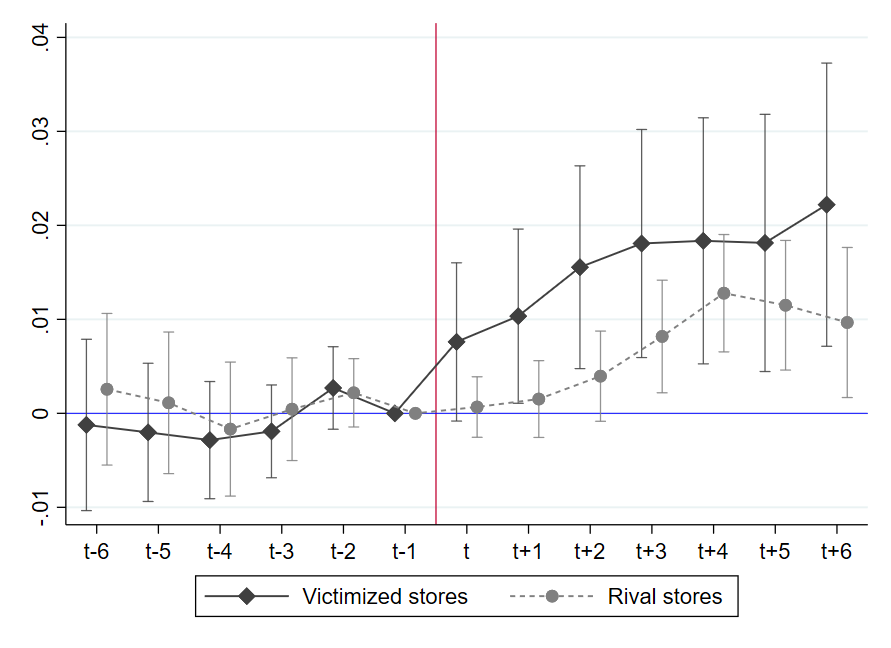}
		\caption{Prices}
    \end{subfigure}
\par
\rule{\textwidth}{0.4pt}
\begin{minipage}[h]{\textwidth}
\medskip
\footnotesize \emph{Notes:} The figures show cumulative treatment effects, ($E_L$), $L$ months after a crime for victimized stores and for rival stores, respectively, along with corresponding 90\% confidence intervals based on standard errors clustered at the store level. Panel A shows the effect on quantity sold when the quantity index is winsorized by 1\%. Panel B shows the effect on prices when the price index is winsorized by 1\%.
\end{minipage}
\end{figure}

To address potential biases from outliers, we re-estimate our main empirical specification using outcome variables winsorized at the 0.5\% and 99.5\% levels. We report the results in Figure \ref{fig:wins}. Considering the store-level quantity index has a standard deviation ten times larger than the price index and 100 times larger than the wholesale cost index (Table \ref{tab:dep_vars}), the quantity index estimates may be particularly sensitive to outliers. Figure \ref{fig:wins} Panel A shows that winsorizing decreases the standard errors of our rival treatment estimates. Notably, winsorizing our data also mitigates the slight pre-trend for the quantity sold at rival stores while leaving the effects at victimized stores unchanged. This suggests that the slightly significant pre-trend in the quantity specification for rival stores in Figure \ref{fig:main_results} is primarily due to a few outliers. Panel B, corresponding to the estimates in Column 4 of Table \ref{tab:robustness}, indicates that winsorizing the store-level price indexes does not influence the estimated price effects of retail crime, further reaffirming our main results.

\subsection{Store closures}\label{sec:closures}

One concern is that the costs associated with retail crime incidents cause victimized stores to go out of business. If the propensity to drop out of the market following a crime is correlated with unobserved store characteristics (e.g. profitability), then our treatment effect estimates may be biased. We view such attrition as unproblematic for two reasons. First, store FE control for time-invariant factors that influence a store's likelihood of dropping out of the sample. Second, we find that victimized stores are not more likely to drop out of the market compared to non-victimized stores. Of the 57 victimized stores in our main specification, eight drop out before the end of the 46-month sample period, a dropout rate of 0.14. For these eight stores, the average duration between a crime incident and dropping out is 15 months. Only three stores drop out within the first 12 months after an incident, which is equal to 5.3\% of victimized stores. For comparison, 94 of the 450 non-victimized stores drop out during the sample period, a rate of 0.21, or 5.5\% on average per year. For comparison, the annual dropout rate for restaurants is estimated to be about 30\% \citep{parsa2005}. Taken together, we view this as suggestive evidence that crime incidents do not lead to a higher dropout rate.

\newpage

\bibliographystyle{apalike}
\bibliography{crime}

\end{document}